\documentclass[]{aa}

\usepackage{times,epsfig,amssymb,amsmath,natbib}
\newcommand{\chandra}{{\it Chandra}}
\newcommand{\xmm}{{\it XMM-Newton}}
\newcommand{\planck}{{\it Planck}}
\newcommand{\nh}{$n_{\rm H}$}

\newcommand{\mos}{MOS}

\usepackage[toc,page]{appendix}
\usepackage{tabularx}
\usepackage{braket}

\usepackage{natbib,twoopt}

\begin{document}

\title{The XMM Cluster Outskirts Project (X-COP): Thermodynamic  properties of the Intracluster Medium out to $R_{200}$ in Abell 2319}
\titlerunning{Abell 2319}

\author{V. Ghirardini\inst{1,2}, S. Ettori\inst{2,3}, D. Eckert\inst{4,5}, S. Molendi\inst{6}, F. Gastaldello\inst{6}, E. Pointecouteau\inst{7,8}, G. Hurier\inst{9,10}, H. Bourdin\inst{11,12}}
\authorrunning{V. Ghirardini}

\institute{
 Dipartimento di Fisica e Astronomia Universit\`a di Bologna, Via Piero Gobetti, 93/2, 40129 Bologna, Italy
 \and INAF, Osservatorio Astronomico di Bologna, Via Piero Gobetti, 93/3, 40129 Bologna, Italy
 \and INFN, Sezione di Bologna, viale Berti Pichat 6/2, 40127 Bologna, Italy
 \and Max-Planck Institut f\"ur Extraterrestrisch Physik, Giessenbachstrasse 1, 85748 Garching, Germany
 \and Department of Astronomy, University of Geneva, ch. d'Ecogia 16, 1290 Versoix, Switzerland
 \and INAF - IASF-Milano, Via E. Bassini 15, 20133 Milano, Italy
\and CNRS; IRAP; 9 Av. colonel Roche, BP 44346, F-31028 Toulouse cedex 4, France
\and Universit\'e de Toulouse; UPS-OMP; IRAP; Toulouse, France
\and Centro de Estudios de Fisica del Cosmos de Aragon, Plaza San Juan 1, Planta-2, 44001, Teruel, Spain
\and Institut d'Astrophysique Spatiale, CNRS (UMR8617) Universit\'e Paris-Sud 11, Batiment 121, Orsay, France
\and Harvard Smithsonian Centre for Astrophysics, 60 Garden Street, Cambridge, MA 02138, USA
\and Dipartimento di Fisica, Universit\`a degli Studi di Roma Tor Vergata, via della Ricerca Scientifica, 1, I-00133 Roma, Italy
}

\mail{vittorio.ghirardini2@unibo.it}
\abstract
{}
{We present the joint analysis of the X-ray and Sunyaev-Zeldovich signals in Abell~2319, the galaxy cluster with the highest signal-to-noise ratio in SZ \planck\ maps and that has been surveyed within our \xmm\ Cluster Outskirts Project (X-COP), a very large program which aims to grasp the physical condition in 12 local ($z<0.1$) and massive ($M_{200}$ $>$ $3\times 10^{14} M_{\odot}$) galaxy clusters out to $R_{200}$ and beyond. 
}
{We recover the profiles of the thermodynamic properties by the geometrical deprojection of the X-ray surface brightness, of the SZ comptonization parameter, and accurate and robust spectroscopic measurements of the gas temperature, out to 3.2 Mpc (1.6 $R_{200}$), 4 Mpc (2 $R_{200}$), and 1.6 Mpc (0.8 $R_{200}$), respectively.
We resolve the clumpiness of the gas density 
to be below 20 per cent over the entire observed volume. 
We also demonstrate that most of this clumpiness originates from the ongoing merger and can be associated to large-scale inhomogeneities 
(the ``residual'' clumpiness).
We estimate the total mass through the hydrostatic equilibrium equation. 
This analysis is done both in azimuthally averaged radial bins and in eight independent angular sectors, enabling us 
to study in details the azimuthal variance of the recovered properties. 
}
{
Given the exquisite quality of the X-ray and SZ datasets, their radial extension and their complementarity, we constrain at $R_{200}$ the total hydrostatic mass,
modelled with a Navarro-Frenk-White profile, with very high precision ($M_{200} = 10.7 \pm 0.5^\text{stat.} \pm 0.9^\text{syst.} \times 10^{14} M_\odot$). 
We identify the on-going merger and how it is affecting differently the gas properties in the resolved azimuthal sectors.
We have several indications that the merger has injected a high level of non-thermal pressure in this system: 
the clumping free density profile is above the average profile obtained by stacking Rosat/PSPC observations; 
the gas mass fraction recovered using our hydrostatic mass profile exceeds the expected cosmic gas fraction beyond $R_{500}$;
the pressure profile is flatter than the fit obtained by the \planck\ collaboration;
the entropy profile is flatter than the mean one predicted from non-radiative simulations; 
the analysis in azimuthal sectors has revealed that these deviations occur in a preferred region of the cluster.
All these tensions are resolved by requiring a relative support of about 40 per cent from non-thermal to the total pressure at $R_{200}$.
}
{}


\maketitle

%----------------------------------------------------------------------------------------
%	ARTICLE CONTENTS
%----------------------------------------------------------------------------------------

\section{Introduction}
Cosmic structures evolve hierarchically from the primordial density fluctuations into larger and larger systems under the action of gravity. Galaxy clusters are the largest bound structures of the universe and the most recent products of structure formation. Baryons fall into the gravitational potential of dark matter halos and heat up to a temperature of the order of few millions Kelvin, emitting in X-rays mostly through \textit{bremsstrahlung} process. In the last few years our knowledge on the physical condition of the  intra cluster medium (ICM) has significantly improved through the study of the Sunyaev-Zel'dovich effect \citep[SZ;][]{SZ}. It arises when cosmic microwave background photons (CMB) are scattered by the free electrons of the ICM. The observed distortion of the CMB spectrum is directly proportional to the thermal electronic pressure integrated along the line of sight. This linear dependence implies that the SZ signal decreases more slowly than the X-ray signal, which depends quadratically on the density.
The assumption that the ICM is fully thermalized and in hydrostatic equilibrium is usually made in several studies \citep[see][for a review]{ettori+13}. However this assumption might not be valid in cluster outskirts, where the relative contribution of non-thermal pressure to the total one might not be negligible \citep[e.g.][]{battaglia+12}.

The matter distribution in the outskirts of galaxy clusters is expected to be clumpy \citep{nagai+11,vazza+13} and asymmetric \citep{eckert+12, roncarelli+13}, with substantial contribution from non-thermal physics, like turbulence and bulk motion \citep{vazza+11}, cosmic rays \citep{pfrommer+07}, and magnetic fields \citep{dolag+99}. Gas clumping plays an important role in the outer parts of galaxy clusters. 
\cite{zhuravleva+13} showed that the density distribution inside a given shell surrounding the cluster center can be described by a log-normal distribution modified by the presence of a high density tail produced by the presence of clumps. It was shown that the median of this distribution coincides with the mode of the log-normal, while the mean is biased high due to the presence of clumps. Observationally \cite{eckert+15} have confirmed this result: they concluded that the median method is able to recover the true gas density profile when inhomogeneities are present.

The XMM cluster outskirts project \citep[X-COP;][]{xcop} is a very large programme on \xmm\ which aims to increase significantly our knowledge on the physical conditions in the outskirts of galaxy clusters. Thirteen local and massive systems have been selected on the basis of their high signal-to-noise ratio (SNR) in the \planck\ survey, and reported in the first catalog \citep{catalog}.

In this paper, we focus on Abell 2319, the most significant SZ detection in the first \planck\ catalog, with a SNR of 49.0 in the second \planck\ catalog \citep{catalog2}.
Abell 2319 is a very hot and massive cluster at low redshift \citep[z=0.0557;][]{struble+99}. Its galaxy distribution indicates this is a merger of two main components with a 3:1 mass ratio, the smaller system being located ~10$'$ north of the main structure \citep{oegerle+95}. The cluster exhibits a prominent cold front SE of the main core \citep{ghizzardi+10} and a giant radio halo \citep{farnsworth+13,storm+15}. 

The paper is organised as follows: in Section~2, we describe the reduction and analysis of X-ray data, from background modeling to spatial and spectral analysis; in Section~3, we present the data reduction and analysis of the \planck\ SZ data; in Section~4, we show the reconstructed profiles of the thermodynamic quantities, describe their properties,and discuss the different methods adopted to solve the hydrostatic equilibrium equation; in Section~5, the analysis in azimuthal sectors is illustrated. The gas mass fraction and the hydrostatic bias are shown in Section~6.
The summary of our main findings and our conclusions are discussed in Section~7.

Throughout this paper, we assume a $\Lambda$CDM cosmology with $\Omega_\Lambda = 0.7$, $\Omega_m = 0.3$ and $H_0 = 70 \; $km/s/Mpc, $E(z) = \sqrt{\Omega_m (1+z)^3 + \Omega_\Lambda}$. 
At the redshift of A2319, 1 arcmin corresponds to approximatively  64.9 kpc. Uncertainties are provided at the 1$\sigma$ confidence level.

In the following, we refer to, and plot as reference, some characteristic radii, $R_{500}$ = 1368 kpc and $R_{200}$ = 2077 kpc, that are defined at the overdensities of $\Delta$ = 500 and 200, respectively, with respect to the critical value $\rho_c = 3H_0^2 \frac{E(z)^2}{8 \pi G}$ and using the hydrostatic mass profile (see Table~\ref{table:fit_NFW} in Section~\ref{sec:mhyd}).

\begin{table*}[t]
\begin{center}
\begin{tabular}{ c  c  c  c  c  c  c  }
\hline
Observation & OBSID & Total [ks] & \mos 1 [ks] & \mos 2 [ks] & \textit{pn} [ks] & inFOV/outFOV  \\
\hline
Center & 0600040101 & 58.3 & 48.3 & 49.3 & 41.1 & 1.215 \\
North & 0744410101 & 36.0 & 23.8 & 24.5 & 19.4 & 1.132 \\
South & 0744410301 & 31.0 & 13.8 & 14.0 & 7.0 & 1.406 \\
East & 0744410401 & 41.9 & 14.4 & 15.4 & 9.5 & 1.346 \\
West & 0744410201 & 37.5 & 23.4 & 25.1 & 9.8 & 1.152 \\
Outside & 0743840201 & 15.0 & 12.1 & 12.3 & 5.7 & 1.261 \\
Outside2 & 0763490301 & 18.0 & 12.9 & 12.8 & 9.0 & 1.253 \\
\hline 
\end{tabular}
\end{center}
\caption{Pointing name, OBSID, total exposure time and clean exposure time for \mos 1, \mos 2, and \textit{pn}, and inFOV/outFOV ratio, for the seven observations used in this work. All the observations are obtained using the medium filter, the full frame science mode for \mos\, and extended full frame for \textit{pn}.}
\label{table:obsid}
\end{table*}

\section{\xmm\ Analysis}
\label{sec:xmm}

X-ray spatial and spectral analysis provide a direct probe of density and temperature of the ICM. 
However, the X-ray background needs to be modelled very accurately if we want to obtain accurate measurements 
in the outskirts of galaxy clusters, where the background dominates over the signal.

\subsection{Data reduction}

The \xmm\ Science Analysis System (XMM-SAS v15.0) and the corresponding calibration files have been used to reduce the X-ray data, following the Extended Source Analysis Software analysis scheme \citep[ESAS;][]{esas}.
The presence of anomalous individual CCDs is also taken into account, removing them from the analysis.
Soft proton flares periods are filtered out using the ESAS tasks \textit{mos-filter} and \textit{pn-filter}, therefore obtaining clean events files. 
The ESAS procedure \textit{cheese} is adopted in order to mask point sources which contaminates the field of view.

Spectra, effective areas and response files (ARF and RMF respectively) for the selected regions are extracted using the ESAS tasks \textit{mos-spectra} and \textit{pn-spectra}.

This procedure is applied to all the seven observations we use in the analysis of Abell~2319: an archival central exposure, 
four offset observations (done specifically for the X-COP program), and two other archival exposures pointing just outside the virial radius, which are used to estimate the local sky background. 
Table~\ref{table:obsid} provides some information regarding these observations, like the OBSID, the total and the clean exposure time, and the level of soft protons contamination, obtained comparing the measured count rate in a hard spectral band in the exposed and unexposed part of the field of view\citep[inFOV/outFOV,][]{lm+08}.

\subsection{Particle background modeling}

We extracted count images from the cleaned event files in the [0.7 - 1.2] keV energy band, where we expect to maximize the signal-to-background ratio \citep[e.g.][]{ettori+10,ettori+11}. In Appendix A we present our method to model the 2D distribution and intensity of the non X-ray background (NXB), distinguishing its different components, and computing the total NXB image in the required energy band. We briefly summarize the main steps here.

The \xmm\ NXB is made of three separate components: the quiescent cosmic-ray induced particle background (QPB), the soft protons (SP), and a stable quiescent component (QC), whose origin is yet unknown \citep{salvetti+17}. To model the QPB, we used the unexposed corners of the EPIC cameras to estimate the QPB level in each observation. We then use filter-wheel-closed observations to model the spatial distribution of the QPB and renormalize the filter-wheel-closed data to match the count rate measured in the unexposed corners. 

The residual contribution after subtraction of the QPB is split between the QC and SP components.  In Appendix A we describe our method to take the relative contribution of these two components into account. Briefly, we measure radial surface brightness profiles for a large sample of 495 blank-sky pointings and we optimize the relative contribution of these components as a function of the estimated SP contamination, imposing that the residual surface brightness profiles be consistent with a flat curve in the energy band of interest. This procedure leads to an accurate modeling of the SP contamination, as shown in Fig. \ref{fig:calibration}. The deviations from a flat profile are found to be at the level of less than 5\%, thus for the remainder of the paper we adopt a a systematic uncertainty of 5\% of the NXB level on the measured surface brightness profile.

\begin{figure}[t]
\centering
\includegraphics[width=0.5\textwidth]{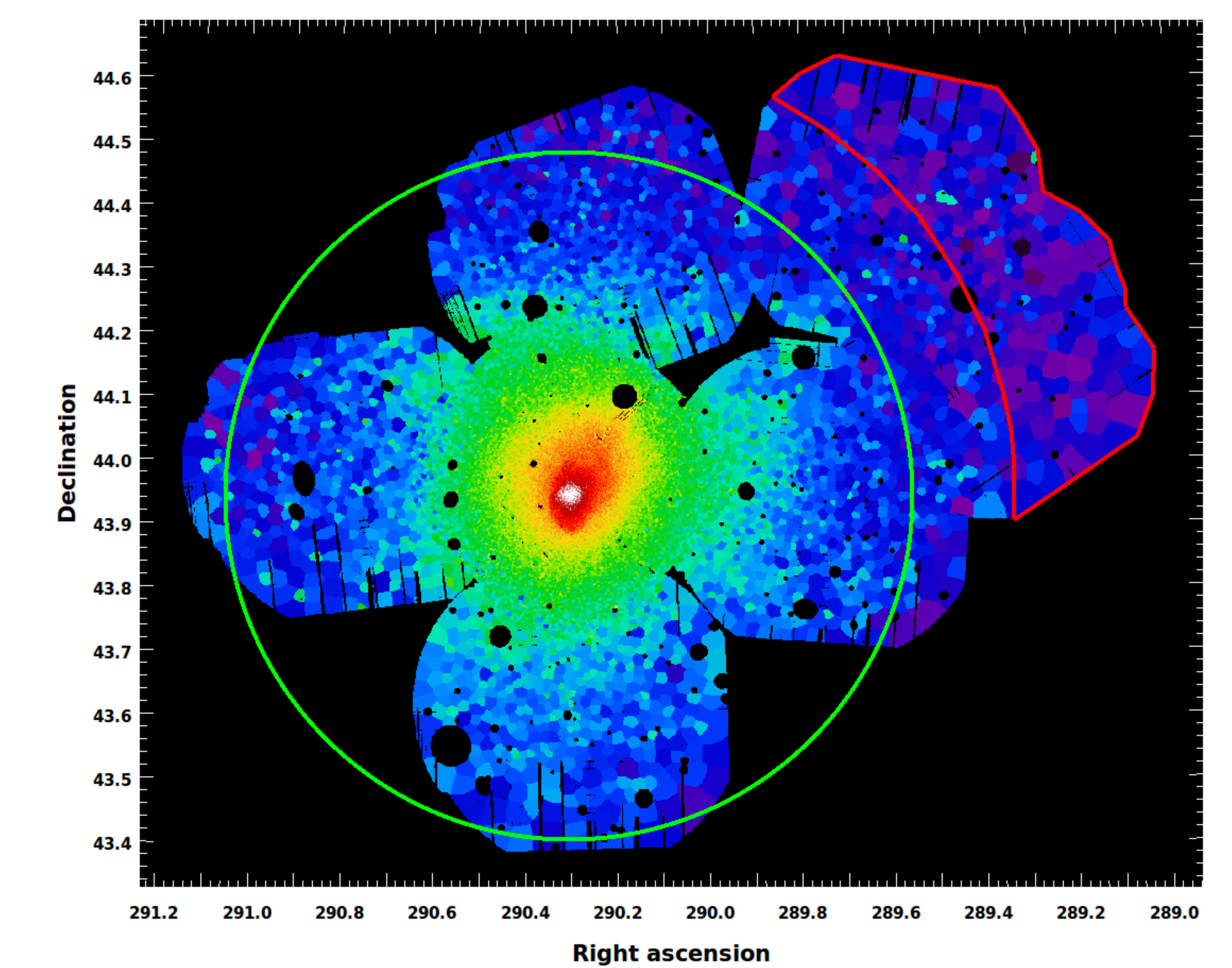}
\caption{Mosaicked and Voronoi tessellated image of A2319, in the energy band [0.7-1.2] keV, corrected for the particle background. The red region is the one chosen for the estimate of the local sky background. The green circle represents the location of $R_{200}$.}
\label{figure:VoronoiA2319}
\end{figure}

\subsection{Spatial analysis}
\label{sec:spatial}
We combine the results from \mos 1, \mos 2 and \textit{pn} and mosaiced all seven observations into one single image.

We filter our image one more time by using the \textit{Chandra} tool \textit{wavdetect}, in order to find the remaining point sources which contaminates the field of view but were missed by the ESAS task \textit{cheese}. 
Indeed this procedure has some difficulties finding some obvious point sources, lying near the gaps of the CCDs, or not found due to the parameters adopted.

A Voronoi tessellation algorithm \citep{voronoi} was applied on the mosaicked count image to create an adaptively binned surface-brightness map with a minimum of 20 counts per bin. The resulting Voronoi tessellated count rate map for A2319 is shown in Fig.~\ref{figure:VoronoiA2319}.

In order to analyze spatially the cluster's image, we choose a background region located as far as possible from the cluster's center in order to have a good estimate of the sky background, minimizing the cluster contamination. We choose all the pixels in the image beyond 42 arcmin from the cluster's center to be the region where we estimate the local sky background (the red region in Fig.~\ref{figure:VoronoiA2319}).
The background level is just the mean count rate in this region: (1.82 $\pm$ 0.06)~$\times$ $10^{-4}$~cts~s$^{-1}$~arcmin$^{-2}$, in the energy band [0.7-1.2] keV
(or, converting in flux using a power law spectral model with photon index 1.41: 1.46 $\pm$ 0.05 ~$\times$ $10^{-15}$~erg~s$^{-1}$~cm$^{-2}$~arcmin$^{-2}$).

\begin{figure}[t]
\centering
\includegraphics[width=0.5\textwidth]{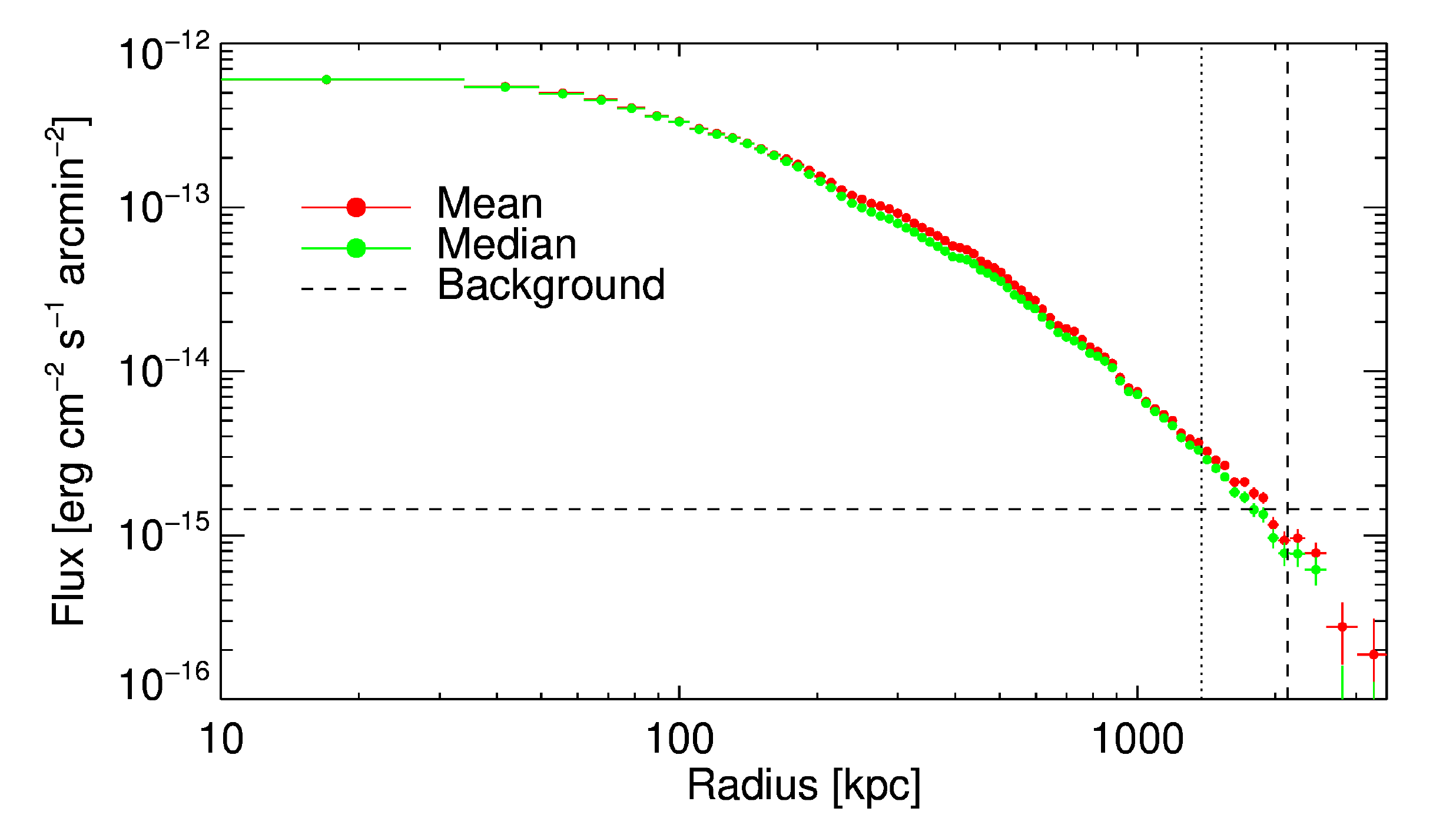}
\caption{Background subtracted surface brightness profiles in the [0.7-1.2] keV energy band, using the mean and median methods, red and green points respectively. The sky background level is shown with a horizontal dashed line. The vertical dotted and dashed lines represents the location of $R_{500}$ and $R_{200}$, respectively.}
\label{figure:SBA2319}
\end{figure}

The background-subtracted surface brightness profile is then computed in annular regions. 
We choose the annuli such that the total amount of net count rate in the [0.7-1.2] keV energy band is the same in all the regions. This choice ensures comparable statistics in all annuli.
Using the ARF and RMF files for \mos 2 (since the combined image was in units of \mos 2), we are able to convert from count rates to fluxes.
As shown in Fig.~\ref{figure:SBA2319}, 
we have also evaluated the surface brightness from both the azimuthal mean and the azimuthal median of the brightness distribution.
Following the analysis in hydrodynamical simulations on the effects of the densest substructures on the average gas density profile \citep{zhuravleva+13,roncarelli+13}, 
\cite{eckert+16} showed that the median is indeed less biased than the mean, since it is a more robust estimator since it is unaffected by compact X-ray substructures filling a small fraction of the total volume, and that the ratio between mean and median can be used to estimate the relative impact of the detected clumps, 
providing an estimate of the level of gas clumpiness. 

The electron density is then recovered using two different techniques:  the onion-peeling technique \citep[e.g.][]{ettori+10}, and the multiscale technique  \citep{eckert+16}.
Both assume the emission to be spherically symmetric. The latter technique requires also a super-parametric functional form for the density profile, decomposing the surface brightness in a very large number of $\beta$-models which can be individually deprojected.
We obtain electron density profiles that are consistent within 0.7$\sigma$, and mean relative deviation of 5\% up to the virial radius (see Fig.~\ref{figure:TKA-2D}).

\subsection{Spectral Analysis}
\label{subsect:spectral}

In order to recover the electron temperature and metal abundance of the X-ray emitting plasma, we perform a spectral analysis by fitting the spectra with an absorbed thermal component
in the energy range [0.5-11.3] keV, and excluding the spectral regions with strong instrumental emission lines ([1.2-2.0] keV for \mos\ and [7.1-9.2] keV for \textit{pn}), using XSPEC \citep{xspec}. 

We extract spectra in 19 concentric annuli, defined in order to reach an approximately constant count rate in the [0.7-1.2] keV energy band. 
The number of net counts in the [0.5-11.3] keV energy band and the signal to background ratio are listed in Table~\ref{table:cashstat}.
We extract spectra also from the background region indicated in Fig.~\ref{figure:VoronoiA2319}, which is the same region used to estimate the local sky background for the spatial analysis and where there is no evidence for cluster emission.

\begin{figure}[t]
\includegraphics[width=0.5\textwidth]{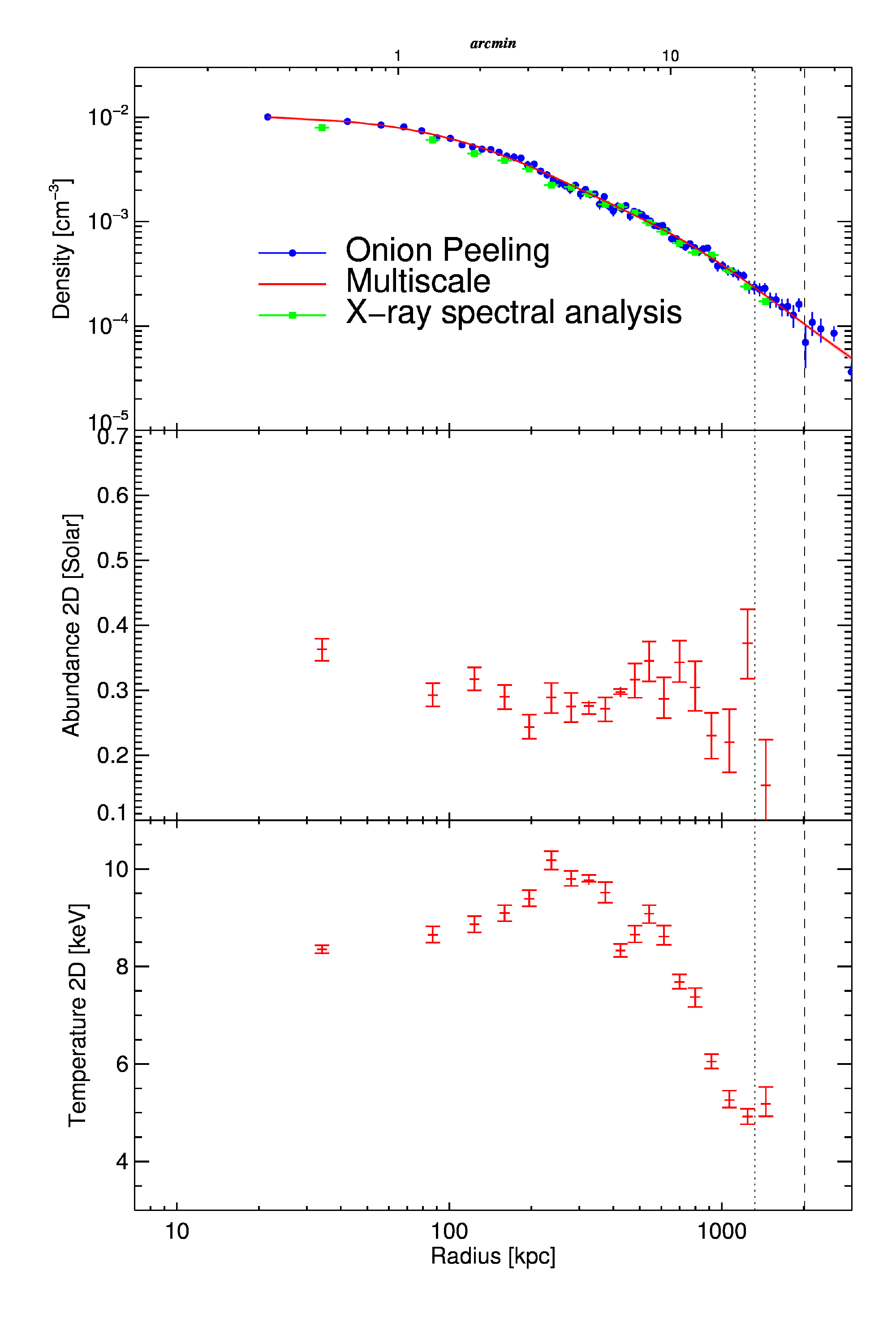}
\caption{(Top) Density profile recovered from the median surface brightness profile, using the multiscale and the onion peeling technique, red line and blue points respectively. The density coming from the spectral analysis is also shown here, green points.
Abundance (Middle) and temperature (Bottom) from the fitting of the spectra in 19 annular regions. The vertical dotted and dashed line indicates the location of $R_{500}$ and $R_{200}$ respectively.}
\label{figure:TKA-2D}
\end{figure}

To model spectrally the NXB component, we follow \cite{lm+08}, modeling the spectra from the unexposed region of the instruments, the QPB component, using a broken power law in the energy range [0.5 - 11.3] keV for \mos\, and in [0.5 - 14.0] keV, excluding the energy bands [7.1 - 9.2] keV where we observe strong instrumental emission lines for \textit{pn}. We fit the background spectra, produced by the ESAS tasks \textit{mos-back} and \textit{pn-back}, which yields the unexposed spectrum representative of the QPB component. This fixes the parameters of the QPB background component.

Then, in the source spectrum, we restrict to a hard band, above 5 keV, and we model the remaining particle background component
using a broken power law with shape parameters (i.e. slopes and break energy) fixed accordingly to the results obtained in other works \citep[see][]{kuntz+08,lm+08}, 
leaving only normalization free. We include in the fit a thermal component ({\tt apec} model in the X-ray spectral fitting package  \cite[XSPEC, version 12.9.1; see][]{xspec}, with only normalization free, and using a temperature of 9.6 keV\citep{temp}, redshift of 0.0557\citep{struble+99} and 0.3 solar abundance)
considering that in the hard band the emission from the cluster is small but not negligible. In this way we fix the parameters describing the quiescent component.

We rescale the model particle background NXB, from the whole field of view to the local sky background region, leaving all the parameters of the models fixed. Normalizations are rescaled according to areas, i.e. if a spectrum comes from half of the field of view, normalizations are halved accordingly. This way the instrumental background and the contamination from soft protons are modelled in the background region. 
Then, we model the local sky background, that is the remaining source of emission in the background region. We construct the model using three different components:
\begin{itemize}
\item the Cosmic X-ray Background (CXB), that is modelled as an absorbed power law with photon index fixed to 1.41 \citep{cxb};
\item the local bubble component, that is modelled as an unabsorbed thermal model with temperature free to vary around 0.11 keV \citep[][]{bubble}, 
redshift equal to 0 and fixed solar elemental abundance;
\item the galactic halo component, that is modelled as an absorbed thermal component  with temperature free to vary around 0.22 keV \citep{halo}, redshift fixed to 0, and fixed solar elemental abundance.
\end{itemize}

Using the emission model \textit{tbabs(apec+powerlaw)+apec}, we fit together all the spectra extracted from the background region and obtain the sky components with normalizations and temperatures listed in Table~\ref{tab:sky}, which provide a flux of 1.7 $\times$ $10^{-15}$ erg/s/cm$^2$/arcmin$^2$
in the energy band [0.7--1.2] keV.

\begin{table}[h]
\begin{center}
\begin{tabular}{ c  c  c  }
\hline
Component & Normalization & Temperature \\
\hline
Galactic halo & $1.0$ $\times 10^{-4}$ cm$^{-5}$ & $0.35$ keV \\
Local bubble & $4.1$ $\times 10^{-4}$ cm$^{-5}$ & $0.15$ keV \\
CXB & $1.1$   $\times 10^{-4}$ $\frac{\rm photons}{\rm keV cm^2 s}$ at 1 keV & / \\
\hline 
\end{tabular}
\end{center}
\caption{Components and parameters of the X-ray background adopted in our spectral analysis.}
\label{tab:sky}
\end{table}

\begin{figure}[t]
\begin{center}
\includegraphics[width=0.4\textwidth]{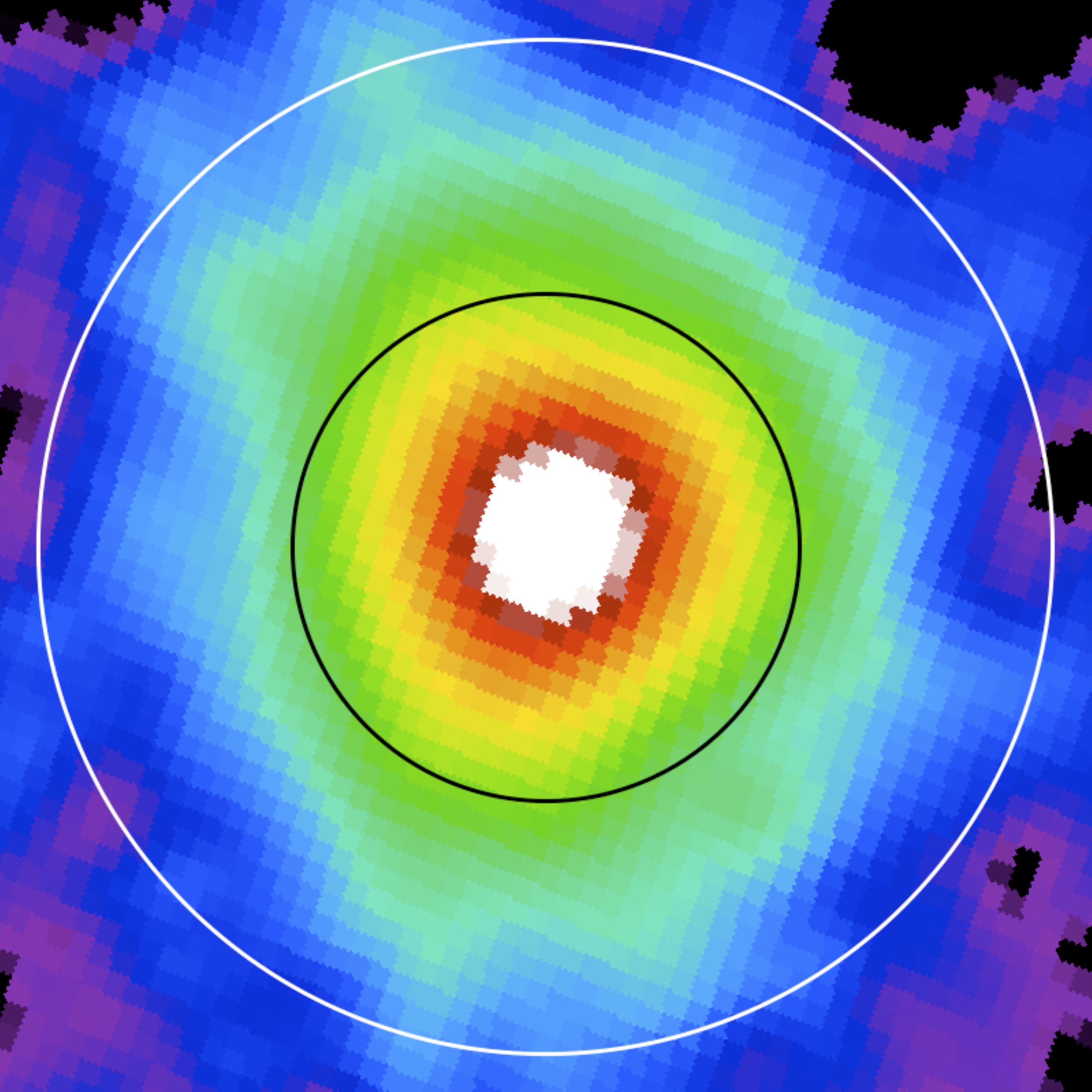}
\end{center}
\caption{Comptonization map of Abell 2319 reconstructed using MILCA \citep[][]{hurier+13}, with an angular resolution of 7  arcmin FWHM. The black and white circles indicates the location of $R_{500}$ and $R_{200}$ respectively.}
\label{fig:ymap}
\end{figure}

Similarly to the background region, 
we extract the spectra in the selected regions, and we rescale the particle background model for the NXB, from the field of of view 
where they were calculated to the specific region of interest with just a change in normalization proportional to the areas. 
We obtain the sky background components from the background regions, rescaled according to the covered areas. 

We fix the particle and sky background. 
The cluster emission is modelled with a thermal component absorbed from our own Galaxy 
(model \textit{tbabs $\cdot$ apec} in XSPEC). 
The gas temperature, abundance, and normalization are free parameters in the spectral fit, whereas the redshift is fixed. 
The galactic hydrogen column density is left 
free to vary between $7.2 \times 10^{20}$ cm$^{-2}$  and $12.8 \times 10^{20}$ cm$^{-2}$,
where the lower value represents the minimum of the Galactic column density due to atomic hydrogen 
\citep[as tabulated in LAB HI Galactic survey in][]{LAB} estimated over the surveyed area, and the higher value indicates the maximum column density over the same area, 
also corrected for molecular hydrogen as suggested in \cite{willingale+13}.

We fitted jointly all the spectra belonging to the same annulus but extracted from different observations using the C-statistics. 
The best-fit parameters are shown in Fig~\ref{figure:TKA-2D} (with goodness of the fit, net counts, signal to background ratio, and best fit \nh\ indicated in Table~\ref{table:cashstat}).

Modelling the ICM emission with a thermal component allows X-ray observations 
to provide a direct probe of the gas electron density, $n_e$. In fact its normalization $K_{apec}$ can be written as:
\begin{equation}
K_{apec} = \frac{10^{-14} \ \text{cm}^{-5}}{4 \pi D_A (1+z)^2} \int_V n_e n_p dV
\label{eq:apecnorm}
\end{equation}
with the proton number density, $n_p$, is proportional to $n_e$ ($n_p \sim 0.8n_e$).

We recover the 3D profiles, temperature and abundance, by adopting the ``onion peeling'' technique \citep[and references therein]{kriss+83, ettori+02}.
Assuming a constant gas density inside each shell, we can rewrite Eq.~\ref{eq:apecnorm} as matrix product
(using ``$\#$'' to indicate it): $K_{apec} \propto V \# n_e^2$, where $V_i^j$ is the geometrical volume of the j$^{th}$ 
shell intercepted by the i$^{th}$ annulus.
By inverting this linear equation, we obtain the electron density inside each shell as $n_e \propto \sqrt{V^{T^{-1}} \# K_{apec}}$.
Values of the temperature and metal abundance in each shell are then obtained as
$Y_{3D} = \frac{V^{Y^{-1}} \# (Y_{2D} \cdot EM)}{V^{Y^{-1}} \# EM}$, where $EM = \int n_e^2 dV$ is the emission measure 
and $Y$ is the quantity of interest \citep[either temperature or metallicity; for a discussion on the systematic effects see][]{ameglio+07}
The errors are estimated through a Monte Carlo process.

\begin{figure}[t]
\includegraphics[width=0.5\textwidth]{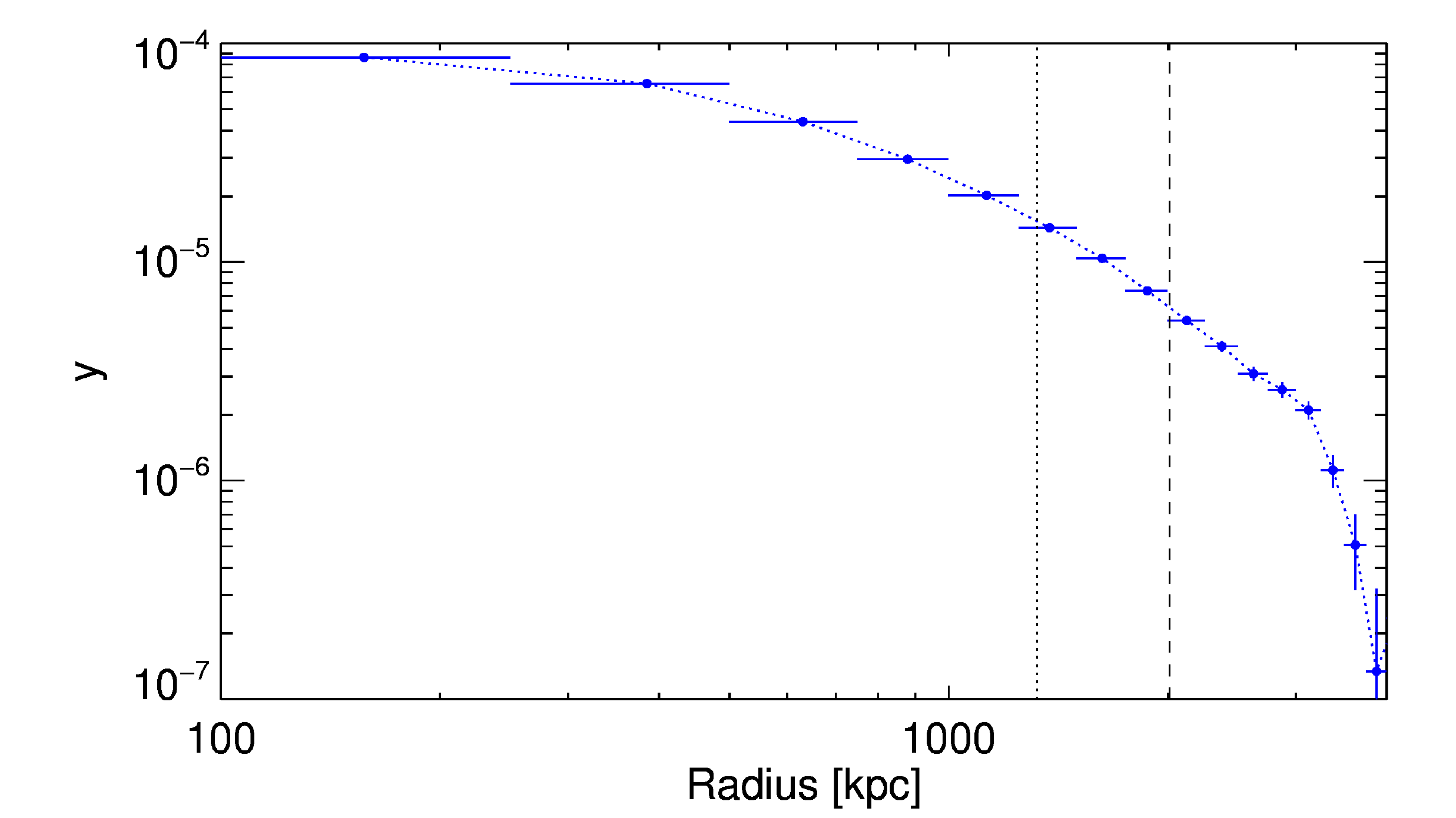}
\caption{Comptonization profile extracted from the SZ map. The vertical dotted and dashed line indicates the location of $R_{500}$ and $2R_{500}$ respectively.}
\label{figure:Planck}
\end{figure}

\begin{figure}[t]
\includegraphics[width=0.5\textwidth]{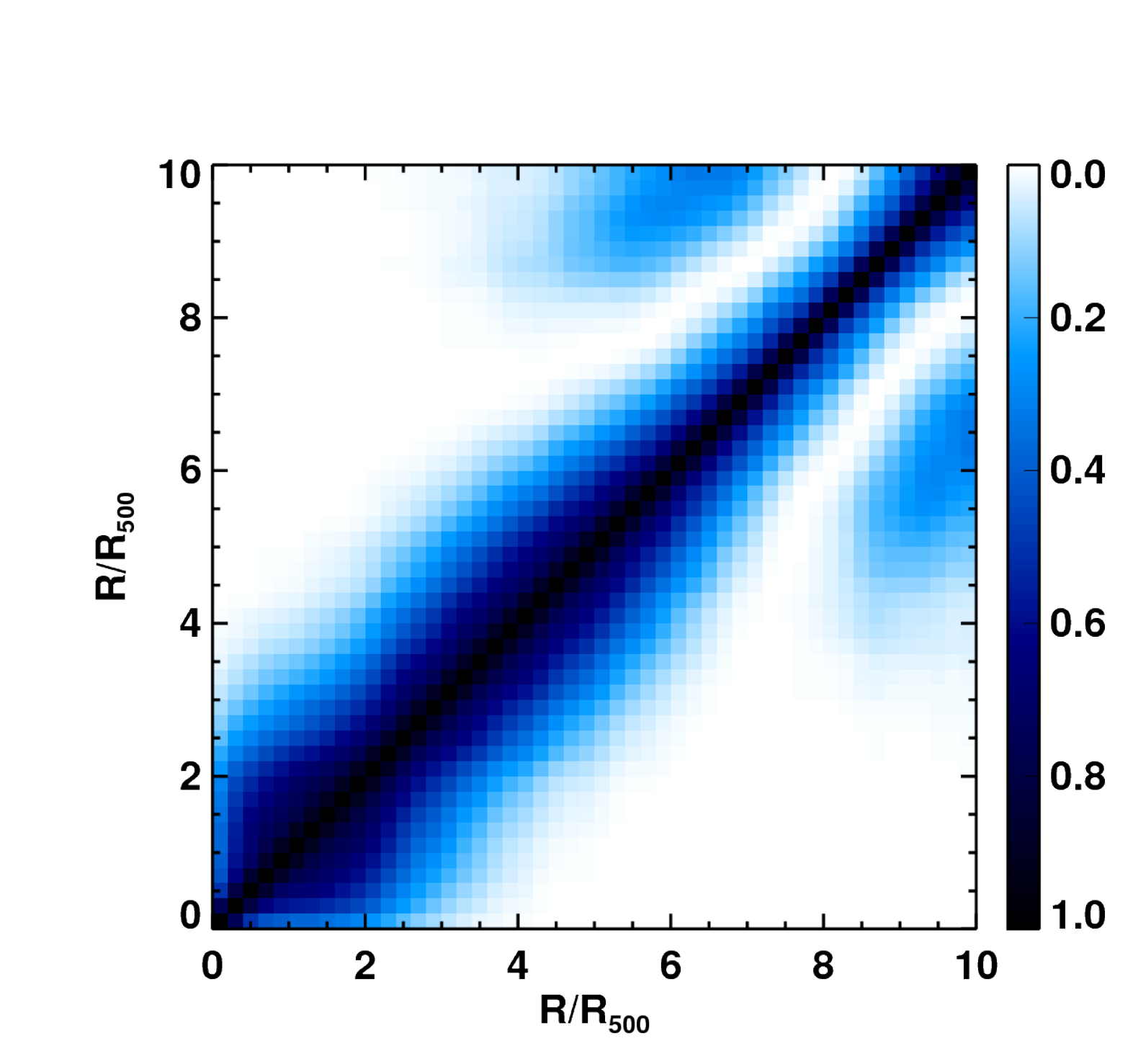}
\caption{\planck\ correlation matrix $\rho_{X,Y}$ for the unbinned comptonization parameter profile.}
\label{fig:planck_corr}
\end{figure}
	
\section{Planck Analysis}

The SZ effect provides a direct measurement of the thermal pressure integrated along the line of sight \citep{SZ}. The dimensionless Comptonization parameter is defined as follow:
\begin{equation}
y(r)=\frac{\sigma_T}{m_ec^2}\int P_e(\ell)d\ell
\label{eq:pressure}
\end{equation}
where the integral is computed along the line of sight, $\ell$, at the radius $r$ from the centre. $\sigma_T$ the Thomson cross section, $m_e$ the mass of the electron, and $c$ the speed of light. 

The pressure profile is recovered from the SZ signal measured in the all-sky survey by the \planck\ mission  \citep{tauber+10, planckI+15}. The SZ signal map is derived from the internal  linear combination  of the six frequency bands of the high frequency instrument \citep[HFI;][]{lamarre+10,planckHFI} on board the \planck\ satellite. More specifically, we made use of the Modified Internal Linear Combination Algorithm \citep[MILCA,][]{hurier+13} which offers the possibility to reconstruct the targeted signal component at various scales contributed differently by the six combined input frequency maps. We therefore reconstruct a $y$-map for A2319 with an angular resolution of 7~arcmin FWHM (see Fig.~\ref{fig:ymap}). 

From the $y$-map, we proceed according to the method used and detailed in \citet{planck+13}.

We extracted the $y$-parameter radial profile of A2319 from our MILCA $y$-map. i.e., the profile is extracted on  a regular grid with bins of width $\Delta\theta/\theta_{500}=0.2$. The local background offset is estimated from the area surrounding the cluster beyond $5\times \theta_{500}=106$~arcmin. The resulting profile is shown in Fig.~\ref{figure:Planck}. 
The pressure profile is then obtained following the real space deconvolution and deprojection regularisation method first described in   \citet{croston+06}, assuming spherical symmetry for the cluster. The correlated errors were propagated from the covariance matrix of the $y$ profile with a Monte Carlo procedure and led to the estimation of the covariance matrix of the pressure profile $P_e(r)$.

Abell~2319 is the highest signal-to-noise ratio SZ  detected cluster in the \planck\ SZ catalogues \citep[$SNR \sim 50$; see][]{catalog, catalog2}. 
Its proximity and its extension makes it fully resolved even at the moderate angular resolution of the \planck\ survey, and its SZ signal extends well beyond $R_{500}$ with at high significance. 
We thereby were able to perform an azimuthal analysis in 8 azimuthally-resolved sectors (see Section~6).
The $y$ and pressure profiles in each sector were obtained as afore-described after masking the $y$-map and its associated error map according to the sector definition.

Due to the moderate angular resolution of the \planck\ survey and the oversampling implied by our sampling of the $y$-map, we introduced co-variance between the individual pixels. It  cascades on the $y$ and pressure profiles computation, hence their respective covariance matrix. 

In Fig.~\ref{fig:planck_corr} we show the correlation matrix between data points, defined as:
\begin{equation*}	
\rho_{X,Y} = \frac{\Sigma(X,Y)}{\sigma^2_X \sigma^2_Y}
\end{equation*}
where $\Sigma$ indicates the covariance matrix.

Consequently, we stress that points of our $y$ and SZ pressure profiles are correlated and that the respective error bars displayed in the figures of this paper represent only the square root of the diagonal of the covariance matrix.
Nevertheless when pressure is used to derive other quantities we make complete use of the whole covariance matrix, and therefore we consider any impact of the \planck\ PSF in our calculations.

\begin{figure}[t]
\includegraphics[width=0.5\textwidth]{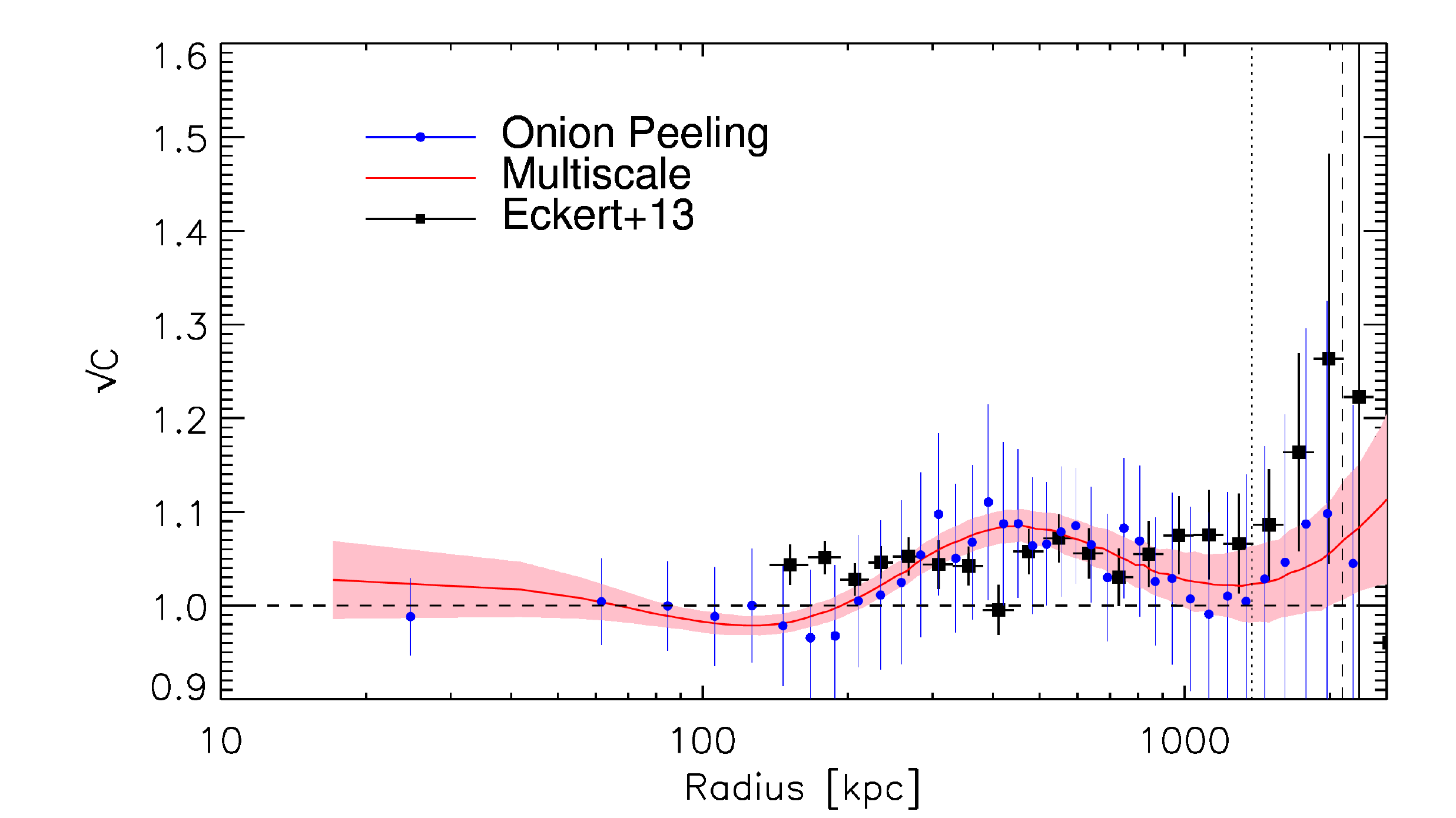}
\caption{Clumping factor radial profile for both the onion peeling (in blue) and the multiscale (in red) techniques. The pink area represents the 1--$\sigma$ confidence interval around the multiscale clumping factor. The black squares represents the observed value for the clumpiness in the work of \cite{eckert+15}. The vertical lines mark the position of $R_{500}$ and $R_{200}$, dotted and dashed respectively.}
\label{figure:clump}
\end{figure}

\section{Joint X-ray/SZ analysis of the thermodynamic properties}
\label{secsz}

The profiles of the electron density estimated from X-ray and of the pressure obtained through SZ
can be combined to recover all the thermodynamic quantities that define the properties of the ICM: 
\begin{itemize}
\item the gas temperature:
\begin{equation}
T = P \cdot n_e^{-1}
\label{eq:temperature}
\end{equation}
\item the gas entropy:
\begin{equation}
K = T \cdot n_e^{-2/3} = P \cdot n_e^{-5/3}
\label{eq:entropy}
\end{equation}
\item the gas mass:
\begin{equation}
M_{\rm gas} (<R) = 4 \pi \int_0^R \rho_g(r') r'^2 dr'
\label{eq:mgas}
\end{equation}
where the gas mass density $\rho_g=(n_e+n_p) m_u \mu$ with $m_u$ being the atomic mass unit and 
$\mu \approx 0.6$ the mean molecular weight in a.m.u.;
\item the hydrostatic gravitating mass: 
\begin{equation}
M_{\rm tot}(<r) = -\frac{r^2}{G \rho_g(r)} \frac{d P_g(r)}{d r}
\label{eq:hee}
\end{equation}
where $G$ is the gravitational constant, and the gas pressure $P_g$ satisfies the ideal gas law $\rho_g kT/(\mu m_	u) = P_g$.
The gas mass fraction is then defined as $f_{\rm gas} = M_{\rm gas} / M_{\rm tot}$.
\end{itemize}

\subsection{Clumpiness profile}
\label{subsect:clumpiness}

X-rays imaging can be directly used to estimate the level of inhomogeneities present in the ICM. 
The clumping factor $C = \braket{n_e^2}/\braket{n_e}^2$ measures the bias that affects the reconstruction of the gas density from the X-ray emission, 
that is directly proportional to $n_e^2$. Since we are considering the X-ray signal collected in a narrow energy range, [0.7--1.2] keV, that is almost 
insensitive to the gas temperature, we can directly use the results from the spatial analysis to estimate the gas clumping factor $C$.

In first approximation, the density distribution inside a volume shell can be described by a log-normal distribution skewed by the presence of denser outliers, clumps \citep{zhuravleva+13,roncarelli+13}. 
Therefore, while the mean of this distribution tends to overestimate the gas density, the median is robust against the presence of clumps \citep{eckert+15}, and
we can estimate $C$ as the ratio of the deprojected X-ray surface brightness profiles obtained from (i) the mean of the azimuthal distribution of the counts in annuli and (ii) the median of the same distribution. The resulting profile is shown in Fig.~\ref{figure:clump} and indicates a $\sqrt{C}$ of about 1.1 at $R_{200}$.

However we can only detect clumps that are resolved by \xmm, i.e. clumps on scales larger than the PSF half energy width ($\sim 17$ arcsec $\approx$ 18.4 kpc, for MOS\footnote{https://heasarc.nasa.gov/docs/xmm/uhb/onaxisxraypsf.html}; see also \cite{read+11}).
This implies that clumped structures below this scale might still bias our measured thermodynamic quantities.

\begin{figure}[t]
\includegraphics[width=0.5\textwidth]{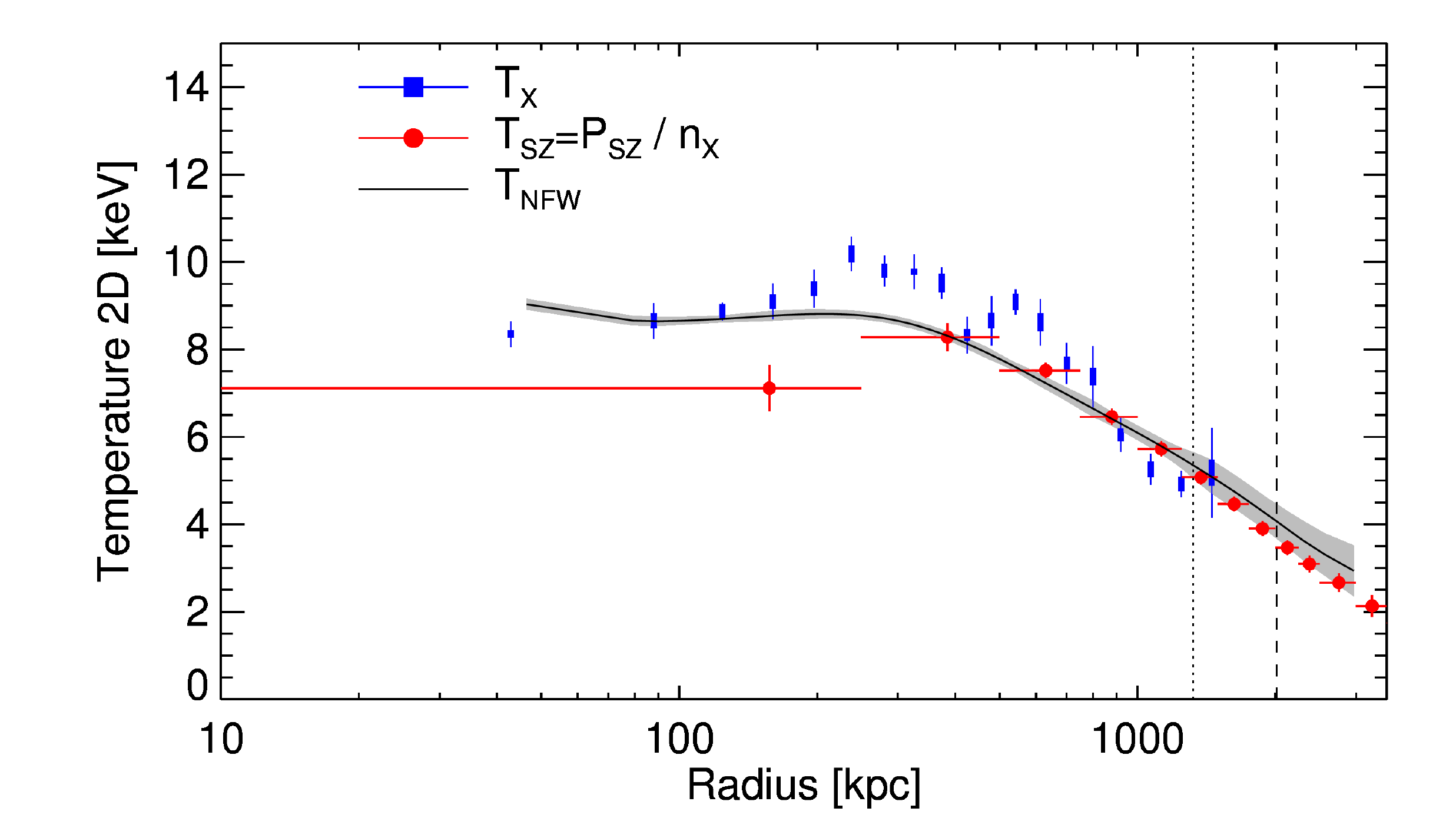}
\caption{Two-dimensional temperature profiles using X-ray spectral data (blue points; thick errorbars represent the systematic uncertainty as estimated in Section~\ref{syst}; thin errorbars indicate the total uncertainties), the pressure from SZ divided by density from X-ray projected on the plane of the sky (red points) and the projection of the reconstructed temperature from the \textit{backward} technique, which makes use of both X-ray and SZ data, on an NFW mass model (black line). 
The grey shaded area is the 1$\sigma$ confidence region around the \textit{backward} result.
The vertical lines mark the position $R_{500}$ and $R_{200}$, dotted and dashed respectively.
}
\label{figure:temperature3d}
\end{figure}

\subsection{Temperature profile}

Similarly to what has been done for the pressure, we can recover the ICM temperature profiles in two different ways: 
(i) from the spectral analysis ($T_{X}$) as detailed in Sect.\ref{subsect:spectral}; 
(ii) by dividing $P_{SZ}$ with the gas density $n_e$ recovered from the deprojection of the X-ray surface brightness ($T_{SZ}$). 
These values can be compared with the profile $T_{\rm NFW}$ that is obtained from the best-fit mass model (see Section~\ref{sec:mhyd})
by requiring that the hydrostatic equilibrium holds between the cluster potential and the observed gas density profile. 
Note that $T_{\rm NFW}$ is not independent from the other two profiles, because the best-fit mass model is obtained 
by fitting both the measured $T_{X}$ and $P_{SZ}$.
In order to obtain a meaningful comparison with $T_X$, we compute an spectroscopic--like projection \citep[see][]{mazzotta+04,morandi+07}
of the three-dimensional quantities $T_{SZ}$ and $T_{\rm NFW}$.
The good agreement among these profiles is shown in Fig.~\ref{figure:temperature3d}.

We notice that, since the pressure gradient in the first point is washed out from the \planck's beam of about 7 arcmin, 
the pressure in this point is underestimated, and therefore also the temperature $T_{SZ}$ is is underestimated with respect to $T_{X}$.

\subsubsection{Systematic uncertainties on the temperature profile}
\label{syst}

\begin{figure}[t]
\includegraphics[width=0.5\textwidth]{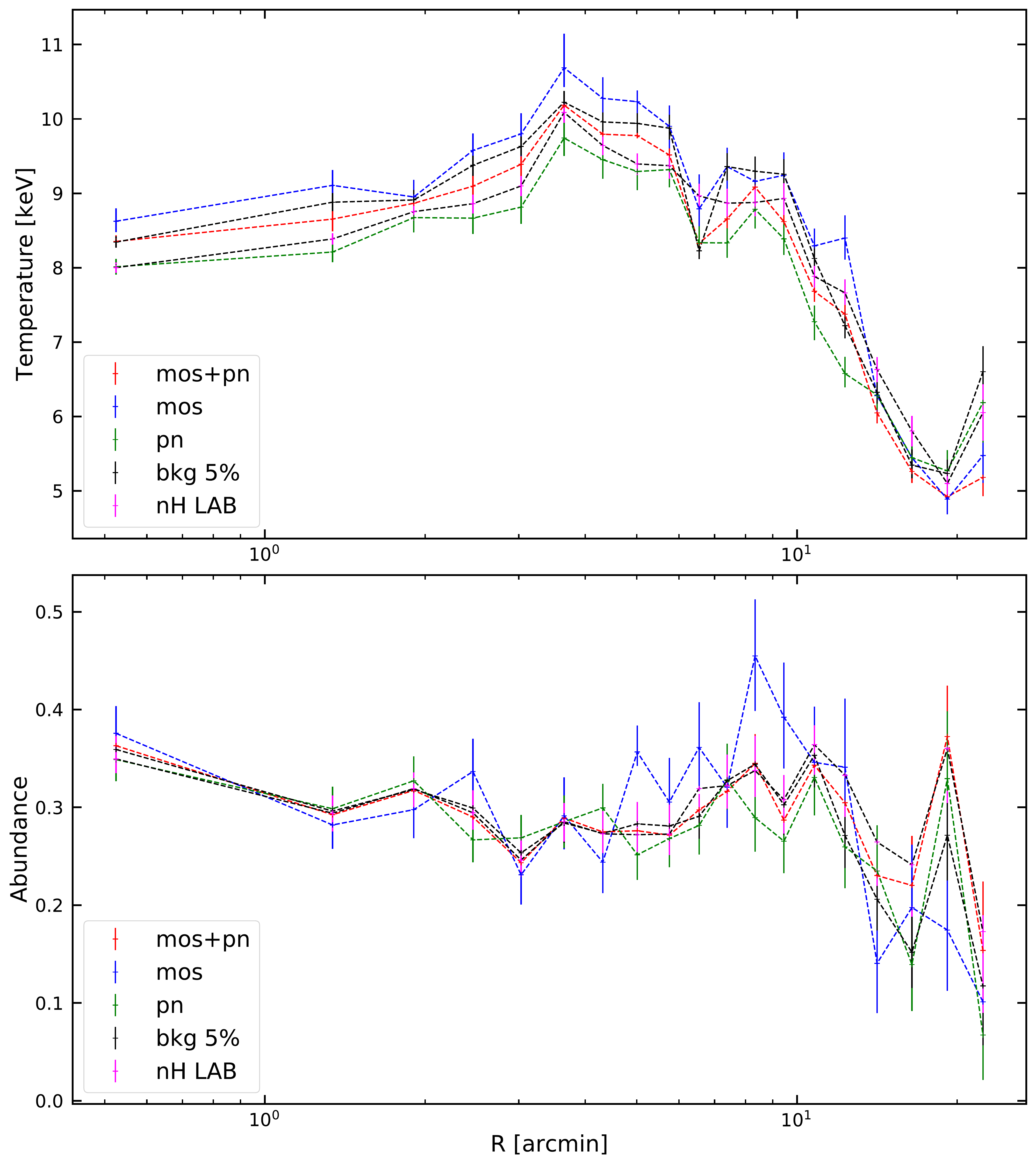}
\caption{Temperature and abundance profile adopting different techniques, this allows to have an estimate of the systematic error affecting our measurement.}
\label{fig:syst_T}
\end{figure}

We constrain the projected spectroscopic temperature (see Sect.\ref{subsect:spectral}) with a relative statistical uncertainty ranging between 1 and 6 per cent (median value: 2\%).
It is thus critical to evaluate the role of possible systematics in our measurements.
In order to calculate some of the most relevant systematic uncertainties affecting our temperature measurements, 
we re-estimate the spectral temperature using several different methods. 
Our reference temperature measurement is the one calculated using both \mos\ and \textit{pn} data, 
leaving \nh\ free to vary within a defined narrow range, and fixing the parameters of the  background model.
By changing all these quantities, one by one, we estimate the level of systematic errors that affect our measurements. 
In details, we calculate the spectral temperatures (i) using only counts collected from \mos\, (ii) only from \textit{pn}, 
(iii) fixing \nh\ to the LAB value \citep{LAB}, and (iv) allowing the background parameters (normalizations) to vary within $\pm 5\%$ of the best-fit values.
We show in Fig.~\ref{fig:syst_T} the results of this procedure. 
Finally, at each radial point, we estimate the systematic error using the standard deviation of the values measured with all the different methods.
This error is then added, in quadrature, to the statistical error and propagated through the entire analysis.
The relative systematic error ranges between 1.4\% and 9.1\%, apart from the outermost radial point where we measure a value of 19\%.

\begin{table*}
\begin{center}
\begin{tabular}{ c c c c c c c c c }
 & $P_0$ & $c_{500}$ & $\gamma$ & $\alpha$ & $\beta$ & $\chi^2$ & d.o.f.\\
  \hline
\cite{arnaud+10} & 8.40 & 1.18 & 0.31 & 1.05 & 5.49 & - & -\\
\cite{planck+13} & 6.41 & 1.81 & 0.31 & 1.33 & 4.13 & - & -\\
SZ+X & $7.7 \pm 2.0$ & $1.34 \pm 0.22$ & $0.47 \pm 0.07$ & 1.05 & $3.80 \pm 0.22$ & 2.62 & 69\\
SZ & $9.6 \pm 5.8$ & $1.10 \pm 0.35$ & $0.23 \pm 0.23$ & 1.05 & $4.50 \pm 0.47$ & 3.47 & 9\\
\end{tabular}
\caption{Best fit parameters of the pressure profile using the functional form introduced by \cite{nagai+07}. ``SZ+X'' refers to the best fit done on the best fit mass model pressure profile(see Sec.~\ref{sec:mhyd}), while ``SZ'' refers to the best fit done only on the $P_{SZ}$. }
\label{table:fit_p}
\end{center}
\end{table*}

\subsection{Pressure profile}

If the galaxy cluster is not affected by an ongoing merger generating shocks through the ICM, the pressure is the thermodynamic quantity that presents a smoother spatial distribution along the azimuth.
It is well described  by an ``universal" form \citep{nagai+07,arnaud+10}:
\begin{equation}
\frac{P(x)}{P_{500}} = \frac{P_0}{(c_{500}x)^\gamma [1+(c_{500}x)^\alpha]^{\frac{\beta-\gamma}{\alpha}}},  
\label{eq:arnaud}
\end{equation}
where 
\begin{equation}
P_{500} = 1.65 \times 10^{-3} \text{\ keV} \text{\ cm}^{-3} \left(\frac{ M_{500} }{ 3 \times 10^{14} M_\odot } \right)^{2/3} E(z)^{8/3}
\label{eq:P500}
\end{equation}
and $x=R/R_{500}$;  $\gamma$, $\alpha$, and $\beta$ are the central slope, the intermediate slope, and the outer slope defined 
by a scale parameter $r_s = R_{500}/c_{500}$ ($R << r_s$, $R \sim r_s$ and  $R >> r_s$ respectively), and $P_0$ is the normalization.
 The values of $R_{500}$ and $M_{500}$ adopted here are presented in Table~\ref{table:fit_NFW} (see Section~\ref{sec:mhyd}).
We list in Table~\ref{table:fit_p} our best-fit values, using all the available radial range to fit.

\begin{figure}[t]
\includegraphics[width=0.51\textwidth]{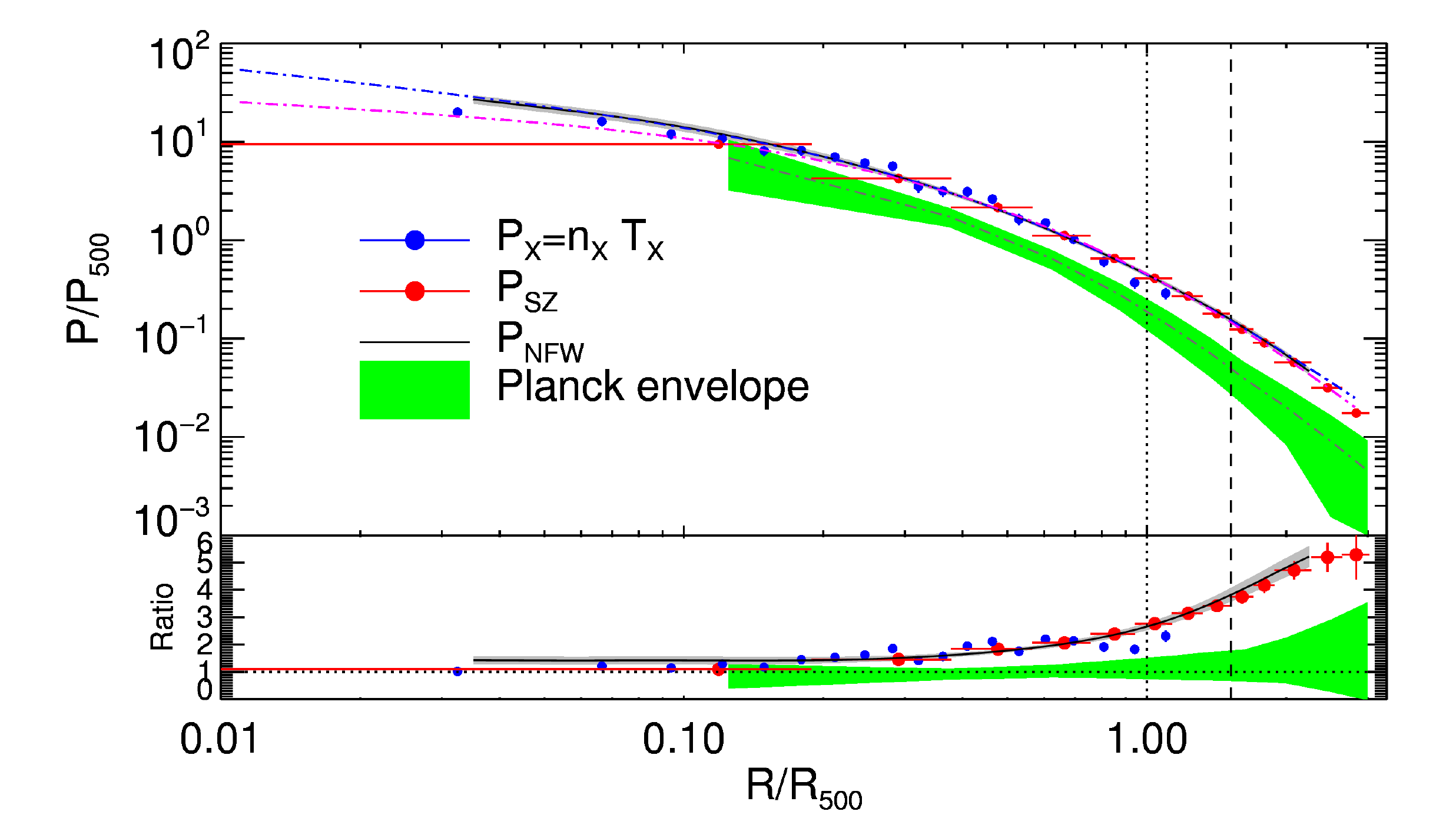}
\caption{
Rescaled pressure profile in units of $R_{500}$. The grey dashed line represents the ``universal'' pressure profile. The blue and the pink lines represents the best fits using the functional form introduced by \cite{nagai+07} done on, $P_{\rm NFW}$ and $P_{SZ}$, respectively.
The dotted and the dashed vertical lines represent the position of $R_{500}$ and $R_{200}$ respectively. 
In the bottom panel we show the ratio of $P_{SZ}$, $P_{X}$, and $P_{\rm NFW}$ with the ``universal'' pressure profile\citep{arnaud+10}.
}	
\label{figure:pressure3d}
\end{figure}

The electronic pressure can be directly recovered both from the comptonization profile (see Eq.~\ref{eq:pressure}; $P_{SZ}$), 
and from deprojection of X-rays measurements of the temperature and density profiles of the emitting electrons ($P_{X}$). 
We can also estimate the pressure profile required from the best-fit mass model to satisfy the hydrostatic equilibrium ($P_{\rm NFW}$, see Section~\ref{sec:mhyd}).
As we show in Fig.~\ref{figure:pressure3d}, these 3D pressure profiles agree well within their statistical errors.

We rescale the pressure profile by $P_{500}$ and fit it with the ``universal'' functional form \citep{nagai+07}.
The best fitting results are tabulated in Table~\ref{table:fit_p}. The comparison with the results of \planck\ \citep{planck+13} and \cite{arnaud+10} is shown in Fig.~\ref{figure:pressure3d}.
We observe that the pressure profile in A2319 is well above the other two profiles, in particular in the outskirts, with values higher by about a factor $\sim$3.5 at $R_{200}$, which is $\sim$2$\sigma$ away from the \planck\ envelope \citep{planck+13}.

We have also adopted a new technique \citep{bourdin+17} in order to evaluate the impact of the anisotropies in the Compton parameter detected in the outskirts of A2319 on the reconstructed pressure profile, and conclude that these anisotropies cannot explain the observed excess.

\begin{figure*}[t]
\hbox{
\includegraphics[width=0.35\textwidth]{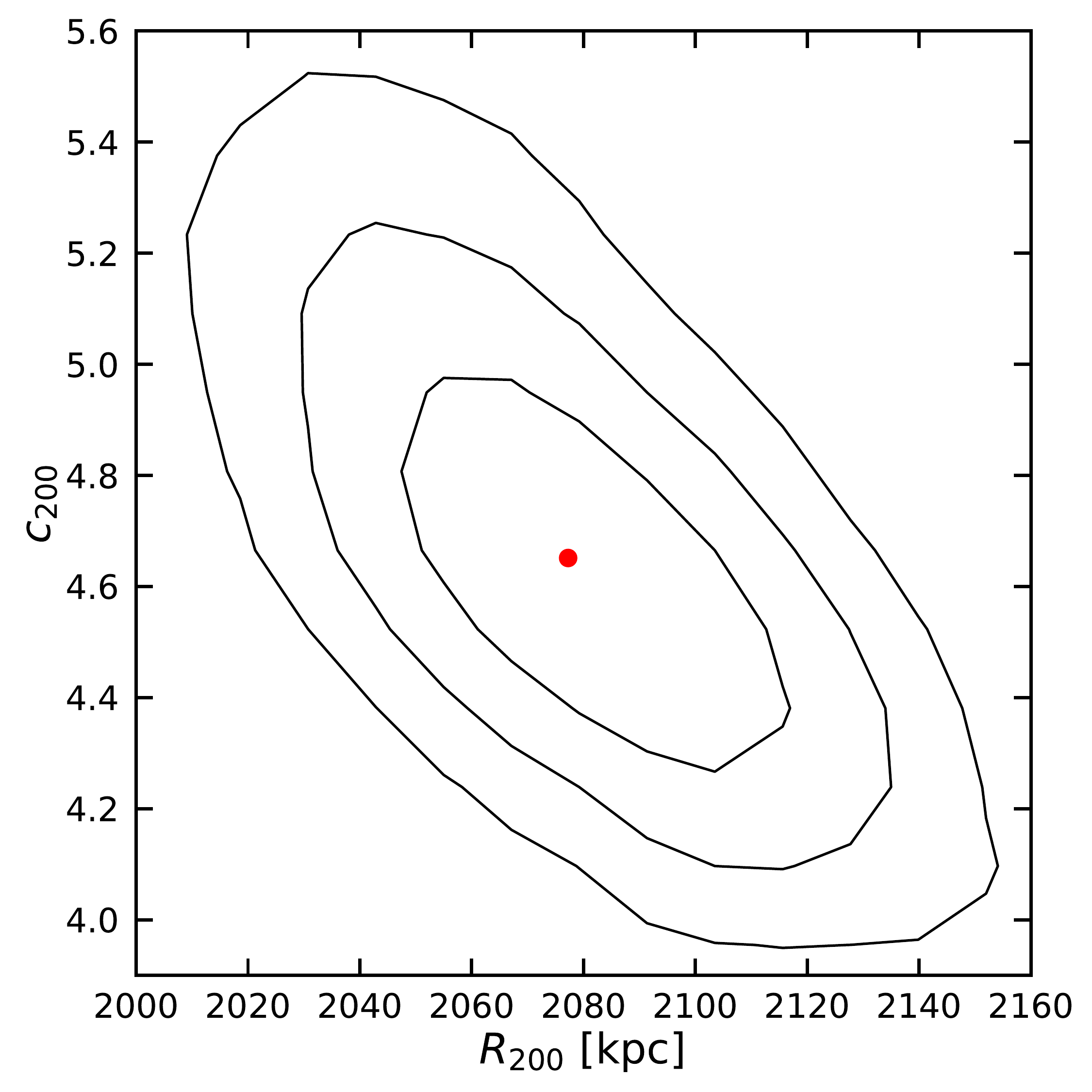}
\includegraphics[width=0.65\textwidth]{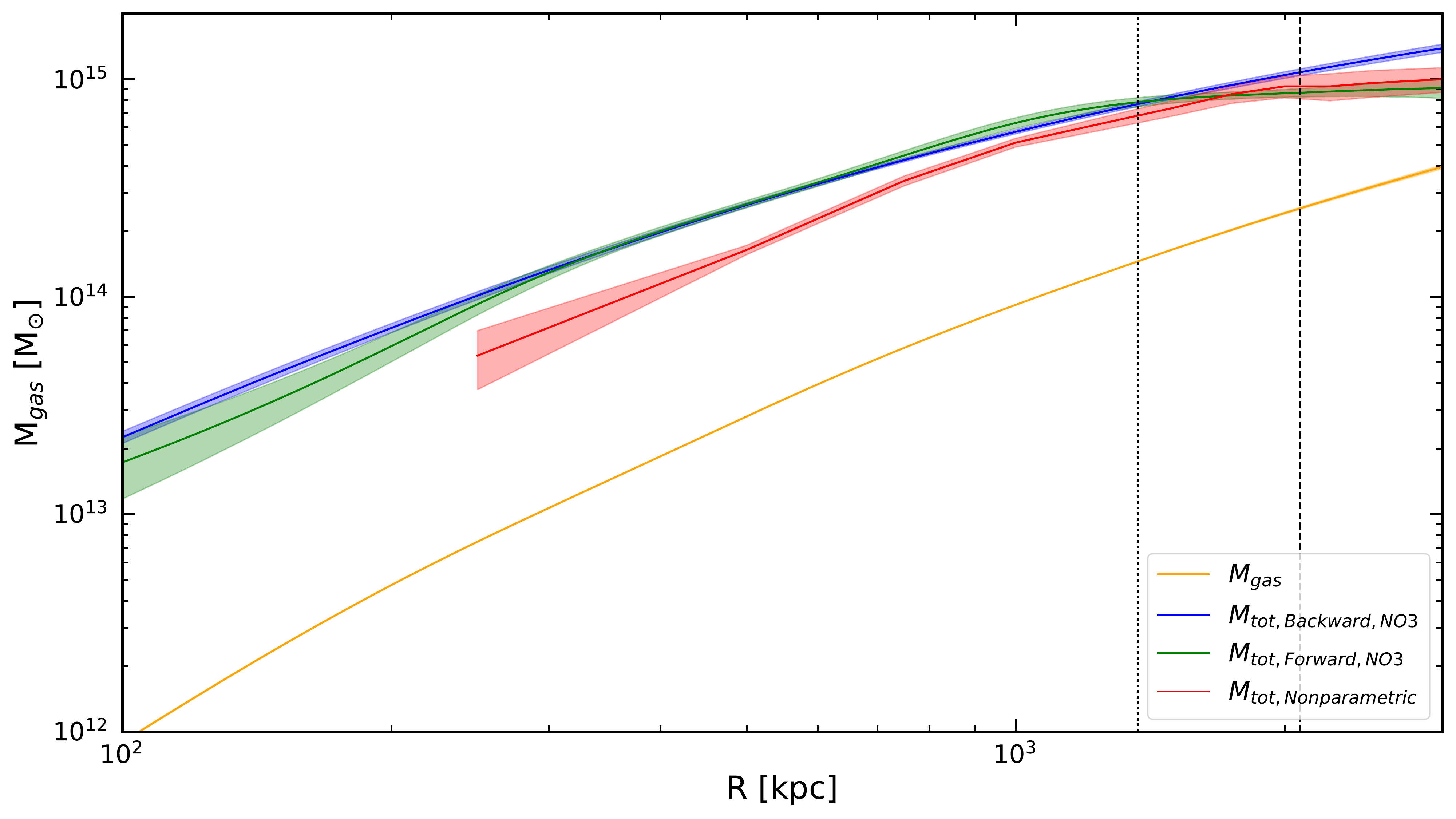}
}
\caption{
(Left)
Contour plot with confidence regions at 1, 2, 3 $\sigma$ (solid lines) applying the \textit{backward} approach to solve HEE in order to constrain the parameters of the NFW mass model; using as inputs the multiscale technique on the median emissivity profile to obtain the density, the pressure from the direct deprojection of the $y$-parameter radial profile, and the temperature from the spectral analysis.
(Right)
Gas mass and total mass profile recovered using \textit{backward} approach (blue and red curves, respectively). The black crosses represent the total mass profile obtained using a non-parametric method and the green one by applying the \textit{forward} method on temperature and density profiles.
The vertical lines mark the positions of $R_{500}$ and $R_{200}$, dotted and dashed respectively.
}
\label{figure:mass}
\end{figure*}

\begin{table*}
\begin{center}
\begin{tabular}{ c c c c c c}
Technique & Data & $M_{200}$ ($10^{14} M_\odot$) &  $R_{200}$ (kpc) & $M_{500}$ ($10^{14} M_\odot$) &  $R_{500}$ (kpc)    \\
\hline
{\bf \textit{backward}} & $\boldsymbol{P_{SZ, NO3}}$ + $\boldsymbol{T_X}$ {\bf - Median} & $\boldsymbol{10.7 \pm 0.5}$ & $\boldsymbol{2077 \pm 33}$ & $\boldsymbol{7.7 \pm 0.4}$ & $\boldsymbol{1368 \pm 17}$\\
\textit{backward} & $P_{SZ}$ + $T_X$ - Median & $10.6 \pm 0.5$ & $2071 \pm 32$ & $7.5 \pm 0.3$ & $1357 \pm 13$\\
\textit{backward} & $P_{SZ, NO3}$ + $T_{X, SYS}$ - Median & $10.3 \pm 0.7$ & $2047 \pm 47$ & $7.4 \pm 0.4$ & $1350 \pm 24$\\
\textit{backward} & $P_{SZ}$ + $T_{X, SYS}$ - Median & $10.5 \pm 0.5$ & $2062 \pm 34$ & $7.3 \pm 0.3$ & $1347 \pm 18$\\
\textit{forward} & $P_{SZ}$ only - Median & $9.4 \pm 0.5$ & $1984 \pm 40$ & $7.4 \pm 0.4$ & $1353 \pm 25$\\
\textit{forward} & $T_{X}$ only - Median & / & / & $7.3 \pm 0.1$ & $1343 \pm 5$\\
\textit{forward} & $P_{SZ}$ + $T_{X}$ - Median & $8.3 \pm 0.3$ & $1906 \pm 20$ & $7.8 \pm 0.2$ & $1375 \pm 11$\\
\textit{forward} & $P_{SZ}$ + $T_{X}$, $\beta$ fixed - Median & $8.5 \pm 0.6$ & $1923 \pm 48$ & $7.7 \pm 0.4$ & $1368 \pm 26$\\
\textit{forward} & $P_{SZ, NO3}$ + $T_{X, SYS}$, $\beta$ fixed - Median & $7.7 \pm 0.7$ & $1859 \pm 59$ & $7.4 \pm 0.6$ & $1354 \pm 37$\\
\textit{forward} & $P_{SZ, NO3, SYS}$ + $T_X$  - Median & $8.3 \pm 0.3$ & $1907 \pm 26$ & $7.8 \pm 0.3$ & $1373 \pm 18$\\
\textit{non parametric} & $P_{SZ}$ - Median & $9.3 \pm 1.1$ & $1979 \pm 78$ & $6.7 \pm 0.5$ & $1307 \pm 33$\\
\\
\textit{backward} & $P_{SZ}$ + $T_X$ - Mean & $10.2 \pm 0.5$ & $2040 \pm 35$ & $7.3 \pm 0.3$ & $1346 \pm 17$\\
\end{tabular}
\caption{Best fitting results on the mass model using different techniques, as specified in the first column. 
 In the second column, the data used to constrain the mass are listed; $P_{SZ}$ and $T_X$ refers to the SZ pressure and the X-ray temperature respectively; the subscript ``NO3'' indicates that the first 3 \planck\ points were not used in the analysis; 
the subscript ``SYS'' indicates that the systematic uncertainties on the X-ray temperature are added in quadrature to the statistical errors in evaluating the $\chi^2$ (see Sec.~\ref{syst}); ``Median'' or ``mean'' refers to how we computed the X-ray emissivity;
``$\beta$ fixed'' indicates that the outer slope of the pressure profile is fixed to the best fit value of the \planck\ collaboration.
In the other four columns, we quote the results on $M_{200}$, $R_{200}$, $M_{500}$, and $R_{500}$ respectively.  
In the first row, we indicate our reference values in the bold font.
The last two rows present the mass reconstructed using the mean density profile, and propagating the statistical error on the temperature profile only (See Section~\ref{syst}).
 $R_{\Delta}$ are defined as $\left( \frac{M(R)}{4/3 \pi \rho_c \Delta} \right)^{1/3} $.
}
\label{table:fit_NFW}
\end{center}
\end{table*}

\subsection{Hydrostatic mass}
\label{sec:mhyd}

The total mass profile of the cluster is reconstructed by solving 
the hydrostatic equilibrium equation~\ref{eq:hee} \citep[HEE,][]{hee}.
In this work, we use three different methods to solve this equation and recover the hydrostatic mass profile \citep[e.g.][]{ettori+13}: 
the \textit{backward} method, the \textit{forward} method and a \textit{non-parametric} method.

The \textit{backward} method follows the approach described in \cite{ettori+10, ettori+17} and, assuming a mass model with few (in general, two) free parameters, 
minimizes a likelihood function by comparing the predicted and observed profiles of some interesting physical quantities (such as the temperature) to constrain these parameters.  
In the present analysis, we assume a Navarro-Frenk-White profile \citep[NFW,][]{nfw+97} for the total mass 
(a more extensive discussion on the best-fitting mass models will be presented in a forthcoming publication), 
and constrain its two parameters, concentration and scale radius (or $R_{200}$), 
using both the  projected temperature profile from X-ray spectral analysis and the thermal pressure profile from the SZ analysis, 
 and maximizing the likelihood described in Appendix~\ref{sec:likelihood}.

In Fig.~\ref{figure:mass}, we show the best fit results obtained using this method to constrain the parameters of the mass model, using the median method and the multiscale technique to obtain the density profile. Very consistent results are obtained by adopting different methods to recover the input profiles of the gas temperature and density (see Table~\ref{table:fit_NFW}). 
We indicate with the subscript "NFW" the thermodynamic quantities corresponding to the best-fit mass model.

In the \textit{forward} method, functional forms are used to fit the thermodynamic quantities, like density, pressure and temperature. 
Then, HEE is directly applied in order to compute the total mass radial distribution. Errors are estimated through a Monte Carlo process. 
As mentioned in Section~\ref{sec:spatial}, we use the multiscale approach \citep{eckert+15} to fit the emissivity profile which yields directly the fitted density functional form. 
We use a 6-parameters functional form \citep{vikhlini+06} to fit the temperature, and a 5-parameters generalized NFW \citep{nagai+07} for the pressure.
We combine in several ways the profiles of the thermodynamic quantities (density, pressure and temperature), as detailed in Table~\ref{table:fit_NFW}, making use 
of a joint likelihood (see Appendix~\ref{sec:likelihood}) when all 3 quantities are fitted together.
It is worth noticing that, while measurements of the gas density and pressure are available up to $\sim R_{200}$,
direct spectral estimates of the temperature are limited to regions below $R_{500}$, defining the radial range where the mass profile is more reliable in this case.

Due to the good quality data both from X-rays and SZ, we can also implement a \textit{non-parametric} method in order to recover the total mass profile. We just insert pressure and density in the HEE, and we calculate the pressure derivative using a three-point quadratic Lagrangian interpolation. We point out that the errors relative to this method are represented by a covariance matrix, since we are using the SZ pressure profile, and therefore what is shown as an errorbar in the plot is just the square root of the diagonal terms.

The recovered mass profiles are shown in Fig.~\ref{figure:mass}. 
They are all compatible within their respective error bars at the characteristic overdensities of 500 and 200.

\subsubsection{Systematic uncertainties on the hydrostatic mass}

In Table~\ref{table:fit_NFW}, only the statistical error on $M_{200}$ (with a relative uncertainty of about 4.7\%) is quoted.
In this section, we evaluate what is the impact of some of the systematic uncertainties that affect the mass reconstruction.

The ability of the particle background model to reproduce a flat surface brightness profile when applied on blank field observations 
is a source of systematic uncertainty caused by the adopted procedure. As we discussed in Section~\ref{sec:xmm}, 
adopting the background modeling described in Appendix~\ref{app_dom}, we are able to reduce the systematic deviation from a flat profile below 5\%.  
We account for this by adding 5\% of the background level as an extra error in the surface brightness profile.

The results obtained applying different methods and techniques are shown in Table~\ref{table:fit_NFW}.
We estimate the level of the systematic uncertainties on the mass measurement at $R_{500}$ and $R_{200}$ of about  3.9\% and 8.4\%, respectively, 
by measuring the relative scatter around the reference value. 

Another source of systematic uncertainty comes from the choice of the background region, defined in an area concentrated to the West of the cluster. 
Considering that A2319 has an angular extension of $\sim 1$ degree, cosmic variance can influence the analysis, especially in the outskirts. 
Using the absorbed thermal model \textit{tbabs(apec)}, and fixing the parameters of the \textit{apec} component, 
we vary the hydrogen column density only, by adopting the values of \nh\ in regions located at North, West, East, and South, 
as far as possible from the center (distance of 33, 55, 36, and 39 $arcmin$ respectively)
and re-measure the conversion factor between the count rate and the surface brightness maps.
This procedure allows to measure a relative deviation of 2\% on the surface brightness, that translates into an effect of about 1.4\% on the gas density and 
1.1\% on the mass measurement.

We therefore estimate that the total systematic uncertainties are at the level of 4.18\% and 8.5\% at $R_{500}$ and $R_{200}$, respectively, implying that the reference values
for the hydrostatic mass are, at  $R_{500}$ and $R_{200}$, respectively:
\[M_{500} = 7.7 \pm 0.4^\text{stat.} \pm 0.3^\text{syst.} \times 10^{14} M_\odot\]
\[M_{200} = 10.7 \pm 0.5^\text{stat.} \pm 0.9^\text{syst.} \times 10^{14} M_\odot\]

\begin{figure}[t]
\includegraphics[width=0.5\textwidth]{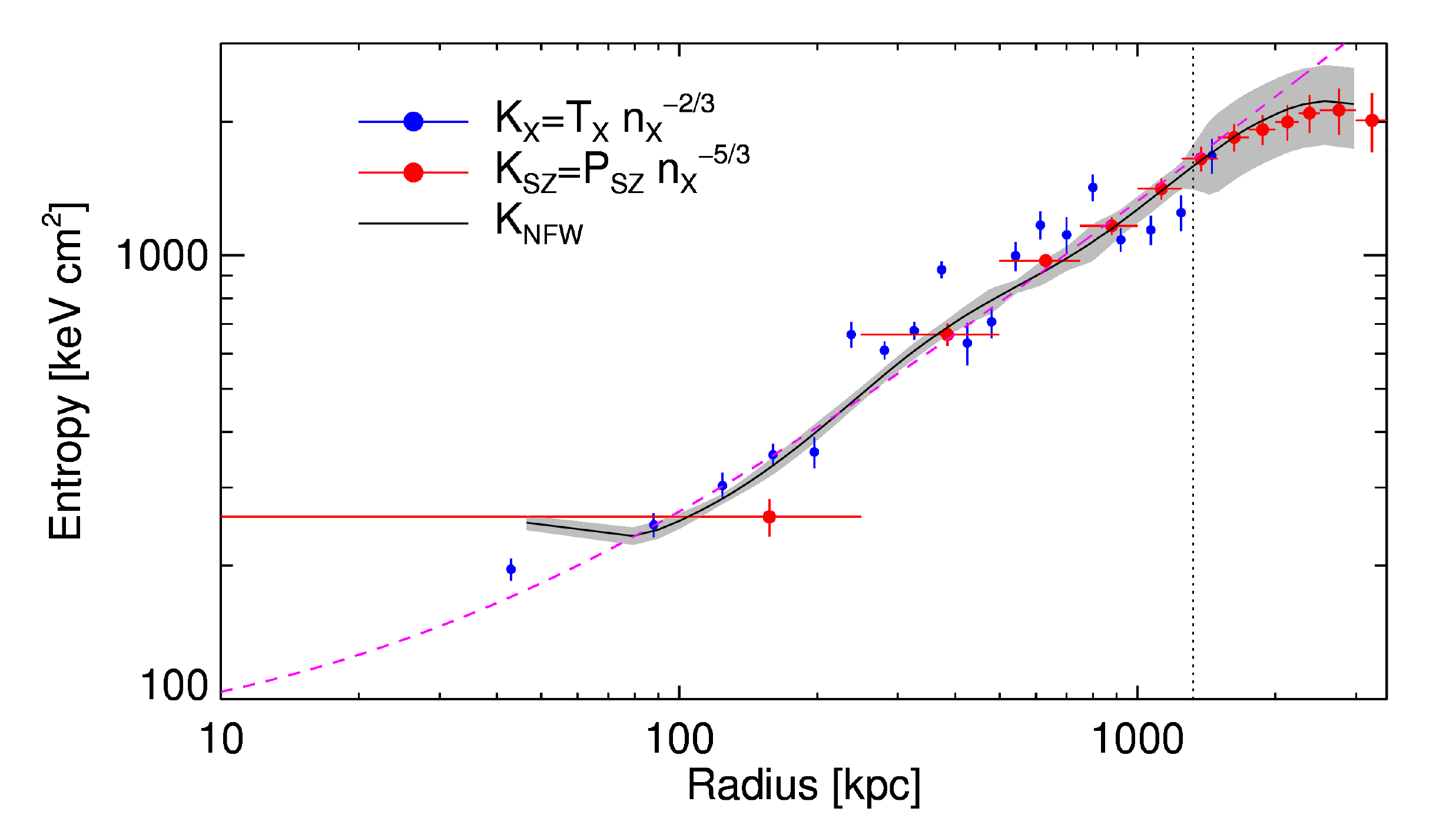}
\includegraphics[width=0.5\textwidth]{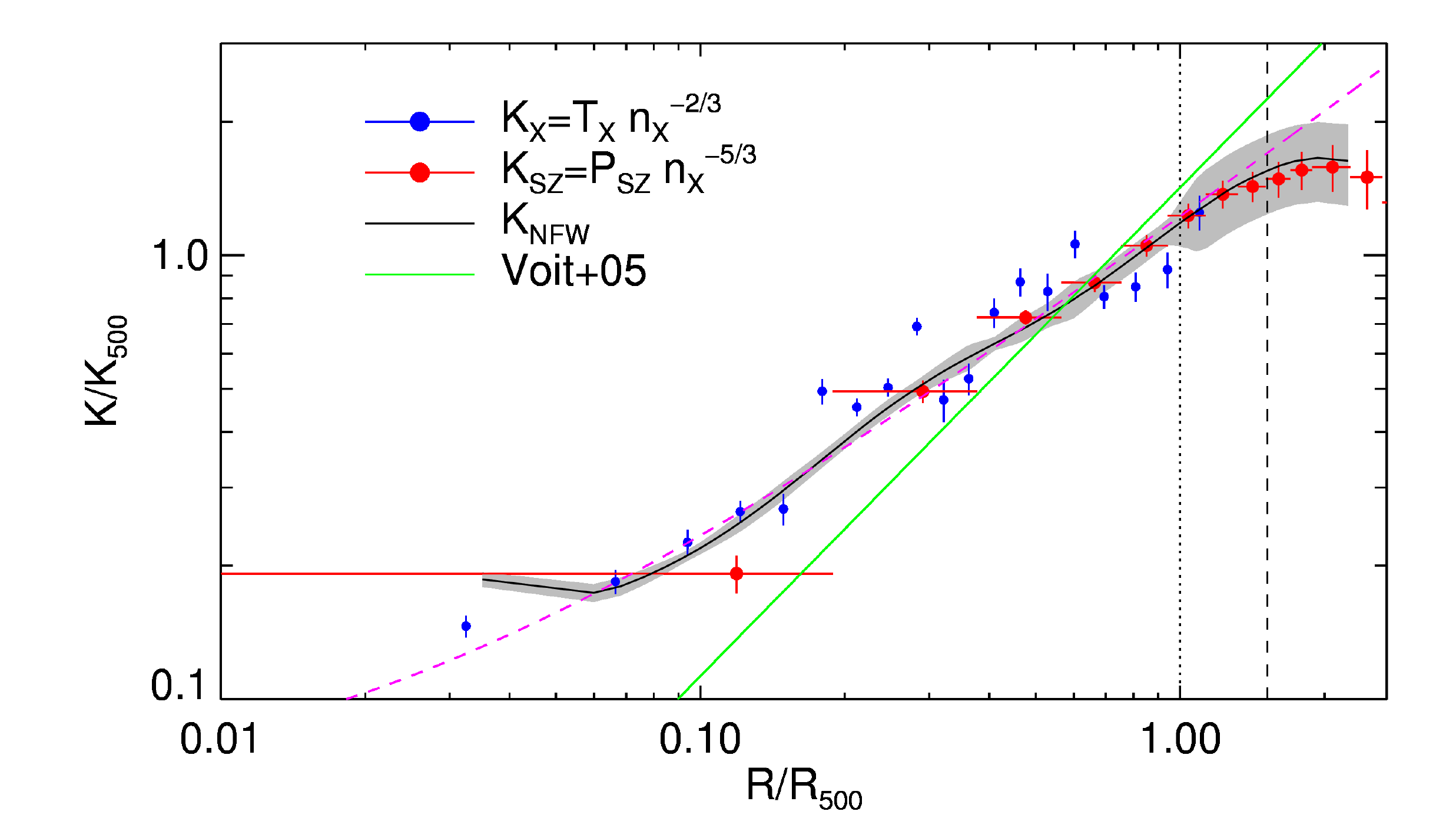}
\caption{
(Top) Entropy profiles obtained from the three different methods described in Sec.~\ref{secK}. The dashed magenta line represents the best fit obtained on the $K_{\rm NFW}$ data using Eq.~\eqref{eq:cavagnolo}.
(Bottom) Entropy profiles rescaled by $K_{500}$. The dashed magenta line represents the best fit obtained on the $K_{\rm NFW}$ data using Eq.~\eqref{eq:powlc_res}.
The green lines represents the prediction from \cite{voit+05}.
The dashed pink lines are the best fit using Eq.~\eqref{eq:cavagnolo} and  \eqref{eq:powlc_res}.
The vertical dotted and dashed lines represents the location of $R_{500}$ and $R_{200}$, respectively.
}
\label{figure:entropy3d}
\end{figure}
\begin{table*}[h]
\begin{center}
\begin{tabular}{ c c c c c c  }
 & $k_0$ & $k_{100 / 500}$ & $\alpha$ & $\chi^2$ & d.o.f.\\ \hline
Eq.~\eqref{eq:cavagnolo} & $75 \pm 13$ & $190 \pm 12$ & $0.82 \pm 0.03$ & 129 & 70\\
Eq.~\eqref{eq:powlc_res} & $0.055 \pm 0.010$ & $1.17 \pm 0.02$ & $0.82 \pm 0.03$ & 124 & 70\\
\end{tabular}
\caption{Best fit results for the model of the entropy profile using the three different rescaling described in the Sec.~\ref{secK}.}
\label{table:fit_k}
\end{center}
\end{table*}

\subsection{Entropy profile}
\label{secK}

The entropy profile is recovered through the gas pressure and temperature profiles via Eq.~\eqref{eq:entropy}.
Entropy is a fundamental quantity to track the thermal history of a cluster: it always rises when a heat flow occurs, 
and in the presence of just non-radiative processes it is expected to follow a power law with characteristic slope of 1.1 \citep{tozzi+01,voit+05}.
Deviations from this power law are observed in the central regions, requiring an entropy ``floor" within $\sim 100$ kpc 
that is expressed through the formula \citep{cavagnolo+09}:
\begin{equation}
K = k_0 + k_{100} \; \left( \frac{R}{100 \ kpc} \right)^	\alpha
\label{eq:cavagnolo}
\end{equation}

The central entropy ($k_0$) measured with the fit in Eq.~\eqref{eq:cavagnolo} is $75 \pm 13$ keV cm$^2$ (see Table~\ref{table:fit_k}), 
suggesting that A2319 does not possess a relaxed, cool core \citep[e.g.][define a CC when $k_0 < 50$ keV cm$^2$]{cavagnolo+09}.

However non-radiative simulations show that the self similar behaviour is reproduced only once entropy is rescaled by a proper quantity 
defined with respect to the critical density \citep{voit+05}:
\begin{equation}
K_{500} =  106 \text{\ keV cm}^2 \left(\frac{M_{500}}{10^{14} M_\odot}\right)^{2/3} E(z)^{-2/3} f_{b}^{-2/3} 
\label{eq:K500}
\end{equation}
where $f_b = 0.15$ is the universal baryon fraction.
Non-radiative simulations \citep{voit+05} predicts that the power law describing the entropy profile is:
\begin{equation}
\frac{K(R)}{K_{500}} =  1.42 \; \left(\frac{R}{R_{500}} \right)^{1.1}
\label{eq:voit}
\end{equation}

In order to accomodate the flattening of the entropy profile observed in many disturbed system we add a constant to a simple power law:
\begin{equation}
\frac{K(R)}{K_{500}} =  k_0 + k_{500} \; \left(\frac{R}{R_{500}} \right)^{\alpha},
\label{eq:powlc_res}
\end{equation}

In Fig.~\ref{figure:entropy3d}, we plot the measured entropy profiles, also rescaled accordingly to Eq.~\ref{eq:powlc_res}.
In Table~\ref{table:fit_k}, we show the best fit results on the data using Eq.\eqref{eq:cavagnolo} and \eqref{eq:powlc_res}. 
We observe that the entropy profile has a shallower slope with respect to what is predicted by simulations\citep{voit+05}.

\section{Analysis in Azimuthal Sectors}

\begin{figure*}[t]
\hbox{
\includegraphics[width=0.49\textwidth]{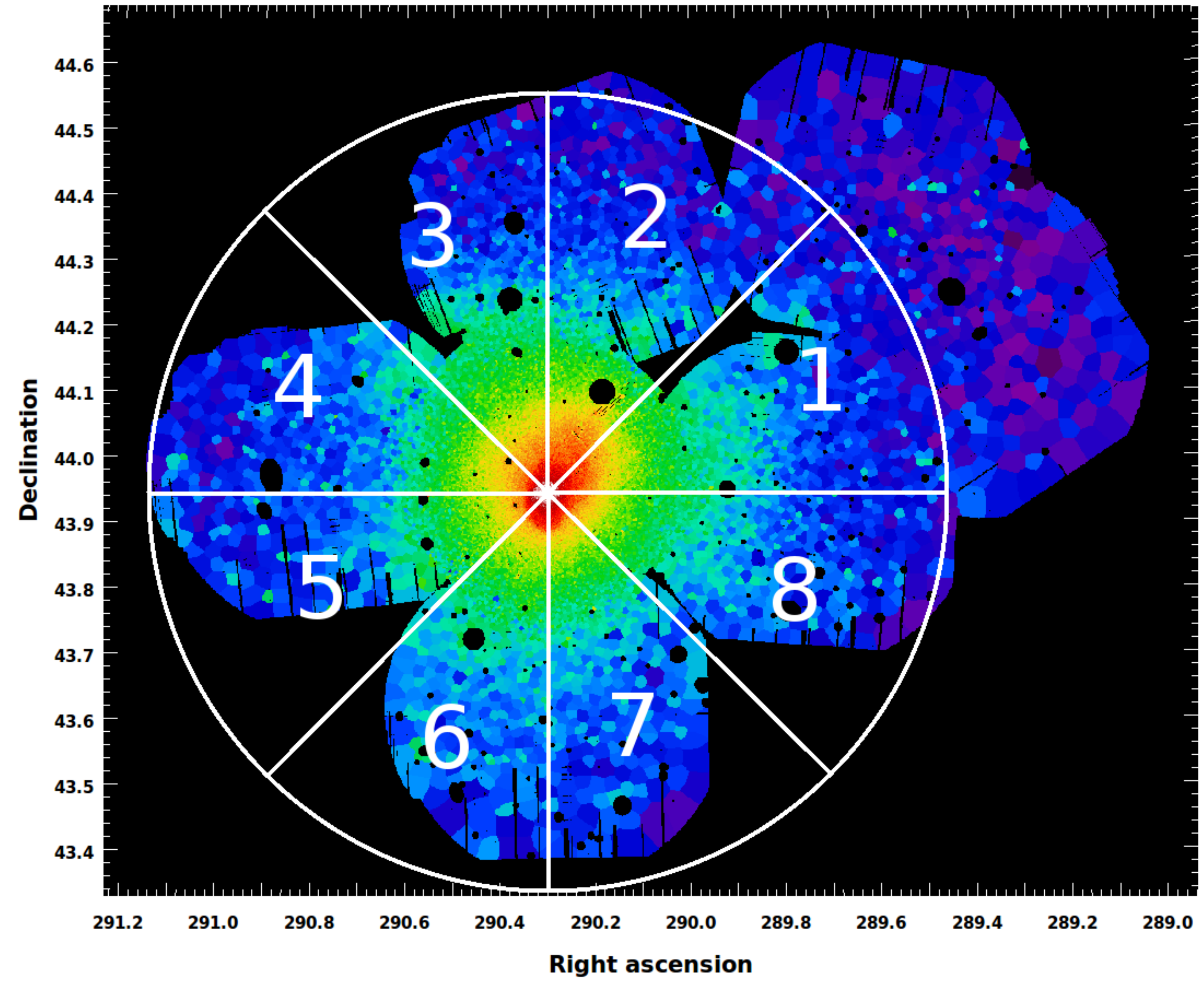}
\includegraphics[width=0.51\textwidth]{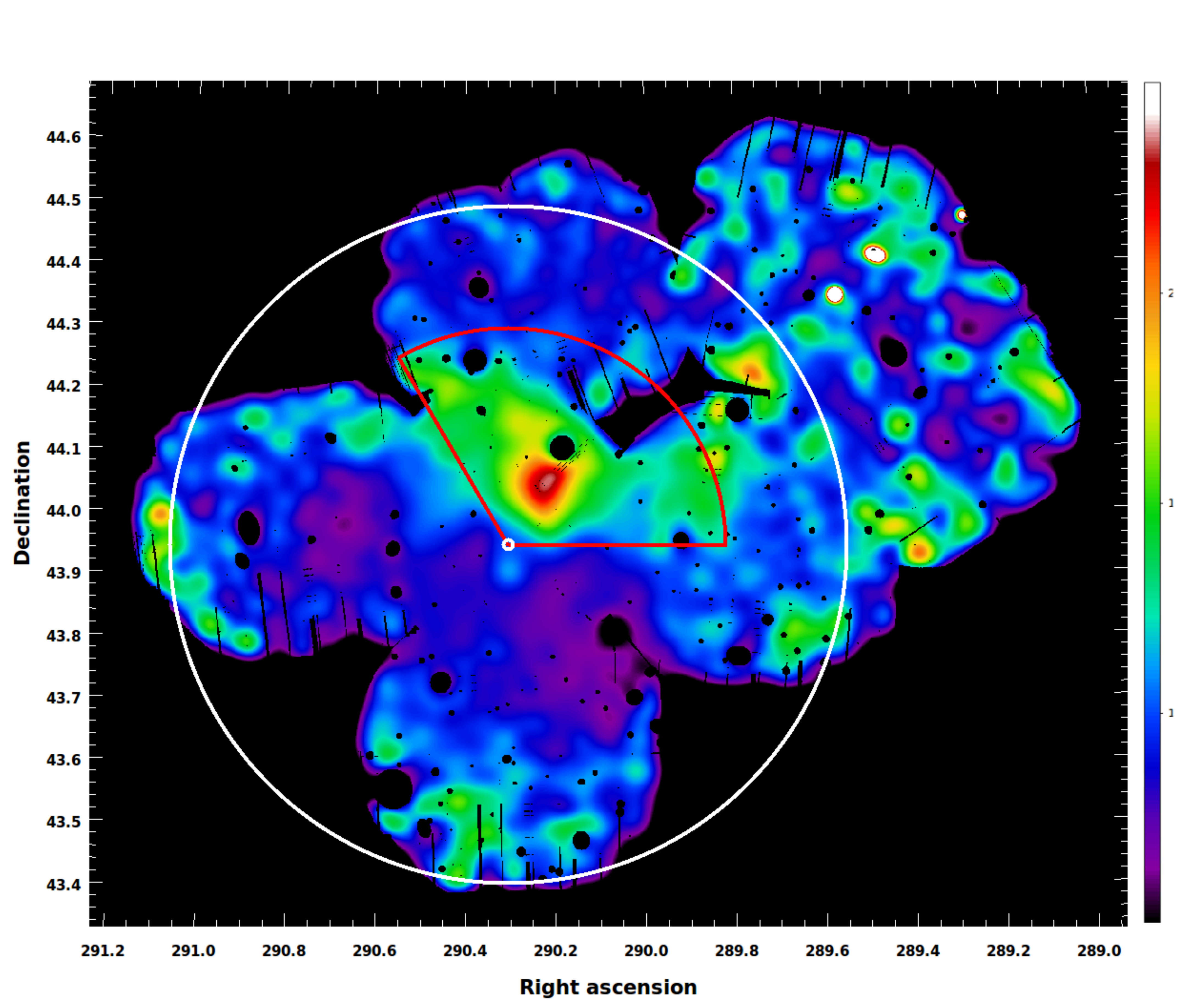}
}
\caption{(Left) 
Same as Fig.~\ref{figure:VoronoiA2319}. The white sectors represents the 8 regions analyzed separetely, each one is marked by a identification number.
(Right)
Residual image obtained by dividing the flux image by the model image reconstructed from the median method. 
The small white circle represent the center of the cluster, and the big white circle represents the position of $R_{200}$.  The red sector represents the region which shows a clear excess in the residual map.}
\label{figure:sectorA2319}
\end{figure*}

\begin{figure*}[th]
\hbox{
\includegraphics[width=0.5\textwidth,page=1]{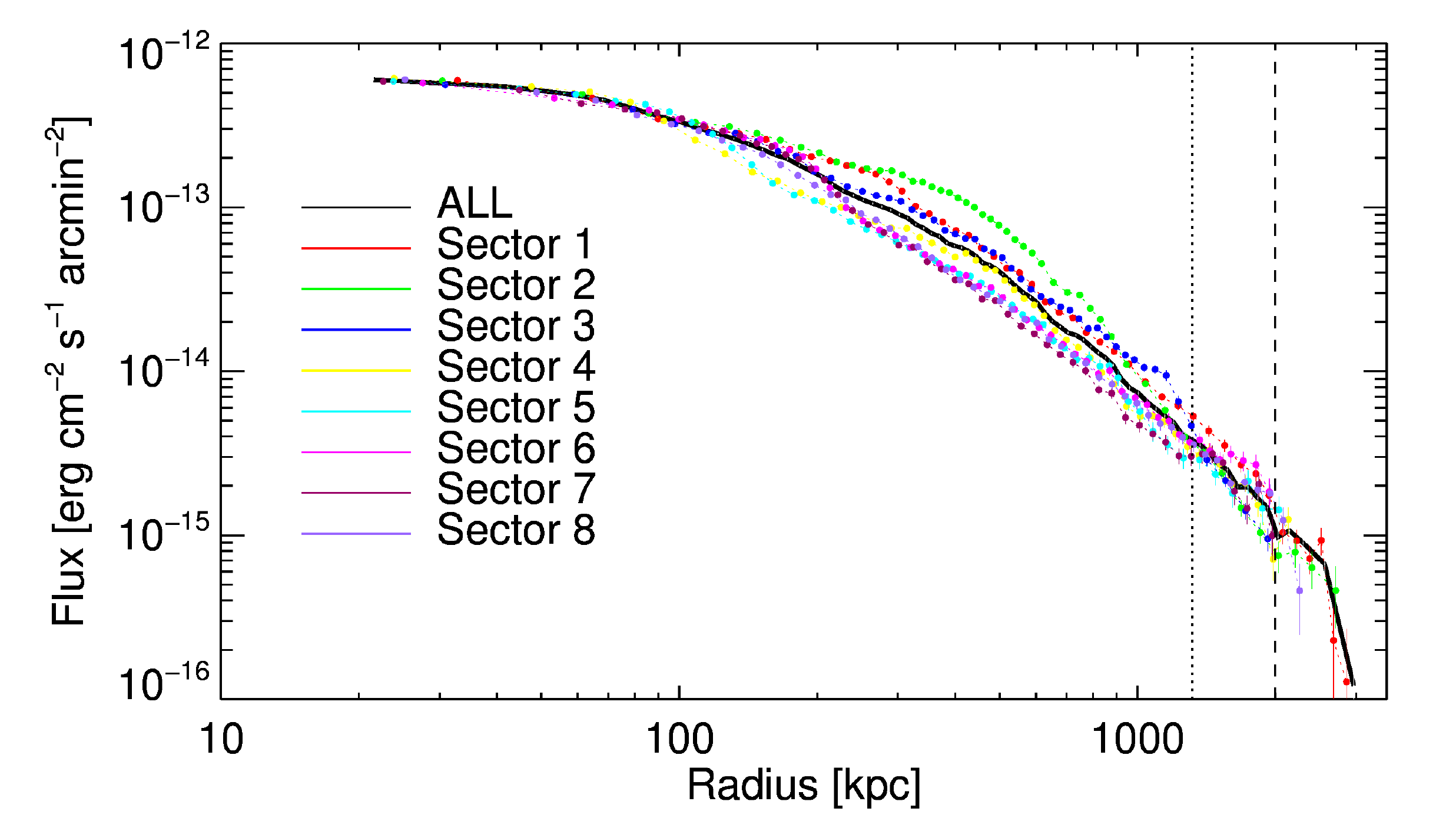}
\includegraphics[width=0.5\textwidth,page=4]{plots/sector.pdf} 
}
\hbox{
\includegraphics[width=0.5\textwidth,page=5]{plots/sector.pdf}
\includegraphics[width=0.5\textwidth,page=6]{plots/sector.pdf}
}
\hbox{
\includegraphics[width=0.5\textwidth,page=7]{plots/sector.pdf}
\includegraphics[width=0.5\textwidth,page=8]{plots/sector.pdf}
}
\caption{(Top-Lef) 
Surface brightness profiles for the 8 sectors using the median method. The thick black line is what we obtain in the whole image analysis.The vertical lines mark the position $R_{500}$ and $R_{200}$, dotted and dashed respectively.
(Top-Right)
SZ pressure profiles for the 8 sectors analyzed on top of the pressure profile for the whole cluster(black line). 
(Center) 2D temperature (left) and abundance (right) profiles for the 8 sectors analyzed. 
(Bottom-Left) Reconstructed entropy profiles for all sectors using the \textit{backward} method. 
(Bottom-Right) Gas fraction profiles recovered applying the {\it backward} technique. 
The thick black line is the result for the azimuthally averaged profile. 
The vertical lines mark the position $R_{500}$ and $R_{200}$, dotted and dashed respectively.
}
\label{figure:sectors}
\end{figure*}

Considering the high signal-to-noise ratio of our X-ray and SZ datasets, we can perform the analysis presented in the previous Sections 
in each of the 8 azimuthal sectors with width of 45$^{\circ}$ that we define in Fig.~\ref{figure:sectorA2319}.
The analysis performed in sectors allows us to measure the azimuthal variance of the physical quantities and to assess
which are the cluster regions more relaxed.
Indeed, by dividing the observed count rate map in Fig~\ref{figure:VoronoiA2319} with a cluster model with perfect spherical symmetry 
and emission equal to the azimuthal median surface brightness profile, we can identify where an excess in the emission due to the ongoing merger 
is located. As shown in Fig.~\ref{figure:sectorA2319}, this excess is concentrated in the NW region (Sectors 1, 2 and 3, in particular).

\begin{figure*}[th]
\centering
\hbox{
\hspace{0.25\textwidth}
\includegraphics[width=0.5\textwidth]{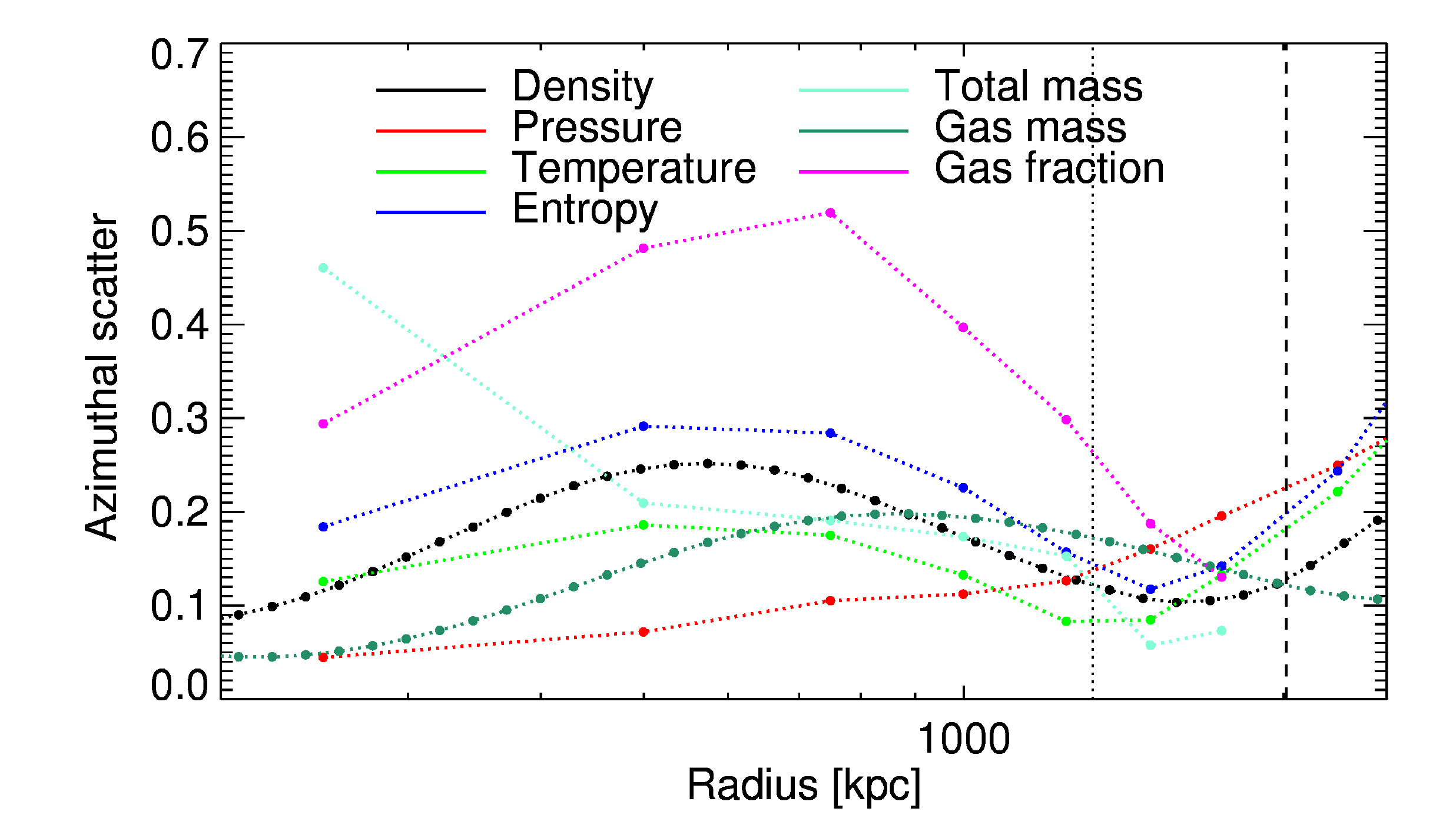}
 }
\hbox{ 
\includegraphics[width=0.5\textwidth]{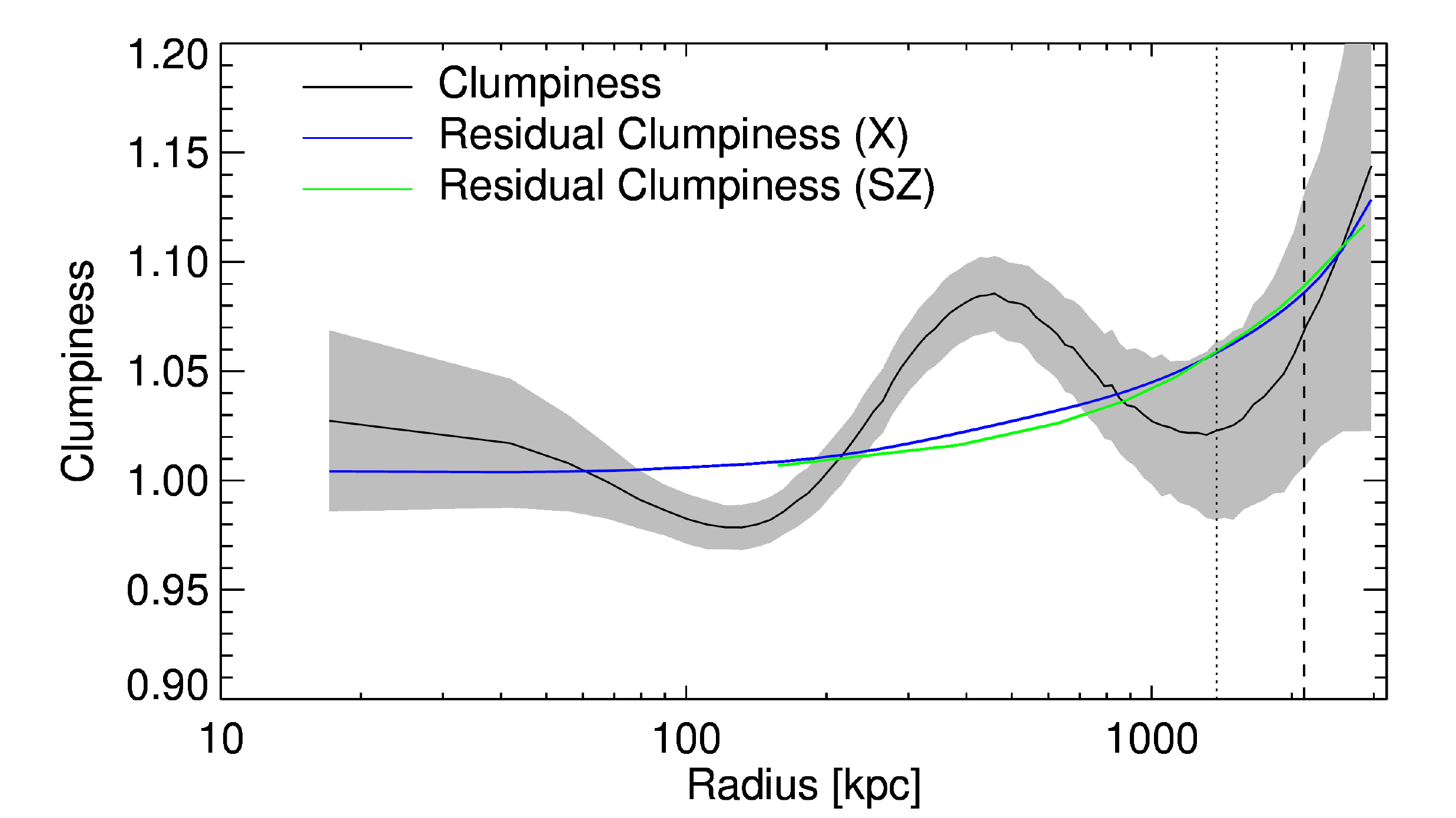}
\includegraphics[width=0.5\textwidth]{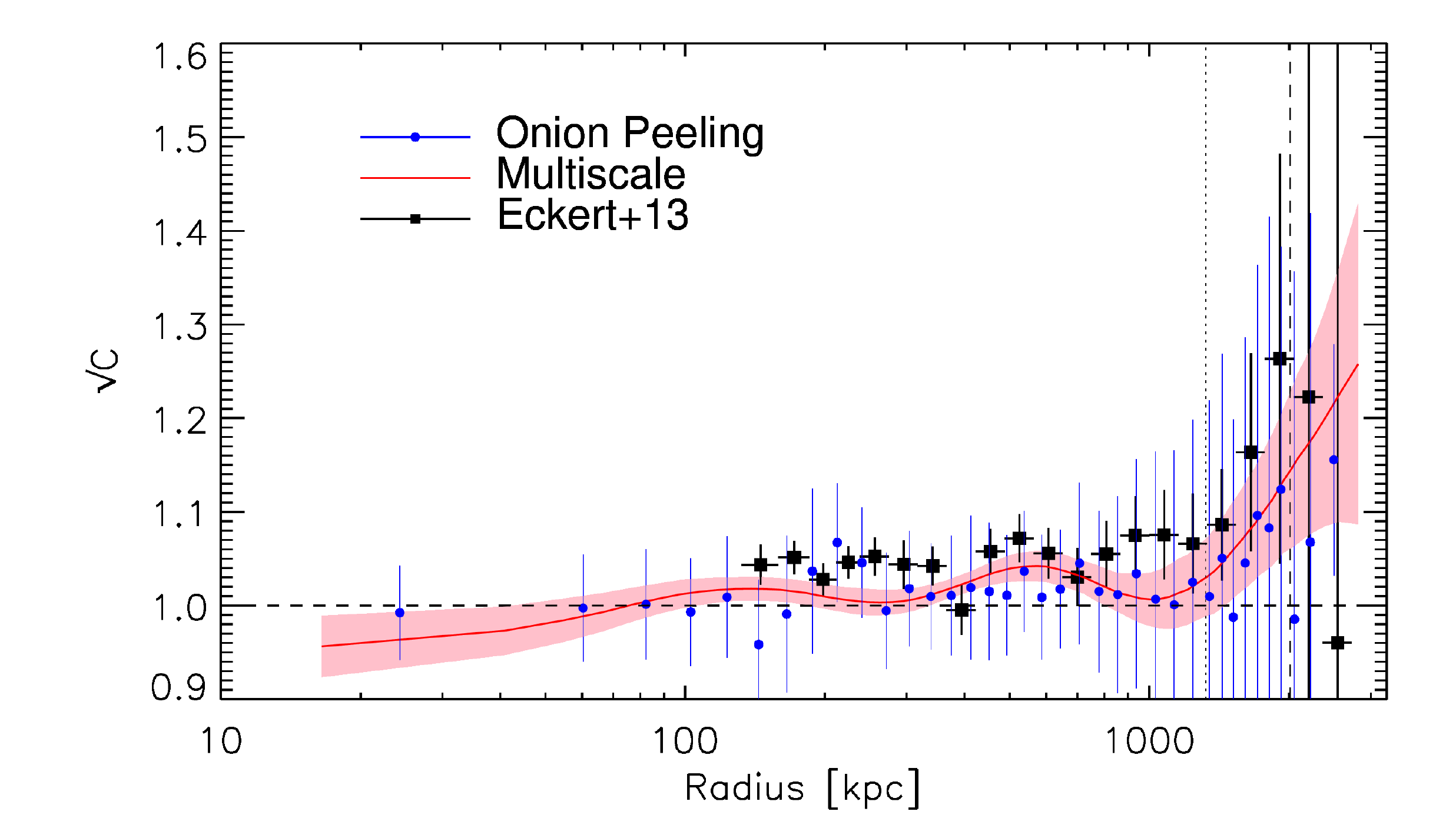}
}
\caption{
(Top) Azimuthal scatter in the thermodynamic profiles:
gas density and gas mass profiles are obtained from the X-ray spatial analysis; the pressure profile is the result of SZ data analysis;
gas entropy and temperature are obtained by combining SZ pressure and X-ray density; 
the total mass is reconstructed by solving the hydrostatic equilibrium equation using the \textit{forward} approach.
The vertical dotted and dashed lines represents the location of $R_{500}$ and $R_{200}$, respectively.
(Center-Left) Total measured clumpiness (see Sec.~5.1; black line; shaded region represents 1 sigma uncertainty) compared with the estimated residual clumpiness using X-ray density (blue line) and SZ comptonization parameter (green line). 
(Center-Right)
Same as Fig.~\ref{figure:clump} but removing the problematic sectors (1, 2 and 3) from the analysis. The features present in the whole clumpiness profile disappear almost completely. 
The vertical lines mark the position of $R_{500}$ and $R_{200}$, dotted and dashed respectively.
(Bottom-Left) Total measured clumpiness (see Sec.~5.1; black line; shaded region represents 1 sigma uncertainty) compared with the estimated residual clumpiness using X-ray density (blue line) and SZ comptonization parameter (green line). 
(Bottom-Right)
Same as Fig.~\ref{figure:clump} but removing the merging region in the problematic sectors (1, 2 and 3) from the analysis. The features present in the whole clumpiness profile disappear almost completely. 
The vertical lines mark the position of $R_{500}$ and $R_{200}$, dotted and dashed respectively.
}
\label{figure:biasetal}
\end{figure*}

We show the profiles of the thermodynamic properties recovered in 8 angular sectors in Fig.~\ref{figure:sectors}.

In the X-ray surface brightness, we identify various features specific in each sector:
\begin{itemize}
\item Sector 1 has an excess in emission starting above 200 kpc with a small radial extent of about 100 kpc. 
This excess is due to a contamination of the merging component in this cluster, located 10 arcmin NW.
\item Sector 2 has also a significant excess in the X-ray emission. 
This excess is located in the region where \cite{oegerle+95} found the merging component in A2319, and has a quite large radial extent from 200 to 800 kpc.
\item Sector 3 has an emission slightly higher than the azimuthally average up to 1 Mpc, where a sharp transition is present reconciling the surface brightness with 
the azimuthally averaged value. This sector shows evidence for a non-negligible contamination from the merger.
\item Sectors 4 and 5 are quite regular, with a behaviour very similar to the azimuthally-averaged profile. 
\item Sector 6 shows the cold front already detected in \cite{ghizzardi+10} and located in the SE region, about 200 kpc $\approx$ 3 arcmin from the cluster's center. 
\item Sectors 7 and 8 are the most regular ones, and reproduces very well the combined surface brightness profile.
\end{itemize}

The pressure profile obtained from the deprojected SZ signal in each sector (see Fig.~\ref{figure:sectors})
shows clearly that this is the quantity least affected by the dynamical history of the cluster. 
For instance, the merging event \citep{oegerle+95} happening in the NW (Sector~2) with mass ratio 3:1 is well resolved 
in the surface brightness/density profile, but it is not evident in the pressure profile (Sector~3 has the highest values in the pressure profile, 
nevertheless Sectors~1 and 2 are slightly below the azimuthally average profile), suggesting that the merger induced some shocks 
that have already propagated through the ICM and, at least partially, thermalized, inducing a reasonably small scatter in the pressure profile at $R_{200}$ (see Fig.~\ref{figure:sectors}).

From the spectral analysis, we observe that, in Sector~2, the gas temperature reaches values below the ones measured in the azimuthally averaged profile between 300 and 800 kpc. In Sectors~1 and 3, the temperature behaves similarly, but over a narrower radial range.
These radial variations can be explained by a low temperature component contaminating Sectors~1, 2 and 3 at intermediate radii. 
This can be associated to the accreting substructure visible in the residual map, see Figure~\ref{figure:sectorA2319}, which is merging with the main cluster halo.
Over the same region, corresponding to the merging component at about 500 kpc in Sector~2, 
we also observe an increase in the metal abundance correlated to the gas at the lower temperature.

In Fig.~\ref{figure:sectors}, we show the entropy profiles obtained by solving the HEE with the \textit{backward} method 
(a comparison between the profiles estimated with different methods is shown in  Fig.~\ref{fig:confrontoK}).
The entropy measured in Sector~2 is well below the mean value estimated in the cluster, while Sector~1 and 3 are just slightly below. 
This suggests that a substructure with a low-entropy gas is still accreting into the cluster's halo, as residual of the ongoing merger.

\subsection{Azimuthal scatter and clumpiness}

The azimuthal scatter of the recovered thermodynamic quantities is defined at each radius $r$ as
\begin{equation}
 \sigma_Q(r)=\sqrt{\frac{1}{N} \sum\limits_{i=1}^N \left( \frac{Q_i(r)-\bar{Q}_i(r)}{\bar{Q}_i(r)} \right)^2 },
\label{eq:azimuthal scatter}
\end{equation}
with $Q$=\{ {\it n, P, T, K}, $M_{tot}$, $M_g$, $f_g$ \}. The profiles of the azimuthal scatter are shown in Fig.~\ref{figure:biasetal}.

As a general trend, we expect that $\sigma_Q(r)$ should increase monotonously with radius, because, moving outward, the considered radial points should be less virialized.
Although this is generally observed, some other features also appear. For instance, at intermediate radii ($\sim$ 600 kpc) there is a clear increase coincidently 
with the clustercentric location where the merger is taking place. 
Moreover, there is a particular radial location between $R_{500}$ and $R_{200}$, where the azimuthal scatter reaches a minimum. 
This point suggests the radial extension of the influence of the merger on the thermodynamic quantities.

Using this information, we can improve the characterization of the properties of the observed clumpiness in the gas density.
As described in \cite{roncarelli+13}, the clumping factor of the gas (see Sect.~\ref{subsect:clumpiness}) is expected to have two major contributors: 
(i) some individual clumps, (ii) large-scale accretion patterns.
The latter is described by the residual clumping $C_R$, that, following \cite{roncarelli+13}, can be estimated as:
\begin{equation}
C_R(r) = 1 + \frac{\sigma}{\sigma_0} + \frac{r}{r_0},
\label{eq:res_clump}
\end{equation}
where $r=R/R_{200}$; $\sigma$ is the azimuthal scatter of the density $n$, or of the comptonization parameter $y$.
$\sigma_0$ and $r_0$ are estimated from simulations \citep{roncarelli+13}

\begin{itemize}
\item ($\sigma_0$, $r_0$) = (16.02, 5.87) for X-ray density
\item ($\sigma_0$, $r_0$) = (2.83, 8.25) for SZ comptonization parameter
\end{itemize}

We compare the estimated clumpiness with the residual clumpiness $C_R$ in Fig.~\ref{figure:biasetal}.
We observe that the measured clumping factor, both X-ray and SZ, only slightly exceeds the estimated $C_R$ over the entire radial range, 
suggesting that large-scale asymmetries account for most of the clumpiness measured.

Moreover, the clumpiness profile in Fig.~\ref{figure:clump} shows a clear excess at intermediate radii. We interpret this excess as the presence of the merger component in the NW direction. We evaluate again the clumpiness, after masking out Sectors~1, 2, and 3 more affected by the presence of the merger. 
As we show in Fig.~\ref{figure:biasetal}, the excess in the clumping factor at intermediate radii disappears and
the total clumpiness at $R_{200}$ decreases to 1.05.

\begin{figure*}[th]
\hbox{
\includegraphics[width=0.5\textwidth]{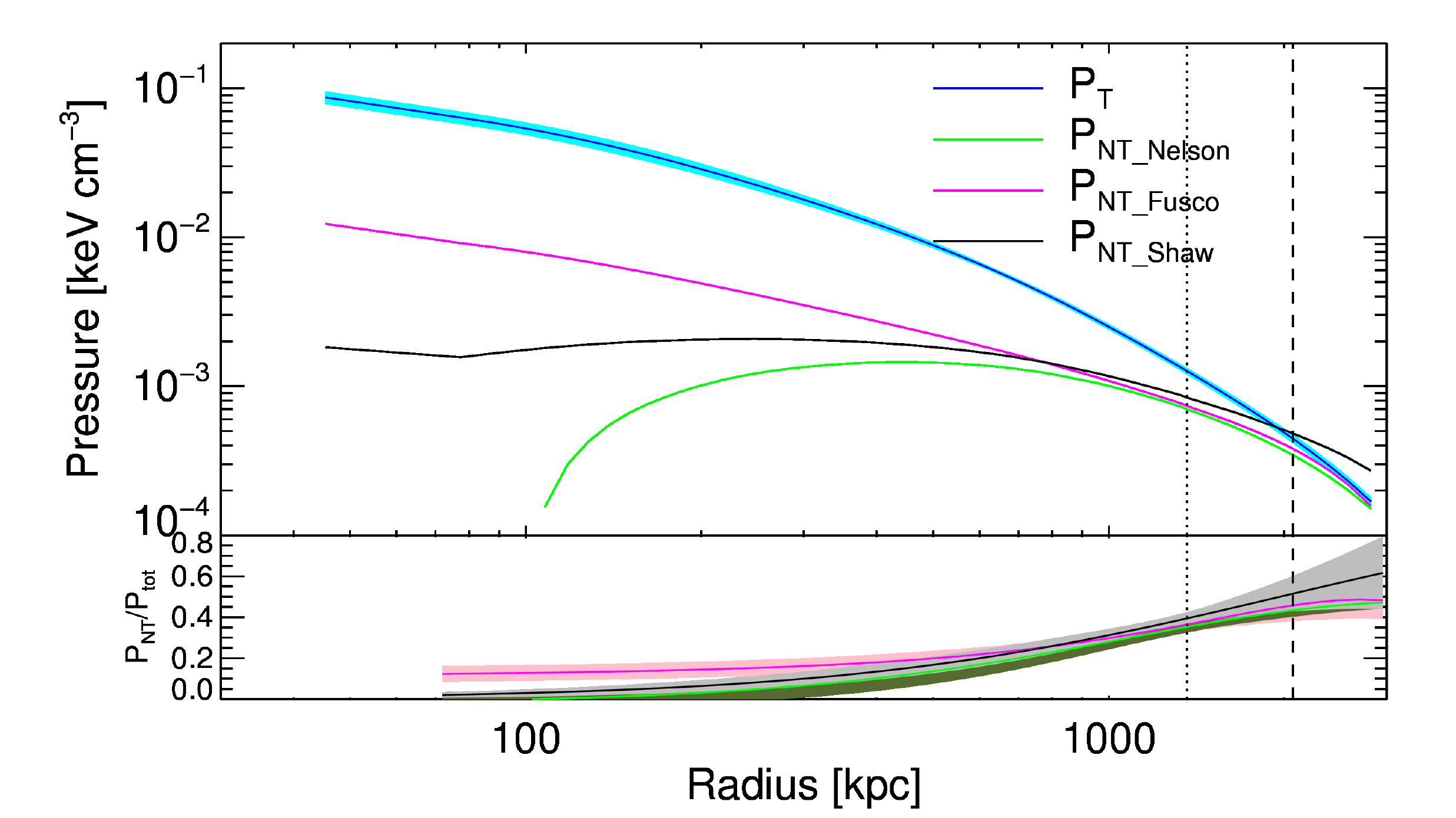}
\includegraphics[width=0.5\textwidth]{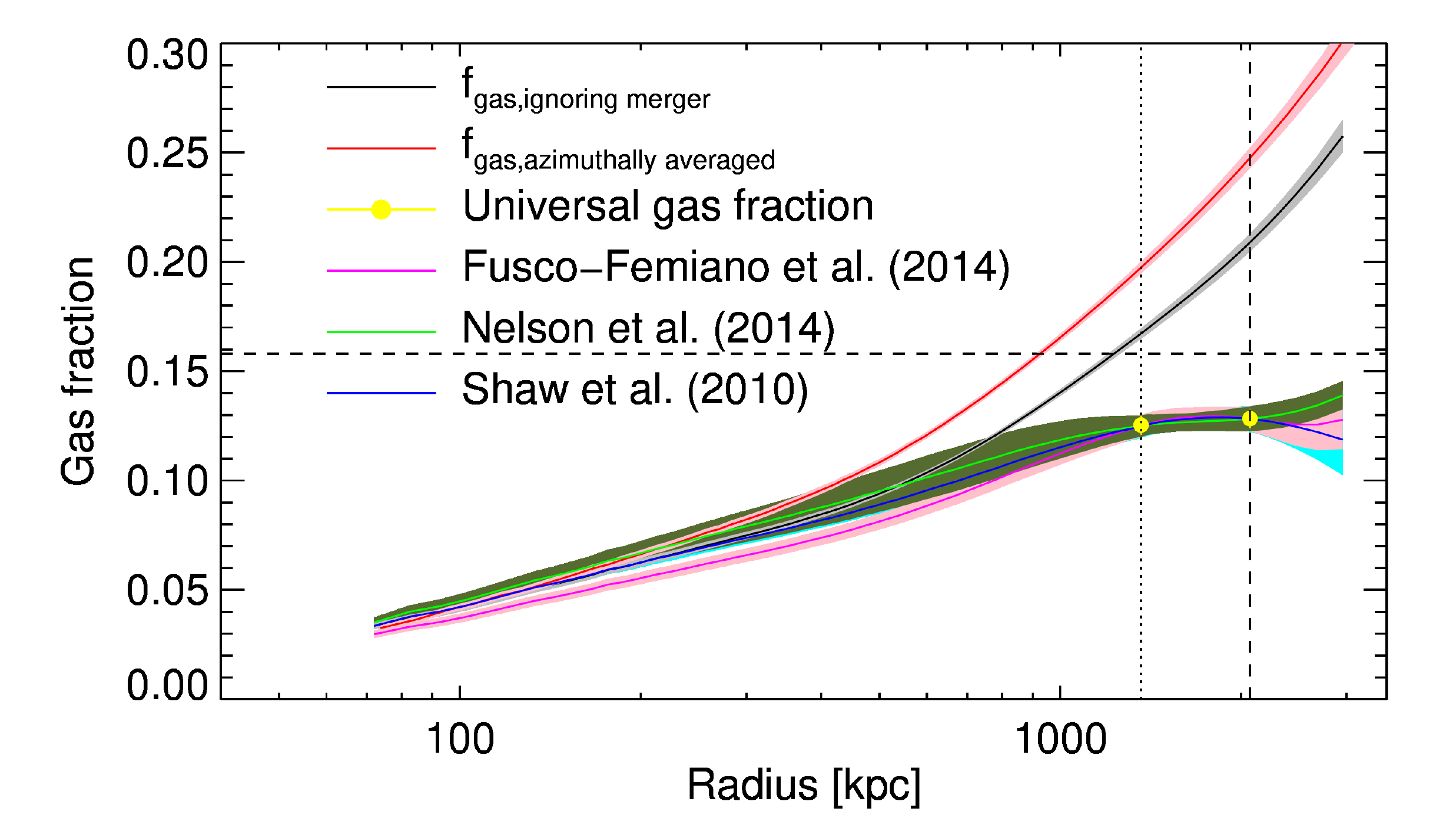}
}
\caption{
(Left)
Thermal pressure compared with non-thermal pressure using three different models \citep[black, pink, and green lines, ][respectively]{shaw+10,fusco+14,nelson+14}.
(Right)
Measured gas fraction profile azimuthally averaged (red line) and ignoring the merging region (black line), and corrected accounting for the contribution of a non-thermal pressure component enabling to match the cosmic gas fraction at $R_{200}$ and $R_{500}$.
The horizontal line represents the ``universal'' baryon fraction \citep{planck+16}, the vertical lines represents the position of $R_{500}$ and $R_{200}$, and the yellow points are the universal baryon fraction depleted by the thermalized gas and by the star fraction. The pink, green and blue line represent the gas fraction we get by using different functional form in order to reduce the observed gas mass fraction to the universal one.
}
\label{figure:fgas_nt}
\end{figure*}

\section{Characterizing the hydrostatic bias}

\subsection{Gas mass fraction and the non-thermal contribution}
\label{sub:fgas}
Since galaxy clusters originate from large regions of the primordial Universe, their baryon fraction is expected to be close to the universal fraction.

The gas mass fraction, $f_g = M_g / M_{\rm tot}$, in massive galaxy clusters represents most of the baryons accreted in the dark matter halo and is a good proxy of the cosmic baryonic budget, which enables us to use galaxy clusters as a cosmological probe \citep[e.g.][]{ettori+02,ettori+09}.
\begin{equation}
 \frac{\Omega_b}{\Omega_m} \cdot b = f_g + f_{\rm star} 
\label{eq:depl} 
\end{equation}
where $\Omega_b$ and $\Omega_m$ are the cosmological baryon and matter density, $b$ is the depletion factor that accounts for the cosmic baryons which thermalize in the cluster's potential, and $f_{\rm star}$ is the stellar mass fraction.
Here, we adopt the cosmological parameters estimated from the \planck\ collaboration, $\Omega_b = 0.045$ and $\Omega_m = 0.3089$ \citep[][]{planck+16}, 
we assume from numerical simulations \citep[e.g.][]{planelles+13} $b =$ 0.85 and 0.87 (with a standard deviation of 0.03) at $R_{500}$ and $R_{200}$, respectively, and 
consider $M_{\rm star} / M_{\rm gas} = 0.069$ from optical measurements in nearby systems \citep{gonzalez+13}.
We predict, thus, a gas mass fraction $f_g$ of 0.125 and 0.128 at $R_{500}$ and $R_{200}$, respectively.

However, we measure a gas fraction, already corrected for the resolved gas clumpiness using the median profile,
that reaches values well above the expected $f_g$ at $r>R_{500}$ (see Fig.~\ref{figure:sectors} and \ref{eq:depl}).
We advocate the role of the non-thermal pressure contribution to the estimate of the total mass in lowering the measured gas fraction.

Indeed, Abell~2319 is in a merging state \citep{oegerle+95}, with the presence of a giant radio halo \citep{farnsworth+13,storm+15} that supports this scenario.
The measured gas fraction
can be then biased high as a consequence of the phenomena (like gas turbulence and bulk motion) 
that occur during a merger and that are not accounted for in the calculation of the hydrostatic mass, 
causing an underestimate of the halo mass.

Before proceeding in quantifying the amount of non-thermal pressure support, we note, from the analysis in azimuthal sectors, 
that the substructure that is merging with the main halo is also able to disturb the system on a much larger scale, by enhancing the measured surface brightness up to $\sim$ 1 Mpc.
The net effect is to increase the gas mass by about 10\% and so the relative amount of non thermal pressure in the outskirts.
To obtain an estimate of the contribution of the non-thermal pressure unbiased from any evident merger, 
we ignore the region where we measure this excess in the surface brightness (see red sector in Figure~\ref{figure:sectorA2319}), and repeat our analysis.
We show the comparison between the results obtained before and after masking the merging region in Table~\ref{table:newfit}.
The  hydrostatic mass remains unchanged, but the gas mass decreases, implying that the gas fraction lowers by 17\% at $R_{200}$, but it is still larger than the cosmological gas fraction predicted from numerical simulations at these radii. We remark that the reconstructed gas fraction is already corrected for  the resolved gas clumping using the median density profile, therefore clumpiness cannot be responsible for the excess gas fraction\citep{simionescu+11}.

\begin{table*}[h]
\begin{center}
\begin{tabular}{ c c c c c}
Region & $M_{200}$ ($10^{14} M_\odot$) &  $R_{200}$ (kpc) & $M_{gas,200}$ ($10^{14} M_\odot$) &  $f_{gas,200}$     \\
\hline
Azimuthally average & $10.7 \pm 0.5$ & $2077 \pm 33$  & $2.54 \pm 0.05$ & $0.237 \pm 0.012$ \\
Ignoring the merging region & $10.7 \pm 0.3$ & $2075 \pm 17$  & $2.22 \pm 0.02$ & $0.207 \pm 0.006$ \\
\end{tabular}
\caption{Comparison between the mass reconstruction at $R_{200}$ using the whole surface brightness image and ignoring the merging component. 
The columns show: the hydrostatic mass by solving HEE (see Eq.~\eqref{eq:hee}), $R_{200}$, the gas mass obtained by integrating the gas density profile 
(Eq.~\eqref{eq:mgas}), and the gas mass fraction defined by $f_{gas} = M_{gas}/M_{tot}$.
}
\label{table:newfit}
\end{center}

\begin{center}
\begin{tabular}{ c c c c c }
Model & Functional form for $\alpha = P_{NT} / P_{T}$ & $a$ & $b$ & $c$ \\
\hline
\cite{nelson+14} & $\left[ a \left( 1 + \exp \left( - \left( \frac{R}{R_{200} b} \right)^c \right) \right) \right]^{-1} -1 $ & $0.52 \pm 0.02$ & 0.52 (fix) & $1.23 \pm 0.27$ \\
\cite{fusco+14} & $ a \exp \left( - \left( \frac{ 1 - R/(2 R_{500}) }{b} \right)^2 \right) $ & $0.91 \pm 0.18$ & $0.706 \pm 0.09$ & --\\
\cite{shaw+10} & $a \left( \frac{R}{R500} \right)^b$ & $0.63 \pm 0.05$ & $1.17 \pm 0.36$ & --
\end{tabular}
\caption{Model, functional form, and best fitting parameters for the three models which describe the ratio between non-thermal and thermal pressure support.}
\label{table:pnt}
\end{center}

\end{table*}

\begin{figure*}[th]
\centering
\hbox{
\includegraphics[width=0.5\textwidth,page=1]{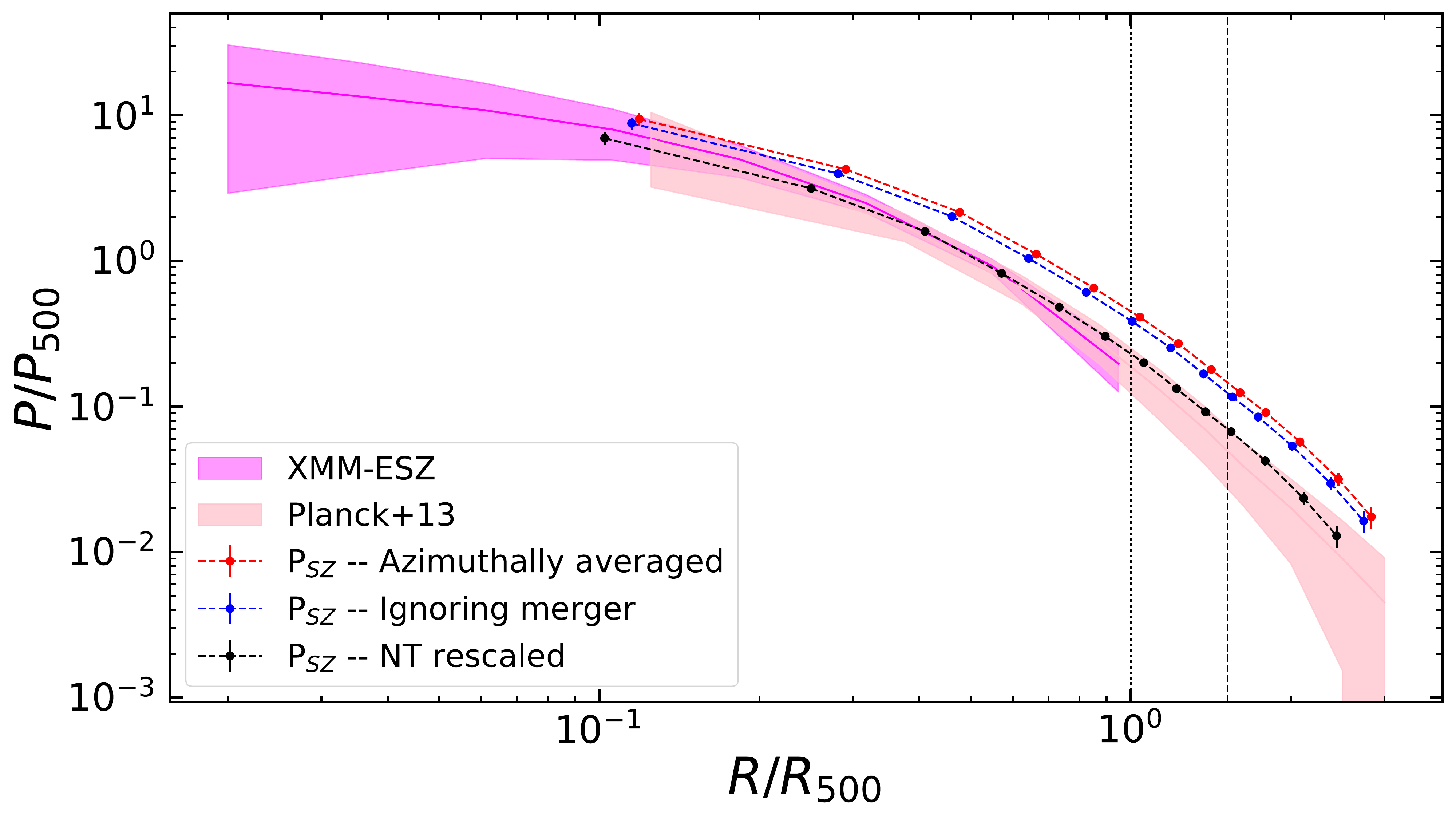}
\includegraphics[width=0.5\textwidth,page=2]{plots/denspres.pdf}
}
\hbox{
\includegraphics[width=0.5\textwidth,page=3]{plots/denspres.pdf}
\includegraphics[width=0.5\textwidth,page=4]{plots/denspres.pdf}
}
\caption{
Rescaled pressure (Top-left) and density (Top-right) profiles considering the azimuthally averaged, ignoring the merger, and ignoring the merging region and consider the $M_{200,tot}$ and $R_{200,tot}$ required to recover the cosmological gas fraction at the virial radius. 
We compare these profiles with the \planck\ envelope \citep{planck+13}, for pressure, and with the universal density profile \citep{eckert+12}, for density.
(Bottom) Rescaled entropy and rescaled entropy corrected by the gas mass fraction, before and after correcting for the true total mass.
}
\label{figure:bcorr}
\end{figure*}	

One possibility to explain this overestimate in the gas fraction is the presence of a substantial non-thermal pressure component
 in the HEE which breaks the hydrostatic equilibrium assumption. 
We modify the HEE in Equation~\ref{eq:hee}, by adding an extra pressure component, that we define as ``non-thermal'' pressure and justify as 
generated by e.g. unresolved gas turbulence, bulk motion, magnetic field, or asphericity. 
This non-thermal component can be modelled, in first approximation, as a constant fraction of the thermal one \citep{loeb+94,zappacosta+06}. 
We add this non-thermal pressure term (indicated with the subscript ``NT'') in the HEE as $P_{NT}(r) = \alpha(r) P_T(r)$, where 
the thermal component has the subscript ``T'', and $\alpha(r)$ is a function of radius.
The HEE is then modified as 
\begin{equation}
 \frac{1}{\rho_g} \left( \frac{dP_T}{dr} + \frac{dP_{NT}}{dr} \right) = - \frac{G}{r^2} \left( M_T + M_{NT} \right).
 \label{eq:hee_nt}
\end{equation}
 and by substituting the non-thermal part we get

By solving the derivatives and readjusting the terms in the equation, we can then write how this propagates into the estimate of the gas mass fraction:
\begin{eqnarray}
f_{g} & = & \frac{M_{g}}{M_{T} + M_{NT}} = \frac{M_{g}}{M_{T}\left( 1 + \frac{M_{NT}}{M_T} \right)} \nonumber \\
& = &  \frac{f_{g,T}}{1+\alpha(r)-\frac{P_T r^2}{GM_T\mu m_p n_e} \frac{d\alpha}{dr}}  \equiv \beta f_{g,T}
\label{eq:fgas_nt}
\end{eqnarray}
with $\beta$ defined as the ratio between the true gas fraction and the measured thermal gas fraction.
This means that in the case of $\alpha = constant$, the real gas fraction is reduced by a factor $1+\alpha$.

By imposing that the observed cluster gas fraction should match the cosmic value in Eq.~\eqref{eq:depl}, and assuming a constant $\alpha$, 
we require $\alpha=0.64$ $(0.32)$ at $R_{200}$ ($R_{500}$), implying that about 39\% (24\%) of the total pressure is in the form of a non thermal component.

In general, $\alpha$ is expected to have a radial dependence.
Numerical simulations \citep[e.g.][]{shaw+10,fusco+14,nelson+14} predict some functional forms for $P_{NT}/P_T$. We can constrain the parameters of these models by requiring that, if we consider the radial dependence of $\alpha$ in HEE, we are able to reproduce the expected gas mass fraction at $R_{500}$ and $R_{200}$.  The errors on the parameters are calculated using Monte Carlo simulations propagating the errors on the gas mass fraction profile, on the measure of $R_{200}$, and on the predicted gas mass fraction points. The non-thermal pressure profiles, and the corresponding gas fraction profiles, obtained using the above mentioned models are shown in Fig.~\ref{figure:fgas_nt}, and in Table~\ref{table:pnt} we provide the three functional form adopted and the best fitting parameters.
We observe that already above 200-300 kpc, the non-thermal pressure support plays a very important role in flattening the gas mass fraction profile. 

Finally, by imposing that the total cluster mass $M_{tot}$ is provided from $M_{T} + M_{NT}$, we can estimate the amount of the hydrostatic bias factor $\beta$ as
\begin{equation}
\beta = \frac{M_T}{M_{tot}} \ \Rightarrow \ M_{tot} = \frac{M_T}{\beta}.
\label{eq:bias}
\end{equation}
Applying Equations~\eqref{eq:hee_nt}, \eqref{eq:fgas_nt} and \eqref{eq:bias}, the cosmological gas fraction at $R_{500}$ and $R_{200}$ is obtained
by requiring 
\[M_{500, tot} = 10.2 \pm 0.4^\text{stat.} \pm 0.4^\text{syst.} \times 10^{14} M_\odot\]
\[M_{200, tot} = 17.3 \pm 0.9^\text{stat.} \pm 1.2^\text{syst.} \times 10^{14} M_\odot\]

Using this mass estimate corrected both by clumpiness and hydrostatic bias, and the value acquired from the \planck\ catalog \citep{catalog2} and based on scaling relations, 
$M_{Y_{SZ}, 500} = 8.74 (\pm 0.12) \times 10^{14} M_\odot$, we infer a \planck\ bias of $1-b = M_{Y_{SZ}, 500} / M_{500, tot} \approx 0.86$. 

\subsubsection{Effects of the hydrostatic bias on the rescaled profiles}

The correction on the mass propagates to the rescale profiles, both directly since $R_{500}$ increases shrinking the x-axis, and indirectly since pressure and entropy, as described from Eq.~\eqref{eq:P500} and \eqref{eq:K500} respectively, follow a rescaling which is mass dependent.

In Fig.~\ref{figure:bcorr}, we show the net effect on the thermodynamic rescaled profiles, that can be summarized in the following statements:
\begin{itemize}
\item the gas pressure profile is now in agreement both with the universal pressure profile \citep{arnaud+10} and with the \planck\ envelope \citep{planck+13};
\item the gas density profile becomes compatible with the stacked density profile presented in \cite{eckert+12};
\item the gas entropy profile shows the least modification before and after this analysis; the profile becomes slightly steeper, however it is still flat in the outskirts, in agreement with the expected impact of any non-thermal pressure support \citep{walker+12}.
\end{itemize}

\cite{pratt+10} have shown that in order to reconcile entropy profile with predictions from non-radiative simulations\citep{voit+05}, the profile has to be corrected by the gas mass fraction
$K \ \Rightarrow \ K \cdot \left( E(z) f_{gas} / f_b \right)^{2/3}$.
Introducing this correction in each entropy profile we consider (i.e. the azimuthally average profile, the profile ignoring the merging region, and the profile required to recover the cosmological gas fraction at $R_{200}$), we obtain the results shown in Fig.~\ref{figure:bcorr}. We observe that only when we include the contribution by the non-thermal pressure we obtain 
a corrected entropy profile that deviates from the numerical predictions, with a flattening above 0.3$R_{500}$ suggesting that turbulence, or non thermal energy at large, has not been yet converted efficiently in heat energy, not allowing the specific entropy of the ICM to rise to the value expected in systems simulated in the absence of non-gravitational processes \citep[e.g.][]{voit+05}.

\section{Summary and conclusions}

The very accurate background modeling of the \xmm\ exposures, and the large extension of of the SZ signal resolved with \planck\ 
allow to combine X-ray and SZ data to study the thermodynamic properties of Abell 2319 over the virial region around $R_{200}$. 
Moreover, since the data quality is very high, we are able to study the properties of this cluster reaching the virial radius in 8 different sectors. 
This enables us to study the azimuthal variance of the thermodynamic properties of the ICM in this merging system for the first time.

The measured clumpiness shows the presence of the merging component with an increase in its value at intermediate radii ($\sim$ 500 kpc). 
This excess disappears when we remove the merging regions from the analysis.
On the other hand, in the outskirts, the clumpiness measured is compatible with the estimated residual clumpiness \citep{roncarelli+13}. 
This means that this cluster has no significative infalling clumps at the virial radius. 

The gas density profile corrected for the resolved clumpiness is then used to recover other fundamental quantities \citep{eckert+15},
together with the gas temperature profile that we measure, from the X-ray spectroscopic analysis, with a median relative statistical uncertainty of 2 per cent and 
with a systematic error that we carefully estimate to be in the order of (median value) 4 per cent, and above 15\% in the outermost radial bin only.
The exquisite quality of these complementary X-ray and SZ datasets, extending across $R_{200}$, enable us to constrain a NFW hydrostatic mass profile 
at very high precision ($M_{200} = 10.7 \pm 0.5^\text{stat.} \pm 0.9^\text{syst.} \times 10^{14} M_\odot$), 
achieving a level where systematic errors dominate over the statistical ones. 

Due to the merging state of this cluster, the recovered entropy profile is flatter than the predicted one by non-radiative simulations \citep{voit+05}.
We observe the most deviations in the first and last few points: in the center this is caused by the fact that this cluster is a well known non cool core cluster \citep{cavagnolo+09} 
with a flat entropy core of $\sim 75$ keV cm$^2$, while some residual non-thermal energy flattens the entropy in the outskirts \citep{walker+12}.

The pressure profile recovered from SZ data is flatter, and above the 1$\sigma$ envelope, than the ``universal'' one measured 
for an ensemble of objects resolved with \planck\ \citep{planck+13}.

The measured gas fraction, corrected by the gas clumpiness using the median density profile, is above 
the value predicted from state-of-art hydrodynamical simulations for the preferred cosmological background 
\citep{planck+16,planelles+13,gonzalez+13}. 
Analyzing the azimuthal variation of the $f_{gas}$ profile (see Fig.~\ref{figure:sectors}), we observe that it is above the average value only in the
sectors most affected by the merger ( i.e. Sectors 1, 2 and 3).
When the region with the ongoing merger and with an estimated higher gas mass is excluded from the analysis,
the gas fraction drops but is still higher than the expectations, indicating a non negligible contribution from a non-thermal pressure support that we quantify 
in the order of 39\% and 24\% of the total pressure at $R_{200}$ and $R_{500}$, respectively.

Once the correction induced by the non-thermal pressure support is propagated through the measurements of $R_{500}$, $K_{500}$, and $P_{500}$, we show that:
(i) the pressure profile matches the mean behaviour of objects resolved with \planck; 
(ii) the gas density profile becomes consistent with the stacked profile obtained from Rosat/PSPC observations in \cite{eckert+12};
(iii) on the contrary, the entropy undergoes a very small change, remaining flatter than the predicted profile.

In forthcoming works, the detailed analysis presented here for A2319 will be extended to the whole X-COP sample \citep{xcop}, 
providing the first ensembled properties of the ICM at $R_{200}$ and above.

\begin{acknowledgements} 
This research has received funding from the European Union's Horizon 2020 Programme under AHEAD project (grant agreement n. 654215).
SE acknowledges the financial support from contracts ASI-INAF I/009/10/0, 
NARO15 ASI-INAF I/037/12/0 and ASI 2015-046-R.0.
\end{acknowledgements} 

\bibliographystyle{aa} 
\bibliography{X-COP-A2319} 

\begin{thebibliography}{75}
\expandafter\ifx\csname natexlab\endcsname\relax\def\natexlab#1{#1}\fi

\bibitem[{{Ameglio} {et~al.}(2007){Ameglio}, {Borgani}, {Pierpaoli}, \&
  {Dolag}}]{ameglio+07}
{Ameglio}, S., {Borgani}, S., {Pierpaoli}, E., \& {Dolag}, K. 2007, \mnras,
  382, 397

\bibitem[{{Arnaud}(1996)}]{xspec}
{Arnaud}, K.~A. 1996, in Astronomical Society of the Pacific Conference Series,
  Vol. 101, Astronomical Data Analysis Software and Systems V, ed. G.~H.
  {Jacoby} \& J.~{Barnes}, 17

\bibitem[{{Arnaud} {et~al.}(2010){Arnaud}, {Pratt}, {Piffaretti},
  {B{\"o}hringer}, {Croston}, \& {Pointecouteau}}]{arnaud+10}
{Arnaud}, M., {Pratt}, G.~W., {Piffaretti}, R., {et~al.} 2010, \aap, 517, A92

\bibitem[{{Battaglia} {et~al.}(2012){Battaglia}, {Bond}, {Pfrommer}, \&
  {Sievers}}]{battaglia+12}
{Battaglia}, N., {Bond}, J.~R., {Pfrommer}, C., \& {Sievers}, J.~L. 2012, \apj,
  758, 75

\bibitem[{{Binney} \& {Tremaine}(1987)}]{hee}
{Binney}, J. \& {Tremaine}, S. 1987, {Galactic dynamics}

\bibitem[{{Bourdin} {et~al.}(2017){Bourdin}, {Mazzotta}, {Kozmanyan}, {Jones},
  \& {Vikhlinin}}]{bourdin+17}
{Bourdin}, H., {Mazzotta}, P., {Kozmanyan}, A., {Jones}, C., \& {Vikhlinin}, A.
  2017, \apj, 843, 72

\bibitem[{{Cavagnolo} {et~al.}(2009){Cavagnolo}, {Donahue}, {Voit}, \&
  {Sun}}]{cavagnolo+09}
{Cavagnolo}, K.~W., {Donahue}, M., {Voit}, G.~M., \& {Sun}, M. 2009, \apjs,
  182, 12

\bibitem[{{Croston} {et~al.}(2006){Croston}, {Arnaud}, {Pointecouteau}, \&
  {Pratt}}]{croston+06}
{Croston}, J.~H., {Arnaud}, M., {Pointecouteau}, E., \& {Pratt}, G.~W. 2006,
  \aap, 459, 1007

\bibitem[{{De Luca} \& {Molendi}(2004)}]{cxb}
{De Luca}, A. \& {Molendi}, S. 2004, \aap, 419, 837

\bibitem[{{Diehl} \& {Statler}(2006)}]{voronoi}
{Diehl}, S. \& {Statler}, T.~S. 2006, \mnras, 368, 497

\bibitem[{{Dolag} {et~al.}(1999){Dolag}, {Bartelmann}, \& {Lesch}}]{dolag+99}
{Dolag}, K., {Bartelmann}, M., \& {Lesch}, H. 1999, \aap, 348, 351

\bibitem[{{Eckert} {et~al.}(2016){Eckert}, {Ettori}, {Coupon}, {Gastaldello},
  {Pierre}, {Melin}, {Le Brun}, {McCarthy}, {Adami}, {Chiappetti}, {Faccioli},
  {Giles}, {Lavoie}, {Lef{\`e}vre}, {Lieu}, {Mantz}, {Maughan}, {McGee},
  {Pacaud}, {Paltani}, {Sadibekova}, {Smith}, \& {Ziparo}}]{eckert+16}
{Eckert}, D., {Ettori}, S., {Coupon}, J., {et~al.} 2016, \aap, 592, A12

\bibitem[{{Eckert} {et~al.}(2017){Eckert}, {Ettori}, {Pointecouteau},
  {Molendi}, {Paltani}, \& {Tchernin}}]{xcop}
{Eckert}, D., {Ettori}, S., {Pointecouteau}, E., {et~al.} 2017, Astronomische
  Nachrichten, 338, 293

\bibitem[{{Eckert} {et~al.}(2015){Eckert}, {Roncarelli}, {Ettori}, {Molendi},
  {Vazza}, {Gastaldello}, \& {Rossetti}}]{eckert+15}
{Eckert}, D., {Roncarelli}, M., {Ettori}, S., {et~al.} 2015, \mnras, 447, 2198

\bibitem[{{Eckert} {et~al.}(2012){Eckert}, {Vazza}, {Ettori}, {Molendi},
  {Nagai}, {Lau}, {Roncarelli}, {Rossetti}, {Snowden}, \&
  {Gastaldello}}]{eckert+12}
{Eckert}, D., {Vazza}, F., {Ettori}, S., {et~al.} 2012, \aap, 541, A57

\bibitem[{{Ettori} {et~al.}(2002){Ettori}, {De Grandi}, \&
  {Molendi}}]{ettori+02}
{Ettori}, S., {De Grandi}, S., \& {Molendi}, S. 2002, \aap, 391, 841

\bibitem[{{Ettori} {et~al.}(2013){Ettori}, {Donnarumma}, {Pointecouteau},
  {Reiprich}, {Giodini}, {Lovisari}, \& {Schmidt}}]{ettori+13}
{Ettori}, S., {Donnarumma}, A., {Pointecouteau}, E., {et~al.} 2013, \ssr, 177,
  119

\bibitem[{{Ettori} {et~al.}(2010){Ettori}, {Gastaldello}, {Leccardi},
  {Molendi}, {Rossetti}, {Buote}, \& {Meneghetti}}]{ettori+10}
{Ettori}, S., {Gastaldello}, F., {Leccardi}, A., {et~al.} 2010, \aap, 524, A68

\bibitem[{{Ettori} {et~al.}(2017){Ettori}, {Ghirardini}, {Eckert}, {Dubath}, \&
  {Pointecouteau}}]{ettori+17}
{Ettori}, S., {Ghirardini}, V., {Eckert}, D., {Dubath}, F., \& {Pointecouteau},
  E. 2017, \mnras, 470, L29

\bibitem[{{Ettori} \& {Molendi}(2011)}]{ettori+11}
{Ettori}, S. \& {Molendi}, S. 2011, Memorie della Societa Astronomica Italiana
  Supplementi, 17, 47

\bibitem[{{Ettori} {et~al.}(2009){Ettori}, {Morandi}, {Tozzi}, {Balestra},
  {Borgani}, {Rosati}, {Lovisari}, \& {Terenziani}}]{ettori+09}
{Ettori}, S., {Morandi}, A., {Tozzi}, P., {et~al.} 2009, \aap, 501, 61

\bibitem[{{Farnsworth} {et~al.}(2013){Farnsworth}, {Rudnick}, {Brown}, \&
  {Brunetti}}]{farnsworth+13}
{Farnsworth}, D., {Rudnick}, L., {Brown}, S., \& {Brunetti}, G. 2013, \apj,
  779, 189

\bibitem[{{Foreman-Mackey} {et~al.}(2013){Foreman-Mackey}, {Hogg}, {Lang}, \&
  {Goodman}}]{emcee}
{Foreman-Mackey}, D., {Hogg}, D.~W., {Lang}, D., \& {Goodman}, J. 2013, \pasp,
  125, 306

\bibitem[{{Fruscione} {et~al.}(2006){Fruscione}, {McDowell}, {Allen},
  {Brickhouse}, {Burke}, {Davis}, {Durham}, {Elvis}, {Galle}, {Harris},
  {Huenemoerder}, {Houck}, {Ishibashi}, {Karovska}, {Nicastro}, {Noble},
  {Nowak}, {Primini}, {Siemiginowska}, {Smith}, \& {Wise}}]{fruscione+06}
{Fruscione}, A., {McDowell}, J.~C., {Allen}, G.~E., {et~al.} 2006, 6270, 62701V

\bibitem[{{Fusco-Femiano} \& {Lapi}(2014)}]{fusco+14}
{Fusco-Femiano}, R. \& {Lapi}, A. 2014, \apj, 783, 76

\bibitem[{{Ghizzardi} {et~al.}(2010){Ghizzardi}, {Rossetti}, \&
  {Molendi}}]{ghizzardi+10}
{Ghizzardi}, S., {Rossetti}, M., \& {Molendi}, S. 2010, \aap, 516, A32

\bibitem[{{Gonzalez} {et~al.}(2013){Gonzalez}, {Sivanandam}, {Zabludoff}, \&
  {Zaritsky}}]{gonzalez+13}
{Gonzalez}, A.~H., {Sivanandam}, S., {Zabludoff}, A.~I., \& {Zaritsky}, D.
  2013, \apj, 778, 14

\bibitem[{{Hurier} {et~al.}(2013){Hurier}, {Mac{\'{\i}}as-P{\'e}rez}, \&
  {Hildebrandt}}]{hurier+13}
{Hurier}, G., {Mac{\'{\i}}as-P{\'e}rez}, J.~F., \& {Hildebrandt}, S. 2013,
  \aap, 558, A118

\bibitem[{{Kalberla} {et~al.}(2005){Kalberla}, {Burton}, {Hartmann}, {Arnal},
  {Bajaja}, {Morras}, \& {P{\"o}ppel}}]{LAB}
{Kalberla}, P.~M.~W., {Burton}, W.~B., {Hartmann}, D., {et~al.} 2005, \aap,
  440, 775

\bibitem[{{Kriss} {et~al.}(1983){Kriss}, {Cioffi}, \& {Canizares}}]{kriss+83}
{Kriss}, G.~A., {Cioffi}, D.~F., \& {Canizares}, C.~R. 1983, \apj, 272, 439

\bibitem[{{Kuntz} \& {Snowden}(2008)}]{kuntz+08}
{Kuntz}, K.~D. \& {Snowden}, S.~L. 2008, \aap, 478, 575

\bibitem[{{Lamarre} {et~al.}(2010){Lamarre}, {Puget}, {Ade}, {Bouchet},
  {Guyot}, {Lange}, {Pajot}, {Arondel}, {Benabed}, {Beney}, {Beno{\^i}t},
  {Bernard}, {Bhatia}, {Blanc}, {Bock}, {Br{\'e}elle}, {Bradshaw}, {Camus},
  {Catalano}, {Charra}, {Charra}, {Church}, {Couchot}, {Coulais}, {Crill},
  {Crook}, {Dassas}, {de Bernardis}, {Delabrouille}, {de Marcillac}, {Delouis},
  {D{\'e}sert}, {Dumesnil}, {Dupac}, {Efstathiou}, {Eng}, {Evesque},
  {Fourmond}, {Ganga}, {Giard}, {Gispert}, {Guglielmi}, {Haissinski},
  {Henrot-Versill{\'e}}, {Hivon}, {Holmes}, {Jones}, {Koch}, {Lagard{\`e}re},
  {Lami}, {Land{\'e}}, {Leriche}, {Leroy}, {Longval},
  {Mac{\'{\i}}as-P{\'e}rez}, {Maciaszek}, {Maffei}, {Mansoux}, {Marty}, {Masi},
  {Mercier}, {Miville-Desch{\^e}nes}, {Moneti}, {Montier}, {Murphy},
  {Narbonne}, {Nexon}, {Paine}, {Pahn}, {Perdereau}, {Piacentini}, {Piat},
  {Plaszczynski}, {Pointecouteau}, {Pons}, {Ponthieu}, {Prunet}, {Rambaud},
  {Recouvreur}, {Renault}, {Ristorcelli}, {Rosset}, {Santos}, {Savini},
  {Serra}, {Stassi}, {Sudiwala}, {Sygnet}, {Tauber}, {Torre}, {Tristram},
  {Vibert}, {Woodcraft}, {Yurchenko}, \& {Yvon}}]{lamarre+10}
{Lamarre}, J.-M., {Puget}, J.-L., {Ade}, P.~A.~R., {et~al.} 2010, \aap, 520, A9

\bibitem[{{Leccardi} \& {Molendi}(2008)}]{lm+08}
{Leccardi}, A. \& {Molendi}, S. 2008, \aap, 486, 359

\bibitem[{{Liu} {et~al.}(2017){Liu}, {Chiao}, {Collier}, {Cravens}, {Galeazzi},
  {Koutroumpa}, {Kuntz}, {Lallement}, {Lepri}, {McCammon}, {Morgan}, {Porter},
  {Snowden}, {Thomas}, {Uprety}, {Ursino}, \& {Walsh}}]{bubble}
{Liu}, W., {Chiao}, M., {Collier}, M.~R., {et~al.} 2017, \apj, 834, 33

\bibitem[{{Loeb} \& {Mao}(1994)}]{loeb+94}
{Loeb}, A. \& {Mao}, S. 1994, \apjl, 435, L109

\bibitem[{{Mazzotta} {et~al.}(2004){Mazzotta}, {Rasia}, {Moscardini}, \&
  {Tormen}}]{mazzotta+04}
{Mazzotta}, P., {Rasia}, E., {Moscardini}, L., \& {Tormen}, G. 2004, \mnras,
  354, 10

\bibitem[{{McCammon} {et~al.}(2002){McCammon}, {Almy}, {Apodaca}, {Bergmann
  Tiest}, {Cui}, {Deiker}, {Galeazzi}, {Juda}, {Lesser}, {Mihara},
  {Morgenthaler}, {Sanders}, {Zhang}, {Figueroa-Feliciano}, {Kelley},
  {Moseley}, {Mushotzky}, {Porter}, {Stahle}, \& {Szymkowiak}}]{halo}
{McCammon}, D., {Almy}, R., {Apodaca}, E., {et~al.} 2002, \apj, 576, 188

\bibitem[{{Molendi} {et~al.}(1999){Molendi}, {De Grandi}, {Fusco-Femiano},
  {Colafrancesco}, {Fiore}, {Nesci}, \& {Tamburelli}}]{temp}
{Molendi}, S., {De Grandi}, S., {Fusco-Femiano}, R., {et~al.} 1999, \apjl, 525,
  L73

\bibitem[{{Morandi} {et~al.}(2007){Morandi}, {Ettori}, \&
  {Moscardini}}]{morandi+07}
{Morandi}, A., {Ettori}, S., \& {Moscardini}, L. 2007, \mnras, 379, 518

\bibitem[{{Nagai} {et~al.}(2007){Nagai}, {Kravtsov}, \& {Vikhlinin}}]{nagai+07}
{Nagai}, D., {Kravtsov}, A.~V., \& {Vikhlinin}, A. 2007, \apj, 668, 1

\bibitem[{{Nagai} \& {Lau}(2011)}]{nagai+11}
{Nagai}, D. \& {Lau}, E.~T. 2011, \apjl, 731, L10

\bibitem[{{Navarro} {et~al.}(1997){Navarro}, {Frenk}, \& {White}}]{nfw+97}
{Navarro}, J.~F., {Frenk}, C.~S., \& {White}, S.~D.~M. 1997, \apj, 490, 493

\bibitem[{{Nelson} {et~al.}(2014){Nelson}, {Lau}, \& {Nagai}}]{nelson+14}
{Nelson}, K., {Lau}, E.~T., \& {Nagai}, D. 2014, \apj, 792, 25

\bibitem[{{Oegerle} {et~al.}(1995){Oegerle}, {Hill}, \&
  {Fitchett}}]{oegerle+95}
{Oegerle}, W.~R., {Hill}, J.~M., \& {Fitchett}, M.~J. 1995, \aj, 110, 32

\bibitem[{{Pfrommer} {et~al.}(2007){Pfrommer}, {Springel}, {Jubelgas}, \&
  {Ensslin}}]{pfrommer+07}
{Pfrommer}, C., {Springel}, V., {Jubelgas}, M., \& {Ensslin}, T.~A. 2007, in
  Astronomical Society of the Pacific Conference Series, Vol. 379, Cosmic
  Frontiers, ed. N.~{Metcalfe} \& T.~{Shanks}, 221

\bibitem[{{Pierre} {et~al.}(2016){Pierre}, {Pacaud}, {Adami}, {Alis},
  {Altieri}, {Baran}, {Benoist}, {Birkinshaw}, {Bongiorno}, {Bremer}, {Brusa},
  {Butler}, {Ciliegi}, {Chiappetti}, {Clerc}, {Corasaniti}, {Coupon}, {De
  Breuck}, {Democles}, {Desai}, {Delhaize}, {Devriendt}, {Dubois}, {Eckert},
  {Elyiv}, {Ettori}, {Evrard}, {Faccioli}, {Farahi}, {Ferrari}, {Finet},
  {Fotopoulou}, {Fourmanoit}, {Gandhi}, {Gastaldello}, {Gastaud},
  {Georgantopoulos}, {Giles}, {Guennou}, {Guglielmo}, {Horellou}, {Husband},
  {Huynh}, {Iovino}, {Kilbinger}, {Koulouridis}, {Lavoie}, {Le Brun}, {Le
  Fevre}, {Lidman}, {Lieu}, {Lin}, {Mantz}, {Maughan}, {Maurogordato},
  {McCarthy}, {McGee}, {Melin}, {Melnyk}, {Menanteau}, {Novak}, {Paltani},
  {Plionis}, {Poggianti}, {Pomarede}, {Pompei}, {Ponman}, {Ramos-Ceja},
  {Ranalli}, {Rapetti}, {Raychaudury}, {Reiprich}, {Rottgering}, {Rozo},
  {Rykoff}, {Sadibekova}, {Santos}, {Sauvageot}, {Schimd}, {Sereno}, {Smith},
  {Smol{\v c}i{\'c}}, {Snowden}, {Spergel}, {Stanford}, {Surdej}, {Valageas},
  {Valotti}, {Valtchanov}, {Vignali}, {Willis}, \& {Ziparo}}]{pierre+16}
{Pierre}, M., {Pacaud}, F., {Adami}, C., {et~al.} 2016, \aap, 592, A1

\bibitem[{{Planck Collaboration} {et~al.}(2016{\natexlab{a}}){Planck
  Collaboration}, {Adam}, {Ade}, {Aghanim}, {Akrami}, {Alves}, {Arg{\"u}eso},
  {Arnaud}, {Arroja}, {Ashdown}, \& et~al.}]{planckI+15}
{Planck Collaboration}, {Adam}, R., {Ade}, P.~A.~R., {et~al.}
  2016{\natexlab{a}}, \aap, 594, A1

\bibitem[{{Planck Collaboration} {et~al.}(2014){Planck Collaboration}, {Ade},
  {Aghanim}, {Armitage-Caplan}, {Arnaud}, {Ashdown}, {Atrio-Barandela},
  {Aumont}, {Aussel}, {Baccigalupi}, \& et~al.}]{catalog}
{Planck Collaboration}, {Ade}, P.~A.~R., {Aghanim}, N., {et~al.} 2014, \aap,
  571, A29

\bibitem[{{Planck Collaboration} {et~al.}(2013){Planck Collaboration}, {Ade},
  {Aghanim}, {Arnaud}, {Ashdown}, {Atrio-Barandela}, {Aumont}, {Baccigalupi},
  {Balbi}, {Banday}, \& et~al.}]{planck+13}
{Planck Collaboration}, {Ade}, P.~A.~R., {Aghanim}, N., {et~al.} 2013, \aap,
  550, A131

\bibitem[{{Planck Collaboration} {et~al.}(2016{\natexlab{b}}){Planck
  Collaboration}, {Ade}, {Aghanim}, {Arnaud}, {Ashdown}, {Aumont},
  {Baccigalupi}, {Banday}, {Barreiro}, {Barrena}, \& et~al.}]{catalog2}
{Planck Collaboration}, {Ade}, P.~A.~R., {Aghanim}, N., {et~al.}
  2016{\natexlab{b}}, \aap, 594, A27

\bibitem[{{Planck Collaboration} {et~al.}(2016{\natexlab{c}}){Planck
  Collaboration}, {Ade}, {Aghanim}, {Arnaud}, {Ashdown}, {Aumont},
  {Baccigalupi}, {Banday}, {Barreiro}, {Bartlett}, \& et~al.}]{planck+16}
{Planck Collaboration}, {Ade}, P.~A.~R., {Aghanim}, N., {et~al.}
  2016{\natexlab{c}}, \aap, 594, A13

\bibitem[{{Planck HFI Core Team} {et~al.}(2011){Planck HFI Core Team}, {Ade},
  {Aghanim}, {Ansari}, {Arnaud}, {Ashdown}, {Aumont}, {Banday}, {Bartelmann},
  {Bartlett}, {Battaner}, {Benabed}, {Beno{\^i}t}, {Bernard}, {Bersanelli},
  {Bhatia}, {Bock}, {Bond}, {Borrill}, {Bouchet}, {Boulanger}, {Bradshaw},
  {Br{\'e}elle}, {Bucher}, {Camus}, {Cardoso}, {Catalano}, {Challinor},
  {Chamballu}, {Charra}, {Charra}, {Chary}, {Chiang}, {Church}, {Clements},
  {Colombi}, {Couchot}, {Coulais}, {Cressiot}, {Crill}, {Crook}, {de
  Bernardis}, {Delabrouille}, {Delouis}, {D{\'e}sert}, {Dolag}, {Dole},
  {Dor{\'e}}, {Douspis}, {Efstathiou}, {Eng}, {Filliard}, {Forni}, {Fosalba},
  {Fourmond}, {Ganga}, {Giard}, {Girard}, {Giraud-H{\'e}raud}, {Gispert},
  {G{\'o}rski}, {Gratton}, {Griffin}, {Guyot}, {Haissinski}, {Harrison},
  {Helou}, {Henrot-Versill{\'e}}, {Hern{\'a}ndez-Monteagudo}, {Hildebrandt},
  {Hills}, {Hivon}, {Hobson}, {Holmes}, {Huffenberger}, {Jaffe}, {Jones},
  {Kaplan}, {Kneissl}, {Knox}, {Lagache}, {Lamarre}, {Lami}, {Lange},
  {Lasenby}, {Lavabre}, {Lawrence}, {Leriche}, {Leroy}, {Longval},
  {Mac{\'{\i}}as-P{\'e}rez}, {Maciaszek}, {MacTavish}, {Maffei}, {Mandolesi},
  {Mann}, {Mansoux}, {Masi}, {Matsumura}, {McGehee}, {Melin}, {Mercier},
  {Miville-Desch{\^e}nes}, {Moneti}, {Montier}, {Mortlock}, {Murphy}, {Nati},
  {Netterfield}, {N{\o}rgaard-Nielsen}, {North}, {Noviello}, {Novikov},
  {Osborne}, {Paine}, {Pajot}, {Patanchon}, {Peacocke}, {Pearson}, {Perdereau},
  {Perotto}, {Piacentini}, {Piat}, {Plaszczynski}, {Pointecouteau}, {Pons},
  {Ponthieu}, {Pr{\'e}zeau}, {Prunet}, {Puget}, {Reach}, {Renault},
  {Ristorcelli}, {Rocha}, {Rosset}, {Roudier}, {Rowan-Robinson}, {Rusholme},
  {Santos}, {Savini}, {Schaefer}, {Shellard}, {Spencer}, {Starck}, {Stassi},
  {Stolyarov}, {Stompor}, {Sudiwala}, {Sunyaev}, {Sygnet}, {Tauber}, {Thum},
  {Torre}, {Touze}, {Tristram}, {van Leeuwen}, {Vibert}, {Vibert}, {Wade},
  {Wandelt}, {White}, {Wiesemeyer}, {Woodcraft}, {Yurchenko}, {Yvon}, \&
  {Zacchei}}]{planckHFI}
{Planck HFI Core Team}, {Ade}, P.~A.~R., {Aghanim}, N., {et~al.} 2011, \aap,
  536, A4

\bibitem[{{Planelles} {et~al.}(2013){Planelles}, {Borgani}, {Dolag}, {Ettori},
  {Fabjan}, {Murante}, \& {Tornatore}}]{planelles+13}
{Planelles}, S., {Borgani}, S., {Dolag}, K., {et~al.} 2013, \mnras, 431, 1487

\bibitem[{{Pratt} {et~al.}(2010){Pratt}, {Arnaud}, {Piffaretti},
  {B{\"o}hringer}, {Ponman}, {Croston}, {Voit}, {Borgani}, \&
  {Bower}}]{pratt+10}
{Pratt}, G.~W., {Arnaud}, M., {Piffaretti}, R., {et~al.} 2010, \aap, 511, A85

\bibitem[{{Read} {et~al.}(2011){Read}, {Rosen}, {Saxton}, \&
  {Ramirez}}]{read+11}
{Read}, A.~M., {Rosen}, S.~R., {Saxton}, R.~D., \& {Ramirez}, J. 2011, \aap,
  534, A34

\bibitem[{{Roncarelli} {et~al.}(2013){Roncarelli}, {Ettori}, {Borgani},
  {Dolag}, {Fabjan}, \& {Moscardini}}]{roncarelli+13}
{Roncarelli}, M., {Ettori}, S., {Borgani}, S., {et~al.} 2013, \mnras, 432, 3030

\bibitem[{{Salvetti} {et~al.}(2017){Salvetti}, {Marelli}, {Gastaldello},
  {Ghizzardi}, {Molendi}, {De Luca}, {Moretti}, {Rossetti}, \&
  {Tiengo}}]{salvetti+17}
{Salvetti}, D., {Marelli}, M., {Gastaldello}, F., {et~al.} 2017, ArXiv e-prints
  [\eprint[arXiv]{1705.04172}]

\bibitem[{{Schellenberger} {et~al.}(2015){Schellenberger}, {Reiprich},
  {Lovisari}, {Nevalainen}, \& {David}}]{schellenberger+15}
{Schellenberger}, G., {Reiprich}, T.~H., {Lovisari}, L., {Nevalainen}, J., \&
  {David}, L. 2015, \aap, 575, A30

\bibitem[{{Shaw} {et~al.}(2010){Shaw}, {Nagai}, {Bhattacharya}, \&
  {Lau}}]{shaw+10}
{Shaw}, L.~D., {Nagai}, D., {Bhattacharya}, S., \& {Lau}, E.~T. 2010, \apj,
  725, 1452

\bibitem[{{Simionescu} {et~al.}(2011){Simionescu}, {Allen}, {Mantz}, {Werner},
  {Takei}, {Morris}, {Fabian}, {Sanders}, {Nulsen}, {George}, \&
  {Taylor}}]{simionescu+11}
{Simionescu}, A., {Allen}, S.~W., {Mantz}, A., {et~al.} 2011, Science, 331,
  1576

\bibitem[{{Snowden} {et~al.}(2008){Snowden}, {Mushotzky}, {Kuntz}, \&
  {Davis}}]{esas}
{Snowden}, S.~L., {Mushotzky}, R.~F., {Kuntz}, K.~D., \& {Davis}, D.~S. 2008,
  \aap, 478, 615

\bibitem[{{Storm} {et~al.}(2015){Storm}, {Jeltema}, \& {Rudnick}}]{storm+15}
{Storm}, E., {Jeltema}, T.~E., \& {Rudnick}, L. 2015, \mnras, 448, 2495

\bibitem[{{Struble} \& {Rood}(1999)}]{struble+99}
{Struble}, M.~F. \& {Rood}, H.~J. 1999, \apjs, 125, 35

\bibitem[{{Sunyaev} \& {Zeldovich}(1972)}]{SZ}
{Sunyaev}, R.~A. \& {Zeldovich}, Y.~B. 1972, Comments on Astrophysics and Space
  Physics, 4, 173

\bibitem[{{Tauber} {et~al.}(2010){Tauber}, {Mandolesi}, {Puget}, {Banos},
  {Bersanelli}, {Bouchet}, {Butler}, {Charra}, {Crone}, {Dodsworth}, \&
  et~al.}]{tauber+10}
{Tauber}, J.~A., {Mandolesi}, N., {Puget}, J.-L., {et~al.} 2010, \aap, 520, A1

\bibitem[{{Tchernin} {et~al.}(2016){Tchernin}, {Eckert}, {Ettori},
  {Pointecouteau}, {Paltani}, {Molendi}, {Hurier}, {Gastaldello}, {Lau},
  {Nagai}, {Roncarelli}, \& {Rossetti}}]{A2142}
{Tchernin}, C., {Eckert}, D., {Ettori}, S., {et~al.} 2016, \aap, 595, A42

\bibitem[{{Tozzi} \& {Norman}(2001)}]{tozzi+01}
{Tozzi}, P. \& {Norman}, C. 2001, \apj, 546, 63

\bibitem[{{Vazza} {et~al.}(2013){Vazza}, {Eckert}, {Simionescu}, {Br{\"u}ggen},
  \& {Ettori}}]{vazza+13}
{Vazza}, F., {Eckert}, D., {Simionescu}, A., {Br{\"u}ggen}, M., \& {Ettori}, S.
  2013, \mnras, 429, 799

\bibitem[{{Vazza} {et~al.}(2011){Vazza}, {Roncarelli}, {Ettori}, \&
  {Dolag}}]{vazza+11}
{Vazza}, F., {Roncarelli}, M., {Ettori}, S., \& {Dolag}, K. 2011, \mnras, 413,
  2305

\bibitem[{{Vikhlinin} {et~al.}(2006){Vikhlinin}, {Kravtsov}, {Forman}, {Jones},
  {Markevitch}, {Murray}, \& {Van Speybroeck}}]{vikhlini+06}
{Vikhlinin}, A., {Kravtsov}, A., {Forman}, W., {et~al.} 2006, \apj, 640, 691

\bibitem[{{Voit} {et~al.}(2005){Voit}, {Kay}, \& {Bryan}}]{voit+05}
{Voit}, G.~M., {Kay}, S.~T., \& {Bryan}, G.~L. 2005, \mnras, 364, 909

\bibitem[{{Walker} {et~al.}(2012){Walker}, {Fabian}, {Sanders}, \&
  {George}}]{walker+12}
{Walker}, S.~A., {Fabian}, A.~C., {Sanders}, J.~S., \& {George}, M.~R. 2012,
  \mnras, 427, L45

\bibitem[{{Willingale} {et~al.}(2013){Willingale}, {Starling}, {Beardmore},
  {Tanvir}, \& {O'Brien}}]{willingale+13}
{Willingale}, R., {Starling}, R.~L.~C., {Beardmore}, A.~P., {Tanvir}, N.~R., \&
  {O'Brien}, P.~T. 2013, \mnras, 431, 394

\bibitem[{{Zappacosta} {et~al.}(2006){Zappacosta}, {Buote}, {Gastaldello},
  {Humphrey}, {Bullock}, {Brighenti}, \& {Mathews}}]{zappacosta+06}
{Zappacosta}, L., {Buote}, D.~A., {Gastaldello}, F., {et~al.} 2006, \apj, 650,
  777

\bibitem[{{Zhuravleva} {et~al.}(2013){Zhuravleva}, {Churazov}, {Kravtsov},
  {Lau}, {Nagai}, \& {Sunyaev}}]{zhuravleva+13}
{Zhuravleva}, I., {Churazov}, E., {Kravtsov}, A., {et~al.} 2013, \mnras, 428,
  3274

\end{thebibliography}

\begin{appendix}

\section{Non X-ray background modeling}
\label{app_dom}

\begin{figure*}[ht]
\hbox{
\includegraphics[width=0.5\textwidth]{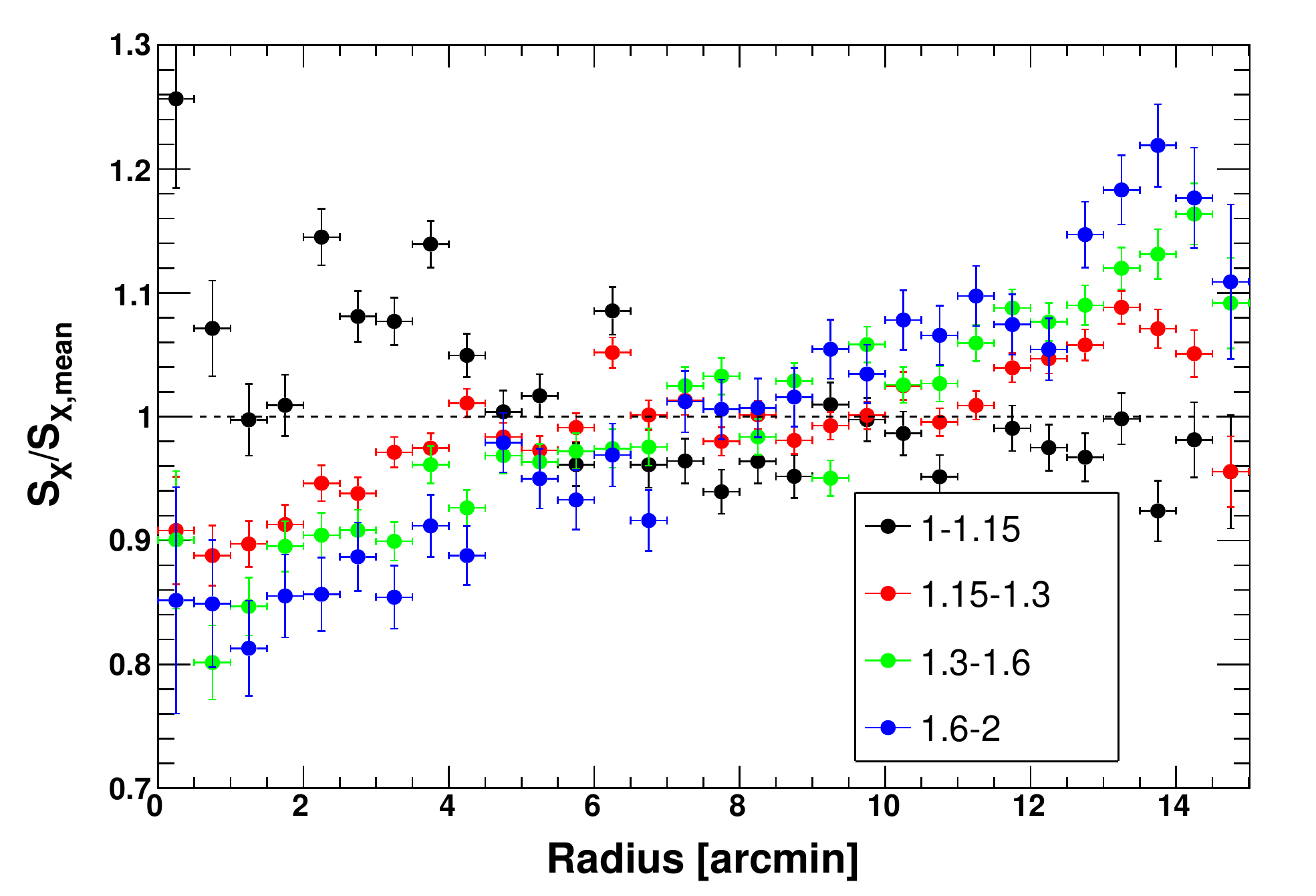}
\includegraphics[width=0.5\textwidth]{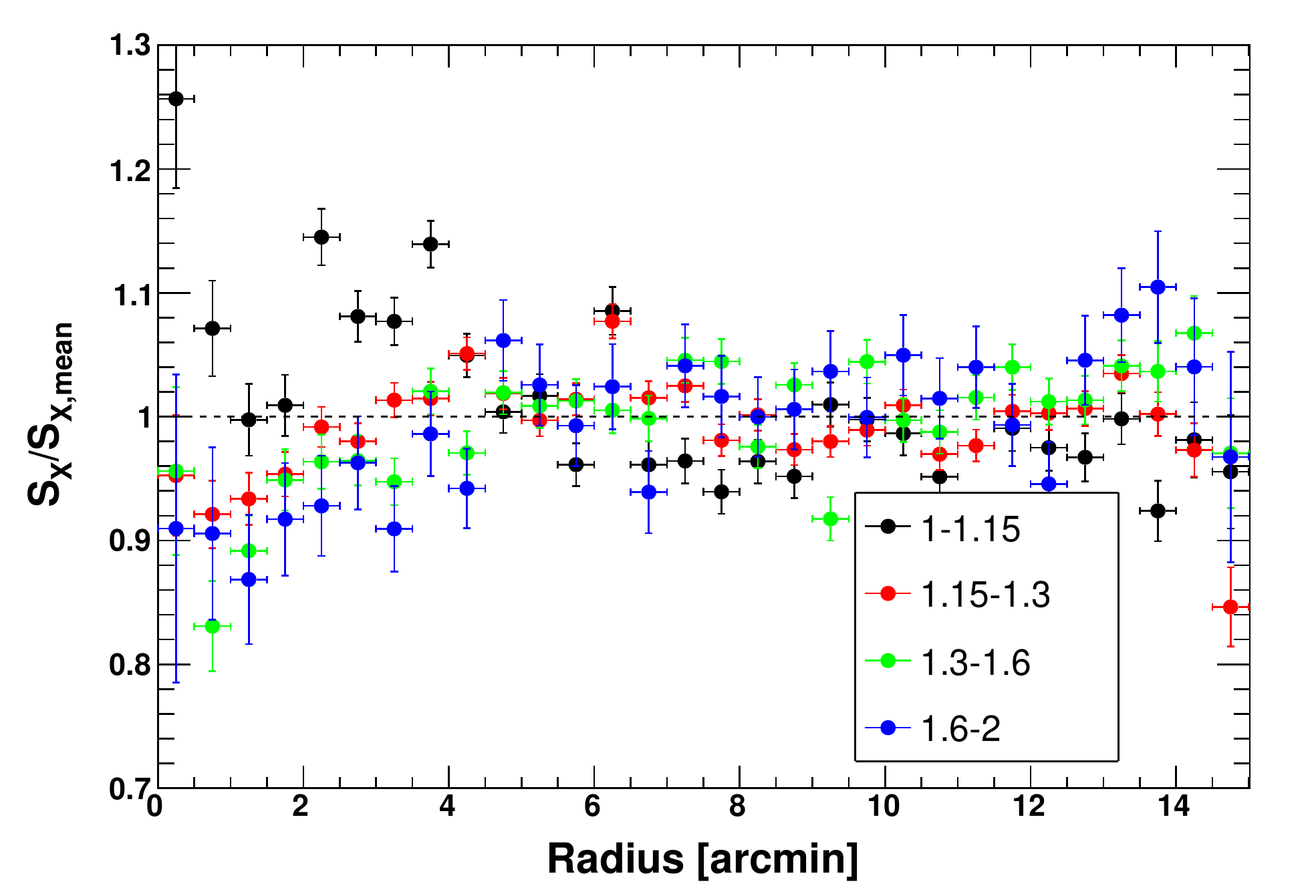}
}
\caption{Stacked EPIC radial profiles of 495 blank-sky pointings, sorted in bins of soft-proton contamination inFOV/outFOV. The black data points show observations with low SP contamination (inFOV/outFOV=1-1.15), whereas the blue points comprise observations that were severely affected by SP contamination (inFOV/outFOV=1.6-2.0). The left-hand panel shows the stacked profiles obtained when subtracting only the QPB component, while in the right-hand panel, the SP and QC components have been taken into account following Eq. \ref{eq:spmodel}.}
\label{fig:calibration}
\end{figure*}

We developed and calibrated a novel technique to model and subtract the non X-ray background (NXB). Our approach builds upon the method devised in \cite{A2142}, however it can be more reliably applied to observations including a significant source emission above 5 keV. Here we describe the main principles of our method and validate it using a large set of blank-sky \xmm\ pointings.

\subsection{Model}

It has long been known that the NXB of \xmm\ is split into two main components, the quiescent particle background (QPB) and the soft protons (SP). Recently, \citet{salvetti+17} analyzed almost the complete \xmm\ archive and showed the presence of an additional stable, low-intensity component within the field of view (FOV) of the \mos 2 instrument, whose origin is yet unknown. As described in Section~\ref{sec:xmm}, a fraction of the area of the \mos\ detectors is located outside the FOV of the \xmm\ telescopes. The outFOV area can be used to estimate the QPB level in each observation by rescaling filter-wheel-closed data to the measured outFOV count rate. The remaining inFOV high-energy count rate can then be decomposed into a variable component (SP) and a quiescent part (QC). We can thus describe the remaining NXB as 

\begin{equation}
{\rm inFOV} - {\rm outFOV} = {\rm SP} + {\rm QC}
\end{equation}
where inFOV and outFOV denote the [7-11.5] keV \mos 2 count rates measured in the exposed and unexposed areas of the detector, respectively. We restrict the measurement to the \mos 2 detector as two of the \mos 1 chips have been lost throughout the mission, and the unexposed area of the pn detector is too small for our needs. 

Importantly, the SP component is expected to show a different spatial signature on the detector compared to the QPB. Indeed, soft protons, which are funneled towards the detector through the telescope, are more spatially concentrated than the QPB and follow a vignetting curve SP$(r)$ that is different from the vignetting curve of the photons \citep{kuntz+08}, where $r$ denotes the distance of each pixel from the aim point. Conversely, given that its origin is currently unclear, the spatial distribution of the QC component is unknown. Here we make the hypothesis that this component is flat over the detector. 

\subsection{Blank-sky dataset and modeling}

To determine the relative contributions of the SP and QC components, we used a large set of 495 \xmm\ blank-sky pointings, most of which from the XXL survey \citep{pierre+16}. Our dataset comprises more than 5 Ms of data. We processed the data using ESAS in the same way as for the A2319 data (see Section~\ref{sec:xmm}). We estimated the QPB component in each observation by measuring the outFOV count rate and rescaling filter-wheel-closed data. We also compute the high-energy inFOV and outFOV count rates for each observation. We then measured the radial profiles in the [0.7-1.2] keV band of the blank-sky pointings from the aim point to the outermost edge of the pointing in annuli of 30 arcsec width. The detected sources were masked and the QPB was subtracted from the data. As already shown in Tchernin et al. (2016), this procedure results in radial profiles that are on average not flat, which indicates the need of modeling additional components (SP and QC). 

We then describe the radial profiles $S_{X}(r)$ as the sum of the SP and QC components following their respective spatial distributions, 

\begin{equation}
S_{X}(r)=C+N_{QC}+N_{SP}({\rm inFOV}-{\rm outFOV} - \bar{QC}){\rm SP}(r),\label{eq:spmodel}
\end{equation}

where $C$ is the sky background intensity at the relevant location, $N_{QC}$ the intensity of the stable QC component, $N_{SP}$ the normalization of the variable SP, and $\bar{QC}=0.023$ counts/s is the mean high-energy count rate of the QC component \citep{salvetti+17}. We then perform a joint fit on all the measured profiles and optimize for the values of $N_{QC}$ and $N_{SP}$. We then used the best-fit values of $N_{QC}$ and $N_{SP}$ to create 2D models of these components and subtract them from the data.

In Fig. \ref{fig:calibration} we show the stacked radial profiles of the full sample. In the left-hand panel we show the stacked profiles obtained when subtracting the QPB component only, whereas in the right-hand panel, the SP and QC components have been modeled using the method described above and subtracted from the data. To investigate the dependence of our results on SP contamination, we grouped the data in bins of increasing SP contamination, which we trace using the inFOV/outFOV ratio (Leccardi \& Molendi 2008). Observations that were mildly affected by SP contamination exhibit a inFOV/outFOV ratio close to one, whereas heavily contaminated observations show high values of the inFOV/outFOV ratio. The effect of SP contamination is evident in the left-hand panel of Fig. \ref{fig:calibration}, where the deviations of the stacked profiles from a straight line progressively increase with increasing SP contamination. Conversely, when applying our SP and QC modeling approach, flat profiles are found in all 4 bins out to the edge of the FOV, indicating that our model accurately reproduces the various NXB components. The excess scatter compared to a straight line is $5\%$, which we adopt as our systematic uncertainty in the subtraction of the NXB.

\section{Results of the spectral fitting}

\begin{table}[h]
\resizebox{0.5\textwidth}{!} {
\begin{tabular}{ c | c | c | c | c | c | c }
radii & C-stat. & PHA bins & C-stat. reduced & net cts & SBR & \nh\ \\
arcmin & - & - & - &  $10^3$  & - & $10^{22}$ cm$^{-2}$\\
  \hline			
0.00 - 1.05 & 2746 & 2603 & 1.05 & 170 & 85 & 0.075\\
1.05 - 1.63 & 2698 & 2591 & 1.04 & 155 & 58 & 0.078\\
1.63 - 2.18 & 2748 & 2552 & 1.08 & 148 & 42 & 0.081\\
2.18 - 2.74 & 2773 & 2575 & 1.08 & 145 & 32 & 0.077\\
2.74 - 3.32 & 2767 & 2484 & 1.11 & 131 & 24 & 0.078\\
3.32 - 3.98 & 2688 & 2573 & 1.05 & 133 & 17 & 0.081\\
3.98 - 4.65 & 2807 & 2582 & 1.09 & 134 & 14 & 0.079\\
4.65 - 5.37 & 2912 & 4005 & 0.73 & 131 & 11 & 0.075\\
5.37 - 6.14 & 2666 & 2387 & 1.13 & 112 & 8.6 & 0.074\\
6.14 - 6.95 & 2811 & 2481 & 1.13 & 101 & 6.8 & 0.077\\
6.95 - 7.83 & 3157 & 4949 & 0.64 & 92 & 5.1 & 0.074\\
7.83 - 8.85 & 3305 & 3866 & 0.85 & 89 & 3.6 & 0.073\\
8.85 - 10.05 & 3697 & 6052 & 0.61 & 82 & 2.5 & 0.074\\
10.05 - 11.51 & 4514 & 3868 & 1.17 & 80 & 1.7 & 0.076\\
11.51 - 13.10 & 4870 & 3583 & 1.36 & 62 & 1.2 & 0.079\\
13.10 - 15.18 & 4893 & 3494 & 1.40 & 46 & 0.9 & 0.077\\
15.18 - 17.70 & 2808 & 1844 & 1.52 & 20 & 0.9 & 0.121\\
17.70 - 20.63 & 2632 & 2175 & 1.21 & 19 & 0.6 & 0.101\\
20.63 - 24.08 & 2098 & 1916 & 1.09 & 12 & 0.4 & 0.113
\end{tabular}
}
\caption{Statistical results of the fitting in the annular regions, with radial extension, C-statistic, number of spectral bins, reduced C-statistic indicated, net number of photons in the energy band [0.5-11.3] keV, signal to background ratio, and best fit \nh.}
\label{table:cashstat}
\end{table}

In Table~\ref{table:cashstat}, we show the spectral fit results in the analysis described in Section~2, indicating the radial extension of the chosen annuli, the C-statistic, the number of the spectral bins, and the reduced C-statistic. We point out that this last quantity is always order of 1, implying high goodness in the fit.

\begin{figure}[t]
\includegraphics[width=0.5\textwidth]{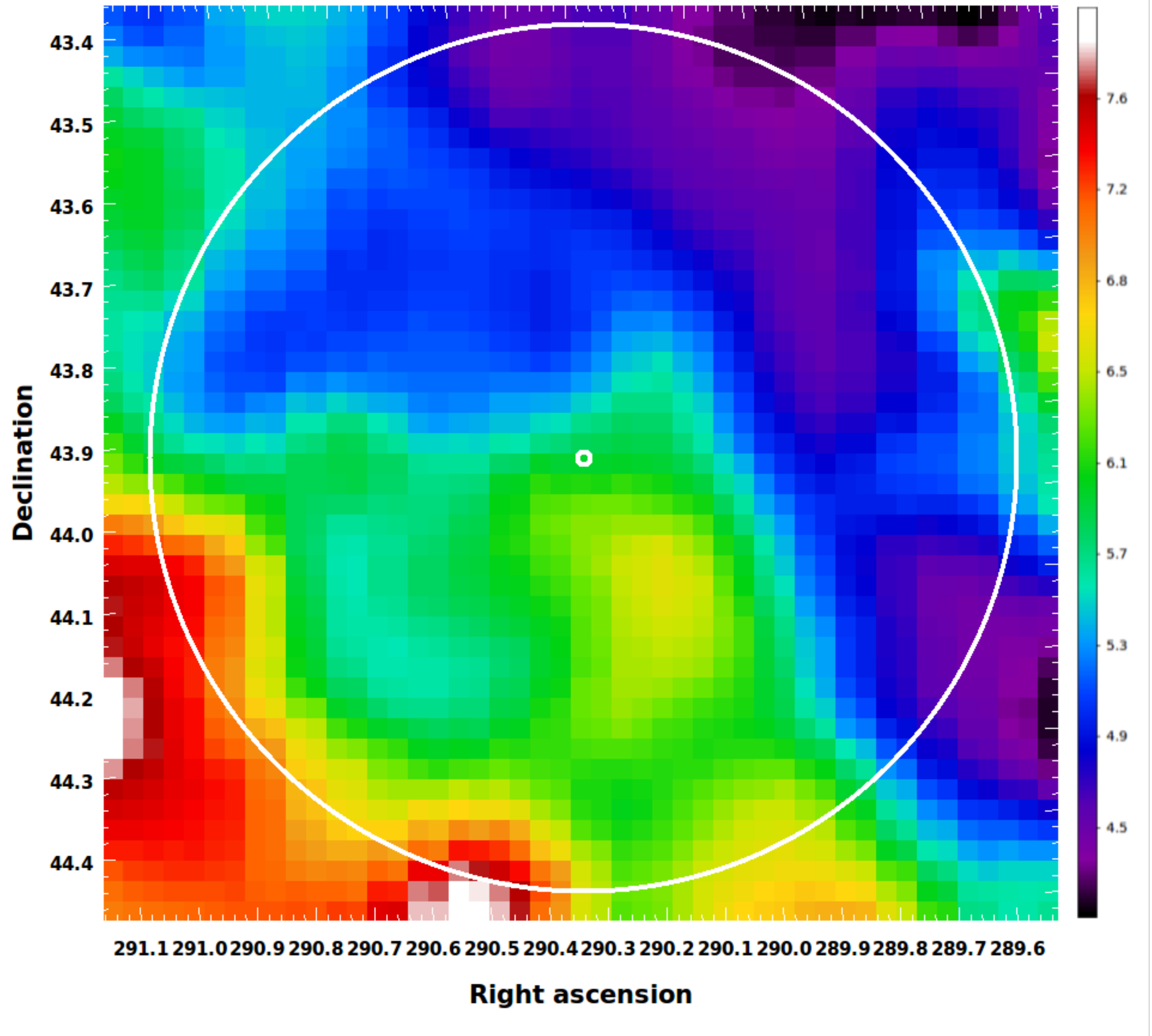}
\caption{IRAS map (minimum--maximum values in the region within $R_{200}$ are 4.22, 7.77 MJy/sr). 
The white external circle represents the location of $R_{200}$, while the small one represents the location of the center of the cluster.}
\label{fig:dust}
\end{figure}

Since A2319 is located at low galactic latitude, $b = +13.5^{\circ}$, 
the choice to leave free \nh\ to vary is reinforced from the azimuthal variation over the cluster's region of the dust emission 
as mapped at 100 $\mu$m by the InfraRed Astronomical Satellite (IRAS; see Fig.~\ref{fig:dust}).
The map shows that the sectors 5, 6, and 7 are the ones expected to have higher Galactic absorption. 
{Indeed the \nh\ in the 8 considered sectors varies according to Table~\ref{nh}, with sector 5, 6, and 7 being $\sim$10\% above the other sectors.

\begin{table}[h]
\resizebox{0.5\textwidth}{!} {
\begin{tabular}{c | c | c | c | c | c | c | c | c }
Sector & 1 & 2 & 3 & 4 & 5 & 6 & 7 & 8\\
  \hline	
\nh\ [$10^{20} cm^{-2}$] & 7.65 & 7.20 & 7.62 & 7.87 & 8.39 & 8.57 & 8.41 & 7.99  \\
\end{tabular}
}
\caption{Best fit \nh\ in the 8 sectors considered.}
\label{nh}
\end{table}
}

\section{Comparison with \chandra\ data}

\begin{figure}[t]
\includegraphics[width=0.5\textwidth]{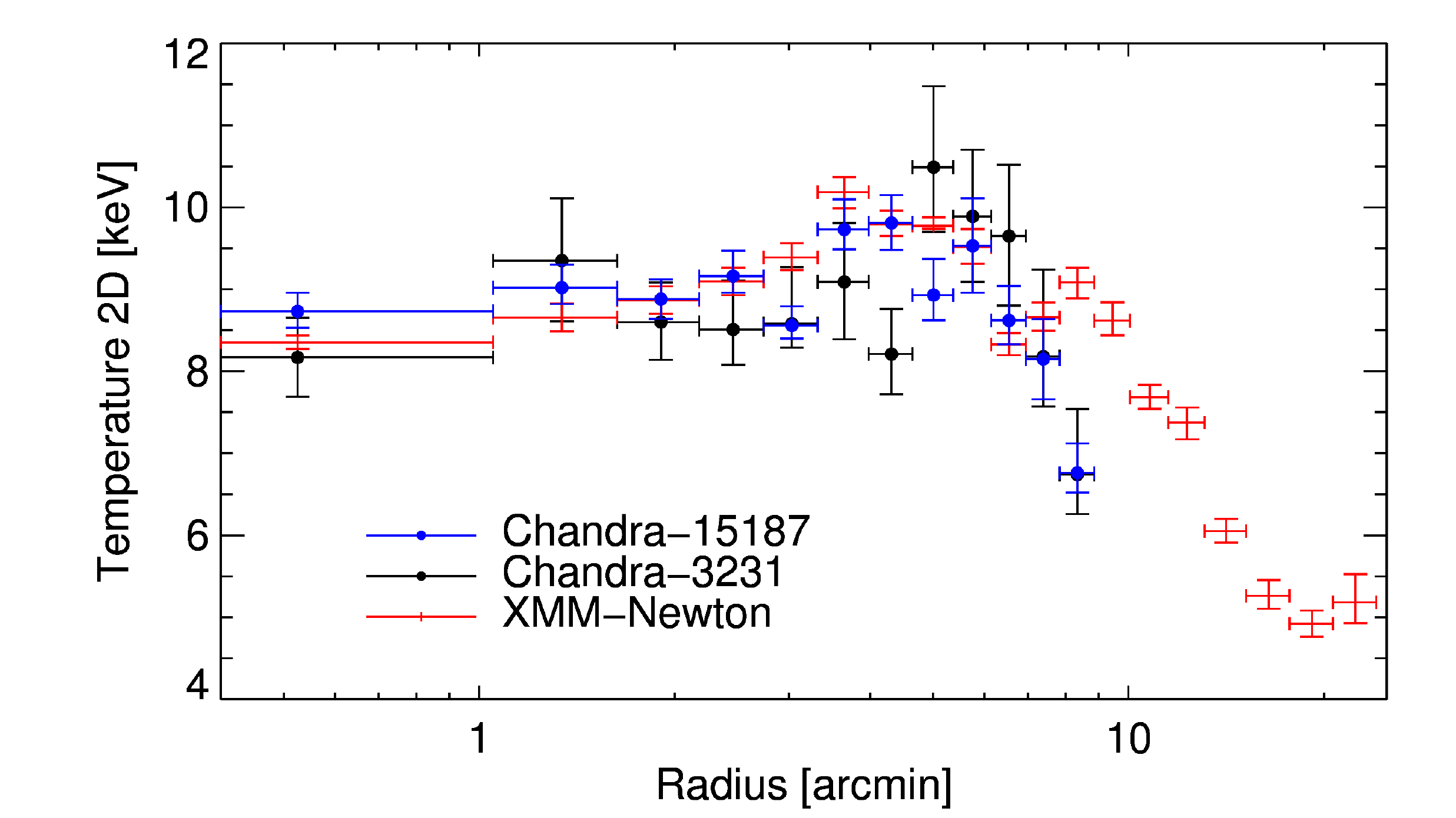}
\caption{Comparison between the spectral temperature obtained using \chandra\ and \xmm. There is a clear excess in the temperature measured by \chandra\ of the order of 2-3 keV up to 7 arcmin.}
\label{fig:cxo}
\end{figure}

We have analyzed two archival \chandra\ observations of the inner region of A2319 (OBSID 15187, with a cleaned exposure time of 75 ksec, and OBSID 3231, with 15 ksec). 
We have processed the two \chandra\ ACIS-I observations of A2319 with a standard pipeline based on CIAO 4.9
\citep{fruscione+06} and CALDB 4.7.4 to create a new events-2 file which includes filtering for grade, status, bad pixels and time intervals for anomalous background levels. The background is estimated through blank sky observations. 
We have extracted the spectra in the same annular regions as for \xmm, and fit them in the identical way, leaving the galactic column density \nh\ 
free to vary within the range $7-13 \times 10^{20}$ cm$^{-2}$ . The  temperature profiles are compared in Fig.~\ref{fig:cxo}. We observe a good agreement among these spectral measurements, despite the claimed, and still debated, cross-calibration issue between \chandra\ ACIS and \xmm\ EPIC  \citep[see e.g.][]{schellenberger+15}, in particular in very hot systems (T>5 keV) as A2163. 
We suggest that leaving free \nh\ plays a determinant role in adjusting the relative impact of the soft part of the spectra, where most of the observed systematic tension has been reported.  
In the present case, \chandra\ prefers systematically higher values of \nh\ ($\sim 1.2-1.3 \times 10^{21}$~cm$^{-2}$) than \xmm\ (see Tab.~\ref{table:cashstat}) 
in all the radial bins. These higher values are more in agreement with the column density corrected for molecular hydrogen as suggested in \cite{willingale+13}.

\section{Likelihood for the mass reconstruction}
\label{sec:likelihood}

We fit our thermodynamic quantities using the MCMC code \textit{emcee} \citep{emcee}, for which we define a likelihood.
We included in the fitting procedure an intrinsic scatter, which is added quadrature on the error of logarithm of pressure such that 
$\log P \sim \log P \pm \sigma_{int} $. By assuming a small value for $\sigma_{int}$ we can write 
\begin{equation*}
\sigma_{P, int} \approx \frac{P \cdot \exp (+\sigma_{int})-P \cdot \exp (-\sigma_{int})}{2} = P \cdot \sinh \sigma_{int}
\end{equation*}
Summed to the covariance matrix as follows:

\begin{equation*}
\begin{split}
\Sigma_{tot} = &
\begin{bmatrix}
    \Sigma_{11} & \Sigma_{12} & \Sigma_{13} & \dots  & \Sigma_{1n} \\
    \Sigma_{21} & \Sigma_{22} & \Sigma_{23} & \dots  & \Sigma_{2n} \\
    \vdots & \vdots & \vdots & \ddots & \vdots \\
    \Sigma_{n1} & \Sigma_{n2} & \Sigma_{n3} & \dots  & \Sigma_{nn}
\end{bmatrix} \\
& +
\begin{bmatrix}
   \sigma_{P_1, int}^2 & 0 & 0 & \dots  & 0 \\
    0 & \sigma_{P_2, int}^2 & 0 & \dots  & 0 \\
    \vdots & \vdots & \vdots & \ddots & \vdots \\
    0 & 0 & 0 & \dots  & \sigma_{P_n, int}^2
\end{bmatrix}
\end{split}
\end{equation*}
where $\Sigma_{i,j}$ is the covariance matrix on the measured \textit{Planck} pressure profile.

The intrinsuc scatter is propagated also to the variance on temperature profile, added in quadrature to the measured errors:
\begin{equation*}
\sigma_{tot}^2 = \sigma_{T}^2 + \sigma_{T,int}^2
\end{equation*}
with
\begin{equation*}
\sigma_{T,int} = \frac{P_{model}}{n_{model}} \sigma_{P,int} = T_{model} \cdot \sigma_{P,int}  
\end{equation*}

We remind that in general the likelihood is defined as:
\begin{equation*}
\mathcal{L} = \frac{1}{\sqrt{2\pi \sigma^2} } \exp ( -\chi^2/2 )
\end{equation*}
so that 
\[ \log \mathcal{L} =  -0.5 (\chi^2 + \log \sigma^2 +\log(2\pi)) \]
where the last term is a constant and therefore is usually ignored while maximizing the likelihood, but the term with $\log \sigma^2$ is not.
Finally, by using the subscript ``m'' or ``o'' to describe model predicted or observed quantities respectively, we can explicitly write the logarithm of the likelihood we use to fit:

\begin{equation*}
\begin{split}
\log \mathcal{L} = & 
-0.5 \left[ (P-P_{\rm{m}}) \Sigma_{tot}^{-1} (P-P_{\rm{m}})^T + n \log \left( \det \left( \Sigma_{tot} \right) \right) \right]
\\
 & -0.5  \sum_{i=1}^n \left[ \frac{(T_i-T_{\rm{m},i})^2}{\sigma_{T,i}^2+\sigma_{T,int}^2} + \log	\left( {\sigma_{T,i}^2+\sigma_{T,int}^2} \right) \right] 
\\ 
 &  -0.5 \left[ \sum_{i=1}^n \frac{(\epsilon-\epsilon_{\rm{m},i})^2}{\sigma_{\epsilon,i}^2} \right]
\end{split}
\end{equation*}

We point out that this method is independent on how $P_{model}$ and $T_{model}$ are computed, meaning that this kind of approach is valid both for the \textit{forward} and \textit{backward} methods.

\section{Thermodynamic quantities in azimuthal sectors}

\begin{figure}[h]
\includegraphics[width=0.5\textwidth]{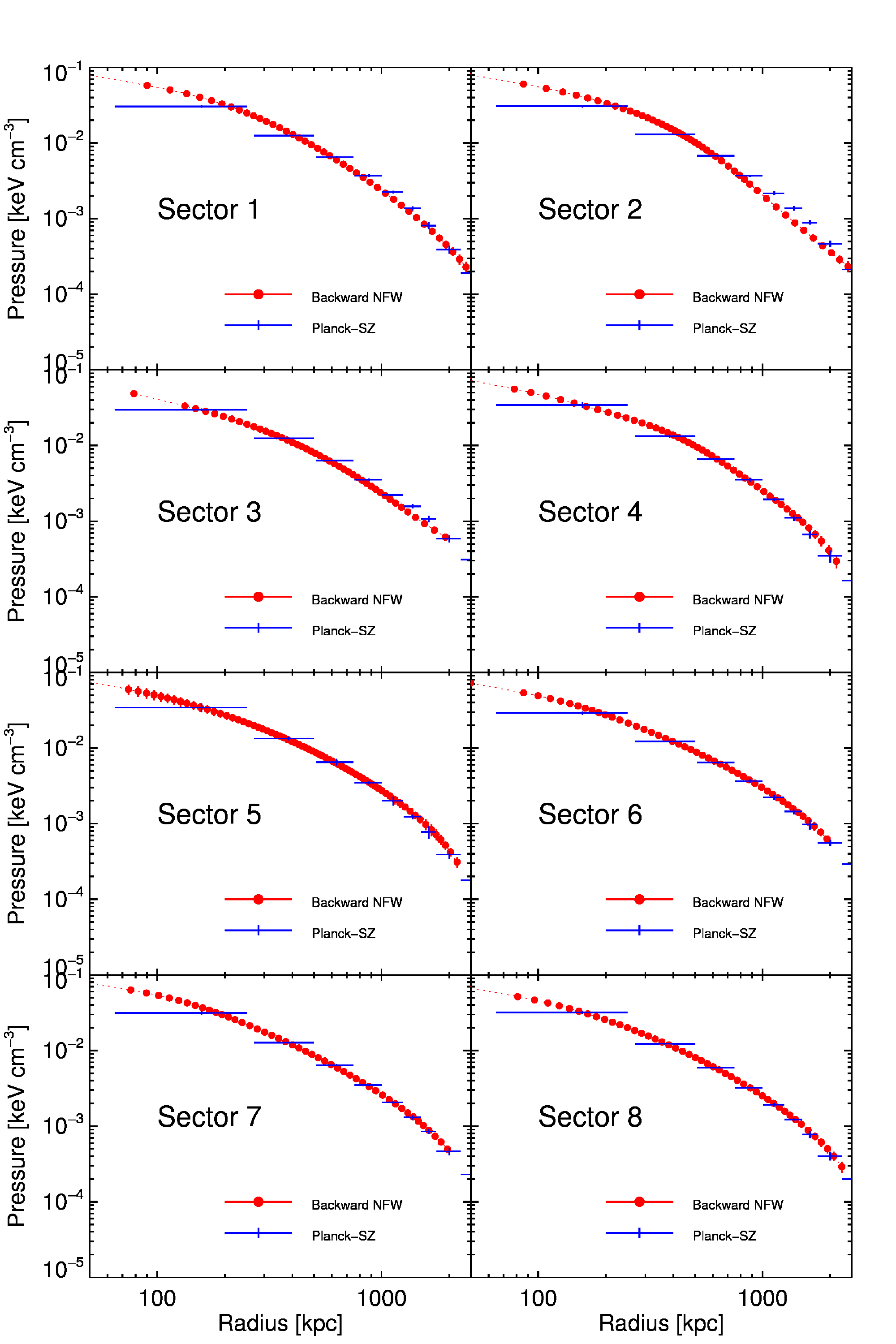}
\caption{Comparison of the observed pressure profile with the one reconstructed by the NFW \textit{backward} best fit.}
\label{fig:confrontoP}
\end{figure}

\begin{figure}[h]
\includegraphics[width=0.5\textwidth]{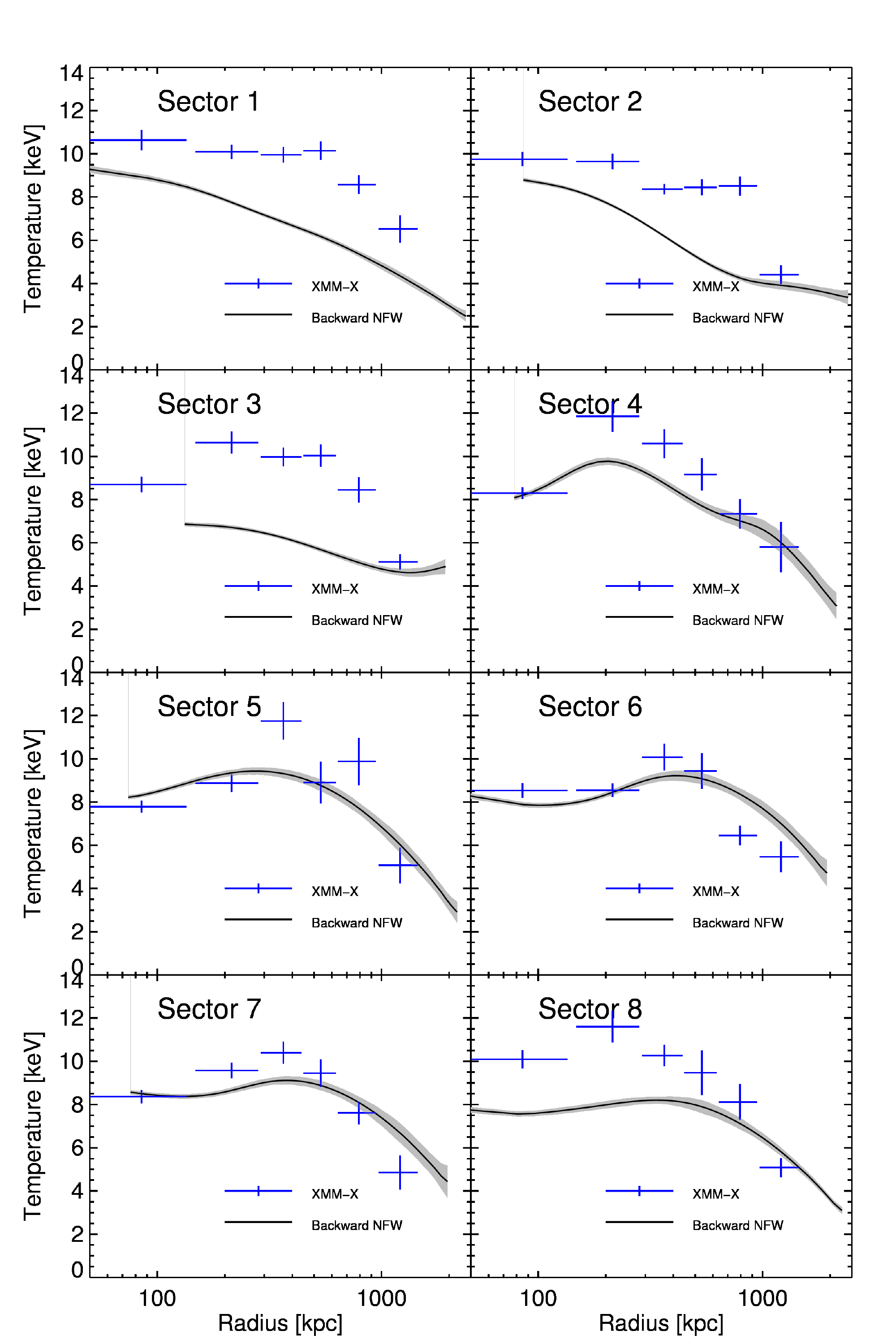}
\caption{Comparison of the observed 2-dimensional temperature profile with the one reconstructed by the NFW \textit{backward} best fit.}
\label{fig:confrontoT}
\end{figure}

The procedure described in Sections 2 to 4 are applied on each azimuthal sector. In summary we deproject surface brightness into density using the multiscale techniqe on the mean profile, we deproject comptonization parameter to retrieve pressure, and we calculate the temperature in 6 spectral annuli. We then apply the \textit{backward} approach on these thermodynamic quantities, in order to find the parameters of a NFW mass model which best reproduce the observables. We compare the observed and reconstructed from the best-fit mass model pressure and temperature profiles sector by sector in Fig.~\ref{fig:confrontoP} and \ref{fig:confrontoT}, respectively.
We observe that the only sectors with an evident discrepancy are the one disturbed the most by the merger event, i.e. Sector 1, 2 and 3.

\begin{figure}[h]
\includegraphics[width=0.5\textwidth]{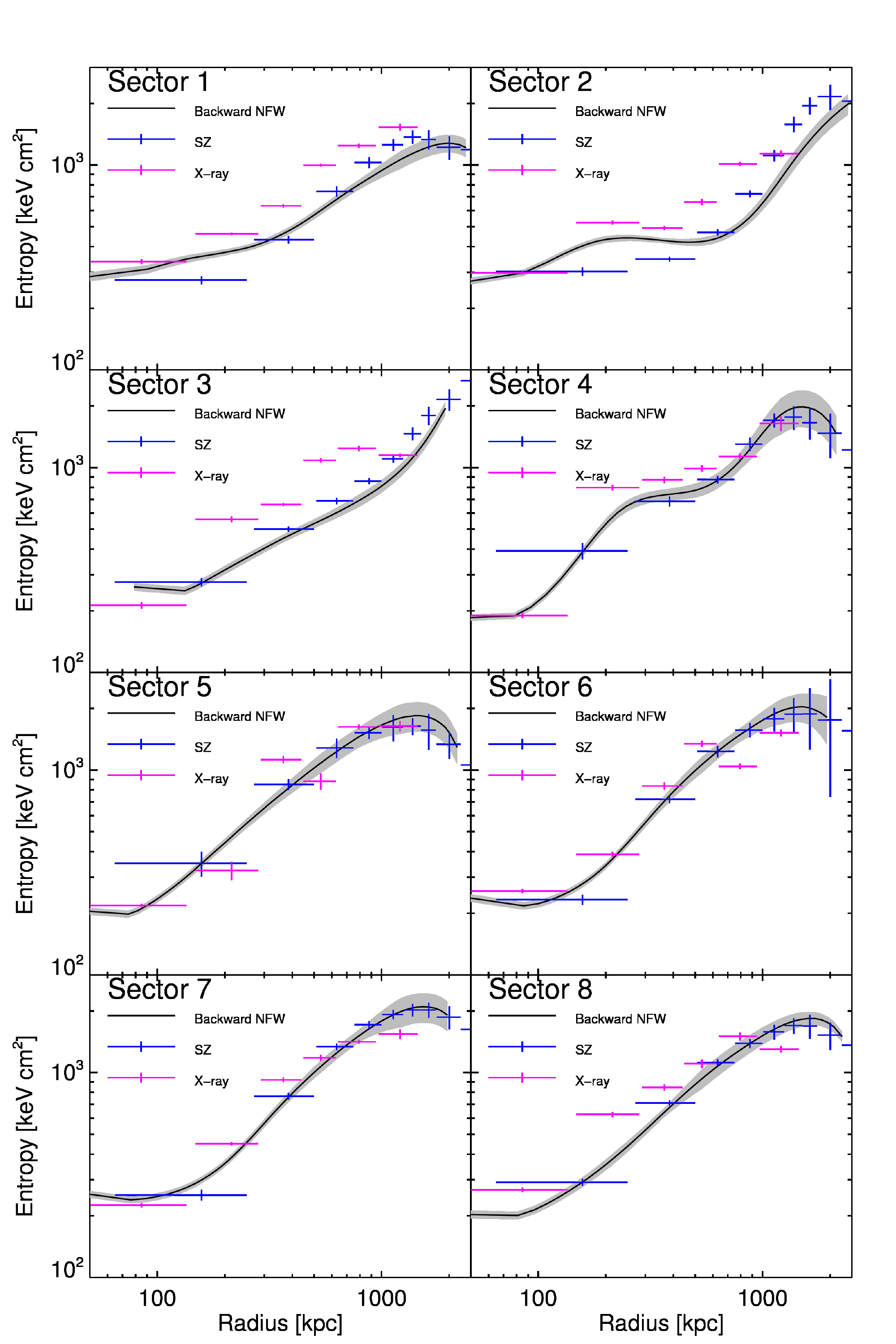}
\caption{Comparison between the entropy profile reconstructed by the NFW \textit{backward} best fit with the entropy coming from the combination of X-ray and SZ and just using X-ray spectral results.}
\label{fig:confrontoK}
\end{figure}

Similarly to what has been done in Section  \ref{secK}, we compare the entropy profile reconstructed by the NFW \textit{backward} best fit, with the entropy recovered from X-ray spectroscopy ($K = kT/n_e^{2/3}$), and with the entropy recovered by combining X-ray density and SZ pressure ($K = P/n_e^{5/3}$); this comparison sector by sector is shown in Fig.~\ref{fig:confrontoK}.

\clearpage
\newpage
\newpage

\end{appendix}
\end{document}